\documentclass[a4paper,11pt]{article}
\pdfoutput=1 % if your are submitting a pdflatex (i.e. if you have
\usepackage{jheppub} % for details on the use of the package, please
\usepackage[T1]{fontenc} % if needed

\allowdisplaybreaks[1]	% So the very long equations can be split in different pages
\def\eqref#1{{Eq.\!~(\ref{#1})}} %References to equations
\def\figref#1{{Fig.\!~\ref{#1}}} %References to figures
\def\secref#1{{Sec.\!~\ref{#1}}} %References to sections
\usepackage{scalerel}  % To change size of subscripts and such
\usepackage{subfigure} % To add subfigures
\usepackage{bbold} %Identity symbol
\usepackage{mciteplus} %Citations
\newcommand{\pperp}{\mathbin{\mathchoice
  {\xperp\scriptstyle}
  {\xperp\scriptstyle}
  {\xperp\scriptscriptstyle}
  {\xperp\scriptscriptstyle}
}}
\newcommand{\xperp}[1]{\vcenter{\hbox{$#1\perp$}}}
\usepackage{mathtools}

\title{\boldmath Initial correlations of the Glasma energy-momentum tensor}

\author[a]{Javier L. Albacete,}
\author[a,b]{Pablo Guerrero-Rodríguez,}
\author[b]{and Cyrille Marquet}

\affiliation[a]{CAFPE \& Dpto. de F\'{\i}sica Te\'orica y del Cosmos, \\ Universidad de Granada, E-18071 Campus de Fuentenueva, Granada, Spain.}
\affiliation[b]{Centre de Physique Th\'eorique, \'Ecole Polytechnique, \\
CNRS, Universit\'e Paris-Saclay, F-91128 Palaiseau, France.}

\emailAdd{albacete@ugr.es}
\emailAdd{pgr@ugr.es}
\emailAdd{cyrille.marquet@polytechnique.edu}

\abstract{

We present an analytical calculation of the covariance of the energy-momentum tensor associated to the gluon field produced in ultra-relativistic heavy ion collisions at early times, the Glasma. This object involves the two-point and single-point correlators of the energy-momentum tensor ($\langle T^{\mu\nu}(x_{\pperp})T^{\sigma\rho}(y_{\pperp})\rangle$ and $\langle T^{\mu\nu}(x_{\pperp})\rangle$, respectively) at proper time $\tau\!=\!0^+$. Our approach is based on the Color Glass Condensate effective theory, which allows us to map the fluctuations of the valence color sources in the colliding nuclei to those of the energy-momentum tensor of the produced gluon fields via the solution of the classical equations of motion in the presence of external currents. The color sources in the two colliding nuclei are characterized by Gaussian correlations, albeit in more generality than in the McLerran-Venugopalan model, allowing for non-trivial impact parameter and transverse dependence of the two-point correlator. We compare our results to those obtained under the Glasma Graph approximation, finding agreement in the limit of short transverse separations. However, important differences arise at larger transverse separations, where our result displays a slower fall-off than the Glasma Graph result ($1/r^{2}$ vs.$\!$ $1/r^{4}$ power-law decay), indicating that the color screening of the correlations in the transverse plane occurs at distances larger than $1/Q_s$ by a logarithmic factor sensitive to the infrared. In the Glasma flux tube picture, this implies that the color domains are larger than originally estimated.

}

\begin{document} 
\maketitle
\flushbottom
%To have slightly smaller spaces before and after equations:
%\setlength{\abovedisplayskip}{9pt} 
%\setlength{\belowdisplayskip}{9pt}

%\begin{figure}[tbp]
%\centering % \begin{center}/\end{center} takes some additional vertical space
%\includegraphics[width=.45\textwidth,trim=0 380 0 200,clip]{img1.pdf}
%\hfill
%\includegraphics[width=.45\textwidth,origin=c,angle=180]{img2.pdf}
% "\includegraphics" is very powerful; the graphicx package is already loaded
%\caption{\label{fig:i} Always give a caption.}
%\end{figure}

\section{Introduction}
\label{sec:intro}

Understanding the dynamical features of the matter produced in the early stages of heavy ion collisions and its eventual thermalization into a Quark Gluon Plasma (QGP from now on) is one of the most pressing questions in the field of heavy ion collisions, both at the experimental and theoretical levels.

%The experimental effort since the beginning of the research program of the Relativistic Heavy Ion Collider (RHIC) \cite{...} down to the most recent heavy ion experiments performed at the Large Hadron Collider (LHC) \cite{...} has given rise to an extensive body of highly complex data that will keep growing in the upcoming years.

The study of the correlations between the detected particles plays a main role for the understanding of the problem of QGP formation --and its characterization-- in heavy ion collisions, since non-trivial correlations provide a clear indication of the collective behavior of the produced medium. However, it has also become clear over the last years that the observed correlations reflect as much the collective dynamics of the produced medium as they do the initial state correlations, namely those dynamically generated during the early stages of the collision (before an eventual thermalization of the system) or already built in the wave function of the colliding nuclei (see e.g. $\!$\cite{0954-3899-41-6-063102}). This observation relates to the very small ratio of viscosity over entropy density extracted from hydrodynamical simulations or, equivalently, to the low dissipation of the dynamics mapping early and late times of the collision \cite{doi:10.1146/annurev-nucl-102212-170540}. Therefore, a detailed and theoretically robust characterization of initial state correlations is mandatory for a proper understanding of the medium transport and dynamical properties. 

Indeed, the existence in the literature of a broad variety of phenomenological models for the description of the initial stages of heavy ion collisions reflects the importance of this kind of studies (for a review see e.g. $\!\!$\cite{ALBACETE20141}). The main practical use of such models is to generate initial conditions of the energy density and velocity profiles for further evolution of the system, typically described by quasi-ideal relativistic hydrodynamics during the QGP phase followed by kinetic transport during the hadronic afterburner. All such models allow for fluctuations of the energy and momentum deposited in the collision area. The dynamical origin and practical description of such fluctuations vary largely from model to model --from the positions of the nucleons in the transverse plane at the collision time to fluctuations of the sub-nucleonic degrees of freedom and their inelasticity density-- and are, in general, subject to a large degree of phenomenological modeling.  Clearly, a higher degree of theoretical control on the description of the initial collision profile is desirable, and adding to it is precisely the main goal of this work.  

The classical approach that we shall follow is embodied in the Color Glass Condensate (CGC) effective theory (see e.g. $\!$\cite{Weigert:2005us,Gelis:2010nm} for a review), arguably the most complete theoretical framework for the description of the early time dynamics in heavy ion collisions. The CGC describes the high density of small-$x$ gluons carried by nuclei as strong color fields whose dynamics obey the classical Yang-Mills equations. The classical approximation is based on the fact that for very large occupation numbers the quantum fluctuations represent a negligible correction to the strong background field. Quantum corrections are incorporated in the CGC framework via the JIMWLK renormalization group equations \cite{Jalilian-Marian:1997gr,Jalilian-Marian:1997dw,Kovner:2000pt,Weigert:2000gi,Iancu:2000hn,Ferreiro:2001qy,Balitsky:1996ub,Kovchegov:1999yj}. They ensure that the physical observables are independent of the arbitrary longitudinal momentum scale at which the separation between slow (dynamical) and fast (static sources) degrees of freedom, on which the CGC effective theory is build up, is performed. 

The properties of the medium produced in heavy ion collisions at early times, dubbed as Glasma, have been extensively studied in a series of works in the CGC framework \cite{LAPPI2006200,LAPPI200611,DUMITRU200891,FUKUSHIMA2012108}. This kind of studies start by solving the classical equations of motion for the produced gluon field in the presence of two external color sources --the valence degrees of freedom of the two colliding nuclei--. The picture that emerges is that of the Glasma as a strongly correlated, maximally anisotropic system dominated by strong classical fields. The fact that the chromo-electric and magnetic fields are parallel to the collision axis immediately after the collision leads to a very peculiar form for the energy-momentum tensor \cite{LAPPI2006200},
$T^{\mu\nu}_{\scaleto{0}{3.8pt}}\!=\![\text{diag}(\epsilon_{\scaleto{0}{3.8pt}}, \epsilon_{\scaleto{0}{3.8pt}}, \epsilon_{\scaleto{0}{3.8pt}}, -\epsilon_{\scaleto{0}{3.8pt}})]^{\mu\nu} $, where $\epsilon_{\scaleto{0}{3.8pt}}$ is the initial average energy density. The most striking feature is that the longitudinal pressure is negative, reminiscent of strings, or flux tubes, stretching in the longitudinal direction. This picture is reinforced by the observation that the correlations of the classical fields extend over long rapidities in the longitudinal direction. In turn, correlations on the plane transverse to the collision axis are expected to be short-range --parametrically of the order of the inverse of the saturation scale $\sim\!1/Q_s$, much smaller than the nucleon size-- since color charges in the projectile (or target) are correlated only over this typical distance, which effectively plays the role of the scale for color neutrality in the nuclear wave function. In this work we shall provide explicit results to quantify the size and extent of the transverse correlations, not entirely supporting the qualitative expectations on its short-range character. How the above-described coherent ensemble of classical flux tubes decays and whether it eventually thermalizes into a QGP is a subject of open debate and intense investigation over the last years and is beyond the scope of this work. For a review we refer the reader to \cite{doi:10.1142/S0218301315300088} or to the more recent works on the matching of the CGC description with effective kinetic theory as an intermediate dynamical step before the hydrodynamization of the system (e.g. $\!$\cite{Kurkela:2018wud}).

Rather, our goal in this work is to further explore the properties of the strictly classical Glasma dynamics by presenting a first analytical calculation of the two-point correlator of the energy-momentum tensor of the Glasma fields right after the collision time. We start from the assumption that the relevant correlations among the fast color charges in the wave function of the colliding nuclei are known. Then we calculate how the collision dynamics, described under the classical approximation, maps such correlations onto correlations of the energy-momentum tensor of the produced gluon field right after the collision. Hence, the only source of fluctuations in our approach is that of the incoming color sources, since the collision dynamics are fully deterministic in the leading-order classical approach. Detailed knowledge about them or, more generally, about the wave function of the colliding nuclei, can be obtained either at dedicated experiments like the proposed Electron Ion Collider \cite{Accardi:2012qut} or, in the absence of direct empiric data, via theoretical modeling sustained by the abundant empiric information on the proton partonic structure provided by the HERA experiment (see \cite{ALBACETE20141}).

Specifically, we perform the analytical calculation of the following covariance:
\begin{equation}\label{Tmunu0}
\text{Cov}[T^{\mu\nu}](\tau=0^+;x_{\pperp},y_{\pperp})\equiv\langle T_{\scaleto{0}{3.8pt}}^{\mu\nu}(x_{\pperp})T_{\scaleto{0}{3.8pt}}^{\sigma\rho}(y_{\pperp})\rangle\!-\!\langle T_{\scaleto{0}{3.8pt}}^{\mu\nu}(x_{\pperp})\rangle\langle T_{\scaleto{0}{3.8pt}}^{\sigma\rho}(y_{\pperp})\rangle,
\end{equation}
where $T^{\mu\nu}_{\scaleto{0}{3.8pt}}\!(x_{\pperp})$ is the energy-momentum tensor (EMT) associated to the gluon field produced over an infinitesimal positive proper time after two heavy ion nuclei with mass numbers $A_{\scaleto{1}{4.2pt}}$, $A_{\scaleto{2}{4.2pt}}$ collide at relativistic speed. We rely on an extended version of the McLerran-Venugopalan (MV) model for the description of the valence color sources of the colliding nuclei \cite{PhysRevD.49.2233}, whereby we assume that they obey Gaussian statistics, as in the original MV model, but we allow for a more general form of the two-point correlator, $\langle\rho^a(x^-\!,x_{\pperp})\rho^b(y^-\!,y_{\pperp})\rangle$, in order to expand the possibilities for phenomenological applications. Our specific modifications consist of relaxing the assumption of local transverse interactions, as well as including an explicit impact parameter dependence that allows the possibility of describing finite, non-homogeneous nuclei. However, for the sake of simplicity, in some sections we shall discuss our results in terms of the original MV model.

Following the approach outlined above, and despite the complexity of the calculation and of the full result, we obtain a remarkably compact expression for the covariance of the EMT in the limit of large transverse separations, $rQ_s\!\gg\!1$ with $r \equiv |x_{\pperp}\!- y_{\pperp}|$:
\begin{equation}\label{ad1}
\lim_{rQ_s\gg 1}\!\text{Cov}[T^{00}_{\scaleto{\text{MV}}{0.12cm}}](0^+;x_{\pperp},y_{\pperp})=\frac{2 \left(N_c^2-1\right)\!\left(4\pi\,\partial^2L(0_{\pperp})\right)^2\!\left(\bar{Q}_{s\scaleto{1}{4pt}}^4Q_{s\scaleto{2}{4pt}}^2+\bar{Q}_{s\scaleto{2}{4pt}}^4Q_{s\scaleto{1}{4pt}}^2\right)}{g^4N_c^2\,r^2} .
\end{equation}
The factors $Q_{s\scaleto{1}{4pt},\scaleto{2}{4pt}}(r_{\pperp},b_{\pperp})$ and $\bar{Q}_{s\scaleto{1}{4pt},\scaleto{2}{4pt}}(b_{\pperp})$  --two definitions of the saturation scales characterizing each nuclei-- will be introduced later along with the factor $L(0_{\pperp})$. \eqref{ad1} is one of the most important results of the paper, as it could challenge the conjectured physical picture of Glasma flux tubes or color field domains \cite{DUMITRU200891} --which basically states that when two sheets of CGC pass through each other, color flux tubes of transverse size $1/Q_s$ are created. In our result, although the transverse correlation length is parametrically of the order of $1/Q_s$, the correlations decrease only according to a $1/r^2$ power-law tail at large distances, extending further in the transverse plane than what was indicated by previous calculations. Such slowly vanishing covariance could potentially have an impact in both physical interpretations and numerical results for any observable built from this quantity.

For instance, the 2-dimensional transverse integral of \eqref{ad1} will be dominated by the upper bound (the infrared cut-off $r\!\sim\!1/m$) rather than the lower bound $r\!\sim\!1/Q_s$, which is what happens under the Glasma Graph approximation \cite{PhysRevD.97.034034} (that features a $1/r^4$ fall-off as we will discuss later), or even in the case of a more naive exponential fall-off. This indicates that the range of the transverse color screening of the correlations, which determines the size of the color domains in the interaction region, is actually bigger than $1/Q_s$, as it features a logarithmic enhancement $\ln(Q_s/m)$ sensitive to the infrared. Similar observations were made in \cite{Lappi:2009xa} in the context of two-particle correlations: a sensitivity of the color domain size to the infrared was observed numerically, with it getting larger as the infrared cut-off was decreased. In the case of EMT correlations, our qualitative discussion also remains to be quantified with numerical calculations.

As an input to hydrodynamical simulations, \eqref{ad1} also has important implications. Indeed, neglecting logarithmic dependencies, we can write
\begin{equation}\label{ad2}
\frac{1}{\langle T^{00}_{\scaleto{\text{MV}}{0.12cm}}(0^+,x_{\pperp})\rangle}
\int d^2 r_{\pperp} \text{Cov}[T^{00}_{\scaleto{\text{MV}}{0.12cm}}](0^+;x_{\pperp},x_{\pperp}-r_{\pperp})
\simeq\frac{\bar{Q}_{s\scaleto{1}{4pt}}^2(x_{\pperp})+\bar{Q}_{s\scaleto{2}{4pt}}^2(x_{\pperp})}{\alpha_s N_c}\ .
\end{equation}
In Monte Carlo Glauber models, where eccentricity fluctuations are created by uncorrelated, small-scale fluctuations in the transverse plane, this quantity is taken as a constant proportional to $\int d^2 x_{\pperp} T^{00}_{\scaleto{0}{4pt}}(x_{\pperp})$ \cite{Blaizot:2014wba}. In our calculation at $\tau\!=\!0^+$, which takes into account sub-nucleonic degrees of freedom (but nevertheless give rise to long-range correlations), that quantity is not a constant. The dimensionless ratio of \eqref{ad2} to the integrated energy density characterizes the strength of the eccentricity fluctuations \cite{BLAIZOT2014166}, and in our calculation is given by $(8\pi/(N_c^2-1))[\bar{Q}_{s\scaleto{1}{4pt}}^2(x_{\pperp})\!+\!\bar{Q}_{s\scaleto{2}{4pt}}^2(x_{\pperp})]/ \int d^2 x_{\pperp} \bar{Q}_{s\scaleto{1}{4pt}}^2(x_{\pperp}) \bar{Q}_{s\scaleto{2}{4pt}}^2(x_{\pperp})$. Therefore, we find this ratio bigger in the middle of the overlap region than near the edge, which brings new insight for the characterization of the initial stage of heavy-ion collisions. This ratio also displays the usual $1/(N^2_c-1)$ suppression characteristic of non-trivial color correlations.

%In the Glasma Graph approximation the  Weizs\" acker-Williams (WW) fields of the colliding nuclei are also assumed to follow Gaussian statistics. In our approach it is only the color sources of the gluon fields that obey Gaussian statistics, while the gluon fields and their correlators follow from the solution of the Yang-Mills equation in the presence of the external sources. The Glasma Graph approximation simplifies notably the color structure of the calculation, in particular it allows to bypass the rather involved calculation of the correlator of four Wilson lines in the adjoint representation presented in Appendix \ref{Kov}, but its applicability seems to be justified only in the short-distance limit. At longer distances, where nonlinearities relating sources and fields become dominant, we find important differences with the full result. 

In more general terms, the calculation presented in this work provides further analytical insight to the dynamics of the classical fields produced in relativistic heavy ion collisions, otherwise also accounted for in the well-known IP-Glasma model \cite{PhysRevC.86.034908,PhysRevLett.108.252301} and related numerical methods, where the classical equations of motion that we discuss here are solved numerically to higher proper times $\tau\!>\!0^+$. However, counting with exact analytical expressions for the description of the initial state could simplify to a large extent the phenomenological analyses of data by reducing the amount of numerical work. 
Our result could be directly applied, for instance, in the multi-parametric fits based on Bayesian statistics aimed to determine the medium properties \cite{PhysRevC.94.024907}. Also, upon the proper spectral decomposition, they may allow to perform mode-by-mode studies of the hydrodynamical propagation of the initial fluctuations as was proposed in \cite{PhysRevC.88.044906,FLOERCHINGER2014407}, or be used to determine the initial eccentricities fluctuations as proposed in \cite{BLAIZOT2014166}.

Our paper is organized as follows. In \secref{sec:setup} we introduce a generalization of the MV model with relaxed transversal locality and explicit impact parameter dependence. In this framework we outline the solution to the Yang-Mills equations with one and two sources at an infinitesimal proper time after the collision $\tau\!=\!0^+$, which acts as boundary condition for the ensuing evolution in the future light-cone. In \secref{sec:t} we calculate the EMT correlator in the previously presented framework. In \secref{sec:tt} we compute the correlator of two EMTs. Using the results of these two sections, we calculate the covariance of the EMT and show the first orders of its $N_c$-expansion, as well as the strict MV model limit. Our final expression for the EMT covariance is presented in \eqref{FullResult}, the main result of this work. We also compare our results with the previously mentioned computation, performed under the Glasma Graph approximation \cite{PhysRevD.97.034034}. Remarkably, throughout this calculation we face a number of outstanding technical challenges such as the calculation of non-trivial projections of the correlator of four Wilson lines in the adjoint representation and the decomposition of correlators of $m$ color sources and $n$ Wilson lines. We analyze these problems in depth on appendices \ref{BigCorrelator} and \ref{Kov}. Finally, in \secref{sec:end} we discuss the physical implications and phenomenological applications of our result, as well as its role in future works.

\section{The classical approach to gluon production in heavy ion collisions}
\label{sec:setup}
In the following section we compute the gluon field generated in ultra-relativistic heavy ion collisions. Although this calculation has been done previously in the literature, we deem it convenient to include this preface as it allows us to introduce our modifications to the MV model and establish the notation used in the rest of the paper. We will follow the derivation steps first presented in \cite{PhysRevD.52.3809}.

In the MV model we represent the high density of small-$x$ gluons carried by each nuclei with gauge fields $A^{\mu}_{\scaleto{1,2}{5.8pt}}(x)$ whose dynamics follow from the classical Yang-Mills equations:
\begin{equation}\label{YMeqs}
\left[ D_{\mu},F^{\mu\nu}\right]=J^{\nu}=J^{\nu,a}\,t^{a}\!.
\end{equation}
The source of the fields is a color current $J^{\nu,a}$ that represents the flow of large-$x$ valence partons. If we assume a nucleus moving in the positive $x^{\scaleto{3}{4pt}}$ direction with a large light-cone momentum $p^+$, we can fix the initial form of $J^{\nu,a}$ based on kinematic considerations:
\begin{equation}
J^{\nu,a}(x^{-},x_{\pperp})=\delta^{\nu+}\rho^{a}(x^{-},x_{\pperp})\approx \delta^{\nu+}\delta(x^-)\rho^{a}(x_{\pperp}),
\end{equation}
where $\rho^a$ is the color charge density. The $\delta^{\nu+}$ factor indicates that the source generates a color current only in the + direction. This suggests a physical picture of the interaction where the fast valence partons do not recoil from their light-cone trajectory as the gluons they continuously exchange with the medium are too soft to affect their motion (eikonal approximation).
%$J^{\nu,a}$ is assumed to be (light-cone) time independent due to the mean lifetime of the emitted small-$x$ gluons being much shorter than that of the valence quarks. These appear to them as forming a static, `frozen' color current.
%Light cone time independent but is only the initial value? no, current conservation implies x^+ independence
As for $\rho^a$, we might factorize its $x^-$ dependence by assuming that the currents are shaped as a Dirac delta on $x^-\!\!=\!0$ (last approximate equality). This approximation is motivated by the Lorentz contraction experienced by the relativistic nuclei%\footnote{In the literature this approximation is also justified with the uncertainty in the position of quarks being of order $\lambda^-\!\!\sim\! 1/p^+$ with large $p^+$.}%Both motivations are correct? They don't seem to be equivalent
. However, we choose not to make any assumptions about the longitudinal structure of the nuclei, thus leaving it undetermined for now.

In the MV model the calculation of gauge-invariant quantities requires performing an average over the background classical field sources, operation that we denote by $\langle ... \rangle$. The physical picture for this average emerges from the process-to-process fluctuations of color charges observed inside the nuclei. We take the spatial configuration of color sources $\rho^a$ as an stochastic quantity with a certain probability distribution $W[\rho]$ associated as weight function. Thus, observables are obtained as expectation values:
\begin{equation}\label{Average}
\langle {\cal O}[\rho] \rangle\!=\frac{1}{{\cal N}}\int[d\rho] W[\rho]{\cal O}[\rho],
\end{equation}
where ${\cal N}$ is a normalization constant equal to $\int[d\rho] W[\rho]$. The main assumption adopted in the MV model is that in nuclei with large mass numbers the valence partons that enter \eqref{YMeqs} through $\rho^a$ emerge from a large number of separate nucleons and therefore are uncorrelated%\footnote{It is assumed that the large transverse momentum scale at which we are probing nuclei results in transverse resolution scales that are very small compared to a fermi. The argument is sometimes extended to nuclei with smaller mass numbers (or even single nucleons) by assuming that on such scales we perceive locally uncorrelated quarks whose charges add together in a random walk in color space.}
. Thus, invoking the central limit theorem, this model approximates $W[\rho]$ with a Gaussian distribution:
\begin{equation}\label{MVAverage}
\langle {\cal O}[\rho] \rangle_{\scaleto{\text{MV}}{0.12cm}}=\frac{\int[d\rho] \exp\left\{ - \int dx^-d^2x_{\pperp} \frac{1}{2\mu^2(x^-)}\text{Tr}\left[\rho^2(x^-,x_{\pperp})\right] \right\} {\cal O}[\rho]}{\int[d\rho] \exp\left\{ - \int dx^-d^2x_{\pperp} \frac{1}{2\mu^2(x^-)}\text{Tr}\left[\rho^2(x^-,x_{\pperp})\right] \right\}}.
\end{equation}
Here $\mu^2(x^-)$ is a parameter proportional to the color source number density that acts as the variance of the Gaussian weight. The main implication of the MV model is the following two-point correlator:
\begin{equation}\label{MV2point}
\langle \rho^{a}(x^-,x_{\pperp})\rho^{b}(y^-,y_{\pperp})\rangle_{\scaleto{\text{MV}}{0.12cm}} = \mu^2(x^-)\delta^{ab}\delta(x^--y^-)\delta^2(x_{\pperp}-y_{\pperp}).
\end{equation}
%Longitudinal position ~ Rapidity (kinematics)
%Because y=0.5*log(p+/p-)=0.5(log(p+)-log(p-))~0.5*log(p+)~0.5*log(1/x-)=-0.5*log(x-)
%which establishes that only interactions with the background field that are local in color, rapidity and transverse position yield a nonvanishing contribution to observables.
However, as we intend to apply a more general approach, we choose to relax some of the approximations implied in \eqref{MV2point} by considering the following, more general, two-point correlator:
\begin{align}
\langle \rho^{a}(x^-,x_{\pperp})\rho^{b}(y^-,y_{\pperp})\rangle &= \mu^2(x^-)h(b_{\pperp})\delta^{ab}\delta(x^--y^-)f(x_{\pperp}-y_{\pperp}) \nonumber\\
&\equiv\lambda(x^-,b_{\pperp})\delta^{ab}\delta(x^--y^-)f(x_{\pperp}-y_{\pperp}),\label{2point}
\end{align}
where we allow the possibility of finite, non-homogeneous nuclei by explicitly introducing an impact parameter ($b_{\pperp}\!\equiv\!(x_{\pperp}\!+y_{\pperp})/2$) dependence as previously done in \cite{PhysRevC.92.064912}. Also, we drop the assumption that interactions are local in the transversal plane by introducing an undetermined function $f(x_{\pperp}\!-y_{\pperp})$ instead of a Dirac delta. This allows to implement the JIMWLK evolution of $W[\rho]$ within the so-called Gaussian truncation \cite{Fujii:2006ab,Marquet:2007vb,Kovchegov:2008mk,Marquet:2010cf}. These extensions of the original MV model might prove especially useful in subsequent phenomenological applications. In section \ref{sec:t} we will go into detail about the specific behavior assumed for both $h(b_{\pperp})$ and $f(x_{\pperp}\!-y_{\pperp})$.

\begin{figure}
\centering
\includegraphics[width=0.47\textwidth]{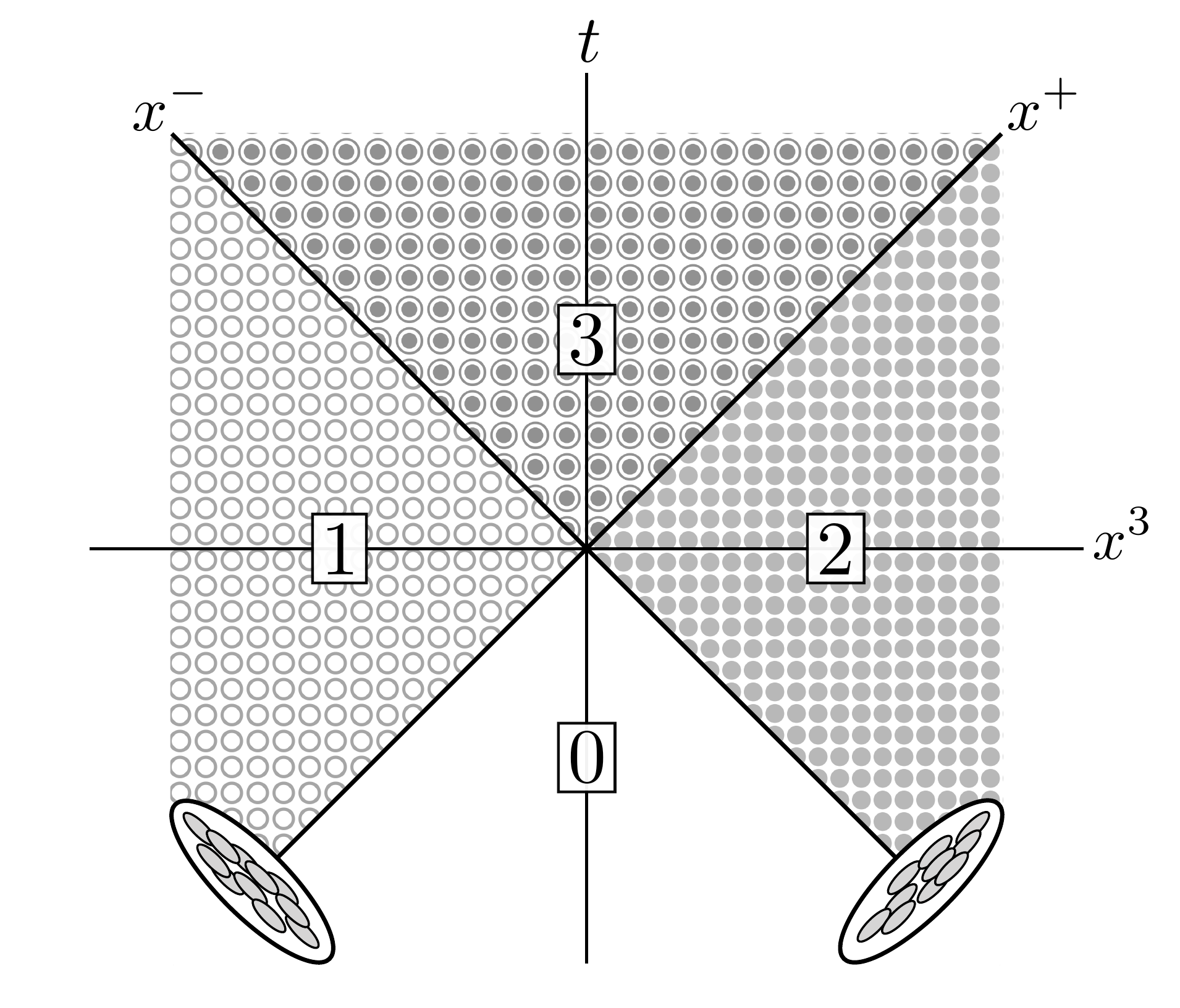}
\caption{Space-time diagram of the collision of two ultra-relativistic heavy ion nuclei. The two diagonal lines represent the trajectory of the nuclei. The points below them (quadrant $0$) represent a region where the projectiles have not yet arrived. By choosing the gauge fields to vanish in the remote past, in this region we have $A^{\mu}\!=\!0$. As quadrants 1 and 2 represent regions where only one of the nuclei has arrived, the dynamics of the gauge fields there are described by the Yang-Mills equations with a single source. However, in quadrant 3 we need to take into account both sources.}
\label{Spacetime}
\end{figure}

When attempting to describe the medium generated in the collision of two nuclei in the framework outlined above, we encounter a crucial problem: there is no general analytical solution for the Yang-Mills equations with two sources. Thus, we need to turn to either analytical or numerical approximations. A good starting point for these methods is the inner surface of the light-cone, $\tau\!=\!0^+$ (i.e.$\!$ an infinitesimal positive proper time after the collision), as in this region it is possible to find an analytical expression of the gauge fields. In order to do so, it is convenient to divide the space-time into four quadrants as indicated in \figref{Spacetime}.
The MV model provides the appropriate framework to calculate the gauge fields that characterize each nuclei before the collision (quadrants 1 and 2). These fields define the boundary conditions for the solution in the future light-cone (quadrant 3). As for $\tau\!<\!0$ the nuclei are located in causally disconnected regions of space-time, we can compute each gauge field independently. Let us take, for instance, a nucleus moving in the positive $x^{\scaleto{3}{4pt}}$ direction (which we indicate with the label 1). By solving \eqref{YMeqs} in the light-cone gauge (see e.g. $\!$\cite{BLAIZOT200413} for a detailed resolution), we obtain:
\begin{align}
A^{\pm}_{\scaleto{1}{4.2pt}}&=0 \\
A^{i}_{\scaleto{1}{4.2pt}}\,&=\theta(x^-)\!\!\int^{\infty}_{-\infty}\!dz^- U_{\scaleto{1}{4.2pt}}^{\dagger}(z^-,x_{\pperp})\frac{\partial^{i}\tilde{\rho}_{\scaleto{1}{4pt}}(z^-,x_{\pperp})}{\nabla^2}U_{\scaleto{1}{4.2pt}}(z^-,x_{\pperp})\equiv\theta(x^-)\alpha_{\scaleto{1}{4.2pt}}^{i}(x_{\pperp}),
\end{align}
which is a non-abelian Weizs\"acker-Williams (WW) field. Here $\tilde{\rho}$ is the color charge density in the covariant gauge\footnote{Providing that we average gauge invariant observables, the specific gauge in which we work does not affect the result of $\langle{\cal O}\rangle$, as both the Gaussian weight $W[\rho]$ and the functional measure $[d\rho]$ are gauge invariant objects.} and $U$ is the Wilson line, an SU($N_c$) element that represents the effect of the interaction with the classical gluon field over the fast valence partons in the eikonal approximation, i.e.$\;$a rotation in color space. $U(x^-,x_{\pperp})$ is defined as a path-ordered exponential:
\begin{align}
U_{\scaleto{1}{4.2pt}}(x^-,x_{\pperp})&=\text{P}^- \exp{\left\{ -ig\!\int^{x^-}_{x_{\scaleto{0}{3pt}}^-}\!\!dz^- \frac{1}{\nabla^2}\tilde{\rho}_{\scaleto{1}{4pt}}(z^-,x_{\pperp})\right\}}\nonumber\\
&=\text{P}^- \exp{\left\{ -ig\!\int^{x^-}_{\scaleto{-\infty}{5pt}}\!\!dz^-\!\!\int dz^2_{\pperp}G(z_{\pperp}-x_{\pperp})\tilde{\rho}_{\scaleto{1}{4pt}}(z^-,z_{\pperp})\right\}},
\end{align}
where $G(z_{\pperp}\!\!-\!x_{\pperp})$ is the Green's function for the 2-dimensional Laplace operator. In the previous expression we show explicitly the definition of the differential operator $1/\nabla^2$, which is the notation we adopt to denote a convolution with $G(z_{\pperp}\!\!-\!x_{\pperp})$. The choice of the integration lower limit $x^-_{\scaleto{0}{3.8pt}}$ is arbitrary, with different choices giving us solutions $A^{i}$ connected by residual, 2-dimensional gauge transformations. We shall adopt $x^-_{\scaleto{0}{3.8pt}}\!=\!-\infty$, which implies that the fields vanish in the remote past (retarded boundary conditions). Likewise, for the nucleus moving in the opposite direction\footnote{We work in a specific gauge that acts as a sort of `mix' of the light-cone gauges of both nuclei: the Fock-Schwinger gauge, defined by the condition $(x^+A^-+x^-A^+)/\tau=0$. Note that, as the separate fields of each nuclei already satisfy this condition, the Fock-Schwinger representation does not introduce any physical assumption that is not already present in the single nucleus characterization.} (indicated with the label 2), we have:
\begin{align}
A^{\pm}_{\scaleto{2}{4.2pt}}&=0\\
A^{i}_{\scaleto{2}{4.2pt}}\,&=\theta(x^+)\!\!\int^{\infty}_{-\infty}\!dz^+ U^{\dagger}_{\scaleto{2}{4.2pt}}(z^+,x_{\pperp})\frac{\partial^{i}\tilde{\rho}_{\scaleto{2}{4pt}}(z^+,x_{\pperp})}{\nabla^2}U_{\scaleto{2}{4.2pt}}(z^+,x_{\pperp})\equiv\theta(x^+)\alpha_{\scaleto{2}{4.2pt}}^{i}(x_{\pperp}),
\end{align}
where:
\begin{align}
U_{\scaleto{2}{4.2pt}}(x^+,x_{\pperp})=\text{P}^+ \exp{\left\{ -ig\!\int^{x^+}_{\scaleto{-\infty}{5pt}}\!\!dz^+\!\!\int dz^2_{\pperp}G(z_{\pperp}-x_{\pperp})\tilde{\rho}_{\scaleto{2}{4pt}}(z^+,z_{\pperp})\right\}}.
\end{align}
Thus, the total gauge field outside the light-cone reads:
\begin{align}
A^{\pm}&=0\\
A^{i}\;&=\theta(x^-)\theta(-x^+)\alpha_{\scaleto{1}{4.2pt}}^{i}(x_{\pperp})+\theta(x^+)\theta(-x^-)\alpha_{\scaleto{2}{4.2pt}}^{i}(x_{\pperp}).\label{FieldsOut}
\end{align}
%where $\partial^{i}\alpha^{i}_{\scaleto{1}{4.2pt}}\!=\!\rho_{\scaleto{1}{4pt}}(x_{\pperp})$ and $\partial^{i}\alpha^{i}_{\scaleto{2}{4.2pt}}\!=\!\rho_{\scaleto{2}{4pt}}(x_{\pperp})$.
%Explanation about why the equation becomes homogeneous is based on the support of /rho? which I don't think I've explained properly on these notes.
The gluon field sources vanish everywhere except at the very light-cone ($\tau\!=\!0$), and thus at $\tau\!=\!0^+$ the Yang-Mills equations become homogeneous. In order to solve them the following ansatz is proposed in \cite{PhysRevD.52.3809}:
\begin{align}
A^{\pm}&\!=\!\pm \,x^{\pm} \alpha(\tau=0^+\!,x_{\pperp})\label{Ansatz1}\\
A^{i}\;&\!=\!\alpha^{i}(\tau=0^+\!,x_{\pperp}),\label{Ansatz}
\end{align}
where we adopted the comoving coordinate system, defined by proper time $\tau\!=\!\sqrt{2x^+x^-}$ and rapidity $\eta\!=\!\frac{1}{2}\ln (x^+/x^-)$.
%The fact that this ansatz does not depend on rapidity... (9502289)
Substituting the above expressions, the separate components of the homogeneous Yang-Mills equations $[D_{\mu},F^{\mu\nu}]\!=\!0$ take the following form \cite{PhysRevD.52.6231}:
\begin{align}
\nu=\tau \longrightarrow\, &\;ig\tau\left[ \alpha, \partial_{\tau}\alpha\right]-\frac{1}{\tau}\left[ D^{i},\partial_{\tau}\alpha^{i} \right]=0\\
\nu=\eta \longrightarrow\,  &\; \frac{1}{\tau}  \partial_{\tau}\frac{1}{\tau}\partial_{\tau}(\tau^2 \alpha) -\left[ D^{i},\left[ D^{i},\alpha \right]\right]=0\\
\nu=j \longrightarrow\,  &\;\frac{1}{\tau}  \partial_{\tau}(\tau \partial_{\tau}\alpha^{j}) -ig\tau^2\left[ \alpha,\left[ D^{j},\alpha \right]\right]-\left[ D^{i},F^{i j}\right]=0.\label{HomogeneousYM}
\end{align}
This system provides the initial conditions for the $\tau$-evolution of the gluon fields in the future light-cone, be it computed via analytical or numerical methods. In order to relate them to the fields prior to the collision (\eqref{FieldsOut}) we invoke a physical `matching condition' that requires Yang-Mills equations to be regular in the limit $\tau\!\rightarrow\!0$
%(i.e. we impose that all singularities that appear in this limit must cancel)
. In doing so, the following relations are obtained:
\begin{align}
\alpha^{i} (\tau=0^+\!,x_{\pperp})&= \alpha_{\scaleto{1}{4.2pt}}^{i}(x_{\pperp})+\alpha_{\scaleto{2}{4.2pt}}^{i}(x_{\pperp})\label{EvolutionBoundaries1}\\
\alpha (\tau=0^+\!,x_{\pperp})&= \frac{ig}{2} \left[ \alpha_{\scaleto{1}{4.2pt}}^{i}(x_{\pperp}) , \alpha_{\scaleto{2}{4.2pt}}^{i}(x_{\pperp}) \right]\!,\label{EvolutionBoundaries}
\end{align}% Write the other conditions obtained? (The ones with the derivatives of the alphas) We don't use them
which act as boundary conditions of the subsequent $\tau$-evolution. Several approaches of both analytical and numerical nature have been applied for this computation in the literature. For instance, in \cite{Fries:2006pv} an analytical approximation based on an expansion of the previous solution in powers of $\tau$ is proposed.
%This method provides a relatively simple way to implement the evolution of our results, as the coefficients of the expansion can be obtained by inserting simple transverse derivatives into the expressions computed in the following sections.
However, this is out of the scope of the work presented in this paper.

\section{The EMT one-point correlator in the classical approximation } 
\label{sec:t}

Using \eqref{Ansatz1}, \eqref{Ansatz} along with the boundary conditions of \eqref{EvolutionBoundaries1}, \eqref{EvolutionBoundaries} we obtain the following expression for the EMT at $\tau\!=\!0^+$ \cite{LAPPI2006200}:
\begin{align}
T^{\mu\nu}_{\scaleto{0}{3.8pt}}(x_{\pperp})\!=\!2\,\text{Tr}\left\{ \frac{1}{4}g^{\mu\nu}F^{\alpha\beta}F_{\alpha\beta}\!-\!F^{\mu\alpha}F^{\nu}_{\;\alpha} \right\}_{\!\!\scaleto{0^+}{5pt}}\!\!\!\!&=\!-g^2(\delta^{ij}\delta^{kl}+\epsilon^{ij}\epsilon^{kl})\text{Tr}\left\{ \![ \alpha_{\scaleto{1}{4.2pt}}^{i}, \alpha_{\scaleto{2}{4.2pt}}^{j}][ \alpha_{\scaleto{1}{4.2pt}}^{k}, \alpha_{\scaleto{2}{4.2pt}}^{l}]\!\right\}\!\!\times\! t^{\mu\nu}\nonumber\\
&=\epsilon_{\scaleto{0}{3.8pt}}(x_{\pperp})\!\times\! t^{\mu\nu}\!,
\end{align}
where $g^{\mu\nu}\!=\![\text{diag}(1,-1,-1,-1)]^{\mu\nu}$, $t^{\mu\nu}\!\equiv\![\text{diag}(1,1,1,-1)]^{\mu\nu}$ (in terms of the Cartesian coordinate system), $F^{\alpha\beta}$ is the field strength tensor, and $\epsilon_{\scaleto{0}{3.8pt}}(x_{\pperp})$ is the energy density at proper time $\tau\!=\!0^+$ in a point $x_{\pperp}$ of the transverse plane.
%new
Note that the characteristic diagonal structure of this tensor is a feature of the specific proper time at which we are setting our calculation. The ensuing time evolution brings non-trivial off-diagonal corrections that largely modify this initial form, as indicated by the higher order terms of the $\tau$-expansion proposed in \cite{Fries:2006pv}. At $\tau\!=\!0^+$, however, the classical approximation yields a remarkably simple, diagonal EMT even at the event-by-event level, prior to the computation of its average over the background fields.
%new

As mentioned earlier, another remarkable aspect of this tensor is the maximum pressure anisotropy denoted by the negative value in the longitudinal direction. The negative pressure slows down the longitudinal expansion of the system, while the remaining components force it to expand in the transverse directions. However, prior to the interpretation of this object we must compute its average over the background fields. We have $\langle T^{\mu\nu}_{\scaleto{0}{3.8pt}}(x_{\pperp})\rangle\!=\!\langle\epsilon_{\scaleto{0}{3.8pt}}(x_{\pperp})\rangle\!\times\!t^{\mu\nu}$, with \cite{LAPPI200611}:
\begin{equation}\label{Epsilon0}
\langle \epsilon_{\scaleto{0}{3.8pt}}(x_{\pperp})\rangle\!=\!-g^2(\delta^{ij}\delta^{kl}+\epsilon^{ij}\epsilon^{kl})\!\left\langle\text{Tr}\left\{[ \alpha_{\scaleto{1}{4.2pt}}^{i}(x_{\pperp}), \alpha_{\scaleto{2}{4.2pt}}^{j}(x_{\pperp})][ \alpha_{\scaleto{1}{4.2pt}}^{k}(x_{\pperp}), \alpha_{\scaleto{2}{4.2pt}}^{l}(x_{\pperp})]\right\}\right\rangle.
\end{equation}
As the trace in this expression is performed over color space, in order to compute it we need to expand the color structure of our fields:
\begin{align}
\alpha^{i}(x_{\pperp})\!=\!\int^{\infty}_{-\infty}dz^-U^{\dagger}(z^-,x_{\pperp})\frac{\partial^{i}\tilde{\rho}(z^-,x_{\pperp})}{\nabla^2}U(z^-,x_{\pperp})\!&=\!\int^{\infty}_{-\infty}dz^-\frac{\partial^{i}\tilde{\rho}^{a}}{\nabla^2}U^{\dagger}t^{a}U\!\nonumber\\
&=\!\int^{\infty}_{-\infty}dz^-\frac{\partial^{i}\tilde{\rho}^{a}}{\nabla^2}U^{ab}t^{b}\equiv \alpha^{i,b}(x_{\pperp}) t^{b}.
\end{align}
Here we used the relation between Wilson lines in the fundamental and adjoint representations $U^{\dagger}t^{a}U\!=\!U^{ab}t^{b}$. Substituting in \eqref{Epsilon0} we get:
\begin{align}
\langle \epsilon_{\scaleto{0}{3.8pt}}\rangle\!&=\!-g^2(\delta^{ij}\delta^{kl}+\epsilon^{ij}\epsilon^{kl})\left\langle\alpha_{\scaleto{1}{4.2pt}}^{i,a}\alpha_{\scaleto{2}{4.2pt}}^{j,b}\alpha_{\scaleto{1}{4.2pt}}^{k,c}\alpha_{\scaleto{2}{4.2pt}}^{l,d}\right\rangle\text{Tr}\left\{[ t^{a}, t^{b}][ t^{c}, t^{d}]\right\}\nonumber\\
&=g^2(\delta^{ij}\delta^{kl}+\epsilon^{ij}\epsilon^{kl})\left\langle\alpha_{\scaleto{1}{4.2pt}}^{i,a}\alpha_{\scaleto{1}{4.2pt}}^{k,c}\alpha_{\scaleto{2}{4.2pt}}^{j,b}\alpha_{\scaleto{2}{4.2pt}}^{l,d}\right\rangle f^{abm}f^{cdn}\text{Tr}\left\{ t^{m} t^{n}\right\}\nonumber\\
&=\frac{g^2}{2}(\delta^{ij}\delta^{kl}+\epsilon^{ij}\epsilon^{kl})f^{abm}f^{cdm}\left\langle\alpha^{i,a}(x_{\pperp})\alpha^{k,c}(x_{\pperp})\right\rangle_{\!\scaleto{1}{4.2pt}}\!\!\left\langle\alpha^{j,b}(x_{\pperp})\alpha^{l,d}(x_{\pperp})\right\rangle_{\!\scaleto{2}{4.2pt}}\!.\label{Epsilon0.1}
\end{align}
%Notice that, as in the previous expressions all indices outside the trace are explicit, we can freely permute the $\alpha^{i,a}$ coefficients in order to factorize the average over color sources $\rho_{\scaleto{1}{4pt}}$ and $\rho_{\scaleto{2}{4pt}}$.
In the last step we factorize the average over color source densities $\tilde{\rho}_{\scaleto{1}{4pt}}$ and $\tilde{\rho}_{\scaleto{2}{4pt}}$, since in the MV model we assume the source fluctuations in each nuclei to be independent of each other:
\begin{align}
\langle {\cal O}[\rho_{\scaleto{1,2}{5.5pt}}] \rangle=\frac{1}{{\cal N}_{\scaleto{1}{4.2pt}}}\frac{1}{{\cal N}_{\scaleto{2}{4.2pt}}}\int[d\rho_{\scaleto{1}{4pt}}] W_{\mu}[\rho_{\scaleto{1}{4pt}}]\!\!\int [d\rho_{\scaleto{2}{4pt}}] W_{\mu}[\rho_{\scaleto{2}{4pt}}]\,{\cal O}[\rho_{\scaleto{1,2}{5.5pt}}].
\end{align}
Thus, the building block of $\langle \epsilon_{\scaleto{0}{3.8pt}}\rangle$ is the average of two gauge fields evaluated in the same transverse coordinate: $\langle \alpha^{i,a}(x_{\pperp})\alpha^{j,b}(x_{\pperp}) \rangle$. Nevertheless, it will prove useful to perform this calculation for different transverse coordinates $x_{\pperp}$, $\,y_{\pperp}$ and eventually take the limit $y_{\pperp}\!\rightarrow\!x_{\pperp}$:
\begin{equation}\label{Correlator1}
\left\langle\! \alpha^{i,a}(x_{\pperp})\alpha^{j,b}(y_{\pperp}) \!\right\rangle\!=\!\!\!\int^{\infty}_{-\infty}\!dz^-dz^{-\prime}\!\left\langle\! \frac{\partial^{i}\tilde{\rho}^{a^{\prime}}(z^-,x_{\pperp})}{\nabla^2}U^{a^{\prime}a}(z^-,x_{\pperp})\frac{\partial^{j}\tilde{\rho}^{b^{\prime}}(z^{-\prime},y_{\pperp})}{\nabla^2}U^{b^{\prime}b}(z^{-\prime},y_{\pperp}) \!\!\right\rangle\!.
\end{equation}
The average in the right hand side of this expression contains, for each transverse coordinate, an infinite product of $\tilde{\rho}$ factors: one external and the rest arranged inside the Wilson lines. Since we are assuming that the color sources obey Gaussian statistics, we can apply Wick's theorem, which states that any correlator can be expressed in terms of products of two-point functions. In our particular case, the only nonvanishing terms of the infinite possibilities available are the ones that correspond to a factorization of the external sources from those inside the Wilson lines (see Appendix \ref{BigCorrelator} for a general analysis of the decomposition of the correlator of $n$ Wilson lines and $m$ external sources):
%Calculate the example of m=n=2 explicitly in the appendices, or just do the big calculation?
\begin{equation}\label{Correlator2}
\left\langle \!\alpha^{i,a}(x_{\pperp})\alpha^{j,b}(y_{\pperp}) \!\right\rangle\!=\!\!\!\int^{\infty}_{-\infty}\!\!dz^-dz^{-\prime}\!\left\langle\! \frac{\partial^{i}\tilde{\rho}^{a^{\prime}}(z^-,x_{\pperp})}{\nabla^2}\frac{\partial^{j}\tilde{\rho}^{b^{\prime}}(z^{-\prime}y_{\pperp})}{\nabla^2}\!\right\rangle\!\left\langle U^{a^{\prime}a}(z^{-},x_{\pperp})U^{b^{\prime}b}(z^{-\prime},y_{\pperp}) \right\rangle\!.
\end{equation}
As the differential operators $1/\nabla^2$, $\partial^{i}$ commute with the average operation, the factor involving the external sources can be calculated via an almost direct application of the two-point correlator. In the MV model (\eqref{MV2point}) this yields a quite simple expression:
\begin{equation}\label{2pointfunction2MV}
\left\langle \frac{\partial^{i}\tilde{\rho}^{a^{\prime}}(x^-,x_{\pperp})}{\nabla^2}\frac{\partial^{j}\tilde{\rho}^{b^{\prime}}(y^-,y_{\pperp})}{\nabla^2}\right\rangle_{\!\!\!\scaleto{\text{MV}}{0.12cm}}\!\!\!=\delta^{a^{\prime}b^{\prime}}\mu^2(x^-)\delta(x^--y^-)\partial^{i}_{x}\partial^{j}_{y}L(x_{\pperp}-y_{\pperp})_{\scaleto{\text{MV}}{0.12cm}},
\end{equation}
with:
\begin{align}
\frac{1}{\nabla^2_x}\frac{1}{\nabla^2_y}\delta^2(x_{\pperp}-y_{\pperp})\!&=\!\!\!\int dz^2_{\pperp}du^2_{\pperp}G(z_{\pperp}\!-x_{\pperp})G(u_{\pperp}\!-y_{\pperp})\delta^2(z_{\pperp}-u_{\pperp})\!\nonumber\\[-0.2em]
&=\!\!\!\int d^2z_{\pperp}G(z_{\pperp}\!-x_{\pperp})G(z_{\pperp}\!-y_{\pperp})\equiv L(x_{\pperp}-y_{\pperp})_{\scaleto{\text{MV}}{0.12cm}}.
\end{align}
However, using our generalized version (\eqref{2point}), we have:
\begin{align}
\frac{1}{\nabla^2_x}\frac{1}{\nabla^2_y}\left(h(b_{\pperp})f(x_{\pperp}\!-y_{\pperp})\right)&=\!\!\int\! dz^2_{\pperp}du^2_{\pperp}G(z_{\pperp}\!-x_{\pperp})G(u_{\pperp}\!-y_{\pperp})h\!\left(\frac{z_{\pperp}+u_{\pperp}}{2}\right)\!f(z_{\pperp}\!-u_{\pperp})\nonumber\\[-0.3em]
&\approx\!h(b_{\pperp})\!\!\int\! dz^2_{\pperp}du^2_{\pperp}G(z_{\pperp}\!-x_{\pperp})G(u_{\pperp}\!-y_{\pperp})f(z_{\pperp}\!-u_{\pperp})\nonumber\\
&\equiv\!h(b_{\pperp})L(x_{\pperp}-y_{\pperp}),\label{Lfunction}
\end{align}
and then:
\begin{align}
\partial^{i}_{x}\partial^{j}_{y}\left(h(b_{\pperp})L(x_{\pperp}-y_{\pperp})\right)\approx h(b_{\pperp})\partial^{i}_{x}\partial^{j}_{y}L(x_{\pperp}-y_{\pperp}),\label{Lfunction1}
\end{align}
yielding:
\begin{equation}\label{2pointfunction2}
\left\langle \frac{\partial^{i}\tilde{\rho}^{a^{\prime}}(x^-,x_{\pperp})}{\nabla^2}\frac{\partial^{j}\tilde{\rho}^{b^{\prime}}(y^-,y_{\pperp})}{\nabla^2}\right\rangle=\delta^{a^{\prime}b^{\prime}}\lambda(x^-,b_{\pperp})\delta(x^--y^-)\partial^{i}_{x}\partial^{j}_{y}L(x_{\pperp}-y_{\pperp}).
\end{equation}
In the same spirit than \cite{PhysRevC.92.064912}, in \eqref{Lfunction} we implicitly make the assumption that the impact parameter profile $h(b_{\pperp})$ introduced earlier is a slowly varying function over lengths of the order of an infrared length scale $1/m$, or smaller. Therefore we take $1/m$ to be an intermediate scale between the inverse saturation scale and the nuclear radius $R_{\scaleto{A}{0.4em}}$:
\begin{align}
\frac{1}{Q_s}\ll \frac{1}{m} \ll R_{\scaleto{A}{0.4em}}\,.\label{Hierark}
\end{align}
One can think of $1/m$ as a cut-off that imposes color neutrality at the nucleon size.
%Thus, we will only be interested in interaction scales that obey $r=\!|x_{\pperp}-y_{\pperp}|\le Q^{-1}_s \ll m^{-1}$.
In addition, in \eqref{Lfunction1} we assume that $f(x_{\pperp}\!\!-y_{\pperp})$ behaves in such a way that its Fourier transform $\hat{f}(k_{\pperp})$ tends to unity in the infrared limit.
%This is a reasonable assumption in the saturation limit.
This requirement, along with the assumed `slow' behavior for $h(b_{\pperp})$, result in this factor being approximately unaffected by the differential operators in both \eqref{Lfunction} and \eqref{Lfunction1}
%. The first correction to this assumption is proportional to $m^{-1}|\partial^{i}h(b_{\pperp})|$
 (see Appendix \ref{GreenDerivatives} for more details about these assumptions).
%By doing this we lose some generality; assuming a `soft' shape for $h(b_{\pperp})$ limits the color charge distributions that can be implemented, potentially excluding complex models like e.g. hot spots. However, it should be good enough for the characterization of a (big) nuclear projectile.
 Substituting in \eqref{Correlator2}, we finally get:
\begin{equation}
\left\langle \alpha^{i,a}(x_{\pperp})\alpha^{j,b}(y_{\pperp}) \right\rangle\!=\!\!\int^{\infty}_{-\infty}\!dz^-\lambda(z^-,b_{\pperp})\partial^{i}_{x}\partial^{j}_{y}L(x_{\pperp}\!-y_{\pperp})\!\left\langle U^{a^{\prime}a}(z^-,x_{\pperp})U^{a^{\prime}b}(z^-,y_{\pperp}) \right\rangle\!,
\end{equation}
where the last factor corresponds to the dipole function in the adjoint representation \cite{1126-6708-2007-06-040}:
%More references?
\begin{align}
\left\langle U^{a^{\prime}a}(x^-,x_{\pperp})U^{a^{\prime}b}(x^-,y_{\pperp}) \right\rangle\!&=\delta^{ab}\exp{\left\{ -g^2\frac{N_c}{2}\Gamma(x_{\pperp}\!-y_{\pperp})\bar{\lambda}(x^-,b_{\pperp})\right\}}\nonumber\\
&\equiv\delta^{ab}C^{(2)}_{\text{adj}}(x^-;x_{\pperp},y_{\pperp}).\label{dipolefun}
\end{align}
Here we introduced the factor
\begin{equation}
\Gamma(x_{\pperp}\!-y_{\pperp})\!=\!2(L(0_{\pperp})-L(x_{\pperp}\!-y_{\pperp}))
\end{equation}
and the integrated color charge density $\bar{\lambda}(x^-,b_{\pperp})\!=\!\int^{\scaleto{x^-}{6pt}}_{\scaleto{-\infty}{4pt}}\!dz^- \lambda(z^-,b_{\pperp})$. Note that in \eqref{dipolefun} we applied the same approximation as in \eqref{Lfunction} in order to obtain the factorization of $h(b_{\pperp})\Gamma(x_{\pperp}\!\!-y_{\pperp})$.
%We are able to factorize $h(b_{\pperp})$ from $\Gamma$ using the fact that we neglect blah blah
%\begin{equation}
%h(x_{\pperp})=h(b_{\pperp})+(x_{\pperp}-b_{\pperp})^{i}\partial^{i}h(b_{\pperp})+...
%\end{equation}
Substituting:
\begin{equation}
\left\langle \alpha^{i,a}(x_{\pperp})\alpha^{j,b}(y_{\pperp}) \right\rangle=\delta^{ab}\!\int^{\infty}_{-\infty}dz^-\lambda(z^-,b_{\pperp})\partial^{i}_{x}\partial^{j}_{y}L(x_{\pperp}-y_{\pperp})C^{(2)}_{\text{adj}}(z^-;x_{\pperp},y_{\pperp}) .
\end{equation}
Now, taking the limit $y_{\pperp}\!\rightarrow\!x_{\pperp}$:
\begin{align}
\left\langle \alpha^{i,a}(x_{\pperp})\alpha^{j,b}(x_{\pperp}) \right\rangle=-\frac{1}{2}\delta^{ab}\delta^{ij}\int^{\infty}_{-\infty}dz^-\lambda(z^-,x_{\pperp})\partial^2L(0_{\pperp})&=-\frac{1}{2}\delta^{ab}\delta^{ij}\bar{\lambda}(x_{\pperp})\partial^2L(0_{\pperp})\nonumber\\
&=-\frac{1}{2}\delta^{ab}\delta^{ij}\bar{\mu}^2h(x_{\pperp})\partial^2L(0_{\pperp}),
\end{align}
where we defined $\bar{\lambda}(b_{\pperp})\!=\!\bar{\lambda}(\infty,b_{\pperp})\!=\!\bar{\mu}^2h(b_{\pperp})$ (in general, we will identify functions integrated in the longitudinal direction from $-\infty$ to $\infty$ by simply omitting their longitudinal dependence) and substituted the following expression:
\begin{equation}
\lim_{r\rightarrow0}\partial^{i}_{x}\partial^{j}_{y}L(r_{\pperp})=\frac{\delta^{ij}}{2}\int\frac{d^2q_{\pperp}}{(2\pi)^2}\hat{f}(q_{\pperp})\frac{1}{q^2}\equiv-\frac{1}{2}\delta^{ij}\partial^2L(0_{\pperp}),
\end{equation}
with $r\!=\!|r_{\pperp}|\!=\!|x_{\pperp}\!-y_{\pperp}|$. Here the double derivative $\partial^2L(0_{\pperp})$ is a model-dependent constant (see Appendix \ref{GreenDerivatives} for details). We apply this result for both nuclei in \eqref{Epsilon0.1}, obtaining:
\begin{align}
\langle \epsilon_{\scaleto{0}{3.8pt}}(x_{\pperp})\rangle &= \frac{g^2}{2}f^{abm}f^{cdm}(\delta^{ij}\delta^{kl}+\epsilon^{ij}\epsilon^{kl})\frac{1}{4}\delta^{ac}\delta^{ik}\delta^{bd}\delta^{jl}\bar{\mu}^2_{\scaleto{1}{4pt}} \,\bar{\mu}^2_{\scaleto{2}{4pt}}\,h_{\scaleto{1}{4pt}}(x_{\pperp})h_{\scaleto{2}{4pt}}(x_{\pperp}) (\partial^{2}L(0_{\pperp}))^2\nonumber\\
&=g^2N_c^2C_{\scaleto{F}{0.4em}}\bar{\lambda}_{\scaleto{1}{4pt}}(x_{\pperp}) \bar{\lambda}_{\scaleto{2}{4pt}}(x_{\pperp}) (\partial^{2}L(0_{\pperp}))^2,\label{onepointT}
\end{align}
%Checked this result with the one in arXiv:1507.03524v1 (eqs 118, 97 and 83) and eq26 of PhysRevC.79.024909 (discrepancy of g^2 due to different two-point function definition)
whose dependence on the transverse position is a consequence of our generalized MV model approach, where we assume finite nuclei. Note that we label both factors $\mu^2$ and $h$ according to the corresponding nucleus, which potentially allows for the use of different nuclear profiles for target and projectile. We absorb these quantities in the definition of the following momentum scale:
\begin{equation}\label{Saturation1}
\bar{Q}_s^2(x_{\pperp})\equiv\alpha_sN_c\,\bar{\lambda}(x_{\pperp}),
\end{equation}
which characterizes each colliding nucleus. Performing this substitution we obtain:
\begin{equation}
\langle \epsilon_{\scaleto{0}{3.8pt}}(x_{\pperp})\rangle =\frac{C_{\scaleto{F}{0.4em}}}{g^2}\bar{Q}^2_{s\scaleto{1}{4pt}}(x_{\pperp})\bar{Q}^2_{s\scaleto{2}{4pt}}(x_{\pperp})\!\left( 4\pi\,\partial^{2}L(0_{\pperp})\right)^2\!.
\end{equation} 

\section{The EMT two-point correlator in the classical approximation } 
\label{sec:tt}

The next step in our calculation is the computation of $\langle T^{\mu\nu}_{\scaleto{0}{3.8pt}}(x_{\pperp})T^{\sigma\rho}_{\scaleto{0}{3.8pt}}(y_{\pperp})\rangle\!\!=\!\!\langle\epsilon_{\scaleto{0}{3.8pt}}(x_{\pperp})\epsilon_{\scaleto{0}{3.8pt}}(y_{\pperp})\rangle\!\!\times\! t^{\mu\nu}t^{\sigma\rho}$. We start by expanding the product of energy densities:
\begin{align}
%\epsilon_{\scaleto{0}{3.8pt}}(x_{\pperp})\epsilon_{\scaleto{0}{3.8pt}}(y_{\pperp})=\frac{1}{4}(E^2_0(x_{\pperp})+B^2_0(x_{\pperp}))(E^2_0(y_{\pperp})+B^2_0(y_{\pperp}))\\
\epsilon_{\scaleto{0}{3.8pt}}(x_{\pperp})\epsilon_{\scaleto{0}{3.8pt}}(y_{\pperp})\!=\!g^4(\delta^{ij}\delta^{kl}\!+\epsilon^{ij}\epsilon^{kl})\text{Tr}\left\{\left[ \alpha_{\scaleto{1}{4.2pt}}^{i}(x_{\pperp}), \alpha_{\scaleto{2}{4.2pt}}^{j}(x_{\pperp})\right]\!\left[ \alpha_{\scaleto{1}{4.2pt}}^{k}(x_{\pperp}), \alpha_{\scaleto{2}{4.2pt}}^{l}(x_{\pperp})\right]\right\}\nonumber\\
\times(\delta^{i^{\prime}j^{\prime}}\!\delta^{k^{\prime}l^{\prime}}\!\!+\epsilon^{i^{\prime}j^{\prime}}\!\epsilon^{k^{\prime}l^{\prime}})\text{Tr}\left\{\left[ \alpha_{\scaleto{1}{4.2pt}}^{i^{\prime}}\!(y_{\pperp}), \alpha_{\scaleto{2}{4.2pt}}^{j^{\prime}}\!(y_{\pperp})\right]\!\left[ \alpha_{\scaleto{1}{4.2pt}}^{k^{\prime}}\!(y_{\pperp}), \alpha_{\scaleto{2}{4.2pt}}^{l^{\prime}}\!(y_{\pperp})\right]\right\}\nonumber\\[-0.4em]
=\!\frac{g^4}{4}(\delta^{ij}\delta^{kl}\!+\epsilon^{ij}\epsilon^{kl})(\delta^{i^{\prime}j^{\prime}}\!\delta^{k^{\prime}l^{\prime}}\!\!+\epsilon^{i^{\prime}j^{\prime}}\!\epsilon^{k^{\prime}l^{\prime}})f^{abn} f^{cdn}f^{a^{\prime}b^{\prime}m}f^{c^{\prime}d^{\prime}m}  \underbrace{\alpha^{i,a}_{\scaleto{1}{4.2pt}} \alpha^{j,b}_{\scaleto{2}{4.2pt}}\alpha^{k,c}_{\scaleto{1}{4.2pt}}\alpha^{l,d}_{\scaleto{2}{4.2pt}}}_{x_{\pperp}} \underbrace{\alpha^{i^{\prime}\!,a^{\prime}}_{\scaleto{1}{4.2pt}}\!\alpha^{j^{\prime}\!,b^{\prime}}_{\scaleto{2}{4.2pt}}\!\alpha^{k^{\prime}\!,c^{\prime}}_{\scaleto{1}{4.2pt}}\!\alpha^{l^{\prime}\!,d^{\prime}}_{\scaleto{2}{4.2pt}}}_{y_{\pperp}}\nonumber\\[-0.25em]
\equiv{\cal A}^{ik;i'k'}_{jl;j'l'} {\cal F}^{ac;a'c'}_{bd;b'd'} \alpha^{i,a}_{\scaleto{1}{4.2pt}\,x} \alpha^{k,c}_{\scaleto{1}{4.2pt}\,x} \alpha^{i^{\prime}\!,a^{\prime}}_{\scaleto{1}{4.2pt}\,y}\alpha^{k^{\prime}\!,c^{\prime}}_{\scaleto{1}{4.2pt}\,y} \alpha^{j,b}_{\scaleto{2}{4.2pt}\,x} \alpha^{l,d}_{\scaleto{2}{4.2pt}\,x} \alpha^{j^{\prime}\!,b^{\prime}}_{\scaleto{2}{4.2pt}\,y}\alpha^{l^{\prime}\!,d^{\prime}}_{\scaleto{2}{4.2pt}\,y}.
\end{align}
Here we defined the transverse and color structure tensors respectively as:
\begin{align}
{\cal A}^{ik;i'k'}_{jl;j'l'}\!\!&=\!(\delta^{ij}\delta^{kl}\!+\epsilon^{ij}\epsilon^{kl})(\delta^{i^{\prime}j^{\prime}}\delta^{k^{\prime}l^{\prime}}\!\!+\epsilon^{i^{\prime}j^{\prime}}\epsilon^{k^{\prime}l^{\prime}})\\
{\cal F}^{ac;a'c'}_{bd;b'd'}\!\!&=\!\frac{g^4}{4}f^{abn} f^{cdn}f^{a^{\prime}b^{\prime}m}f^{c^{\prime}d^{\prime}m},
\end{align}
and adopted a shorthand notation for the gluon fields $\alpha^{i,a}(x_{\pperp})\!\!\equiv\!\!\alpha^{i,a}_{x}$. As the average operation is performed independently for both nuclei, the building block of $\langle \epsilon_{\scaleto{0}{3.8pt}}(x_{\pperp})\epsilon_{\scaleto{0}{3.8pt}}(y_{\pperp}) \rangle$ reads:
\begin{align}
\langle \alpha^{i,a}(x_{\pperp}) \alpha^{k,c}(x_{\pperp}) \alpha^{i^{\prime}\!,a^{\prime}}(y_{\pperp})\alpha^{k^{\prime}\!,c^{\prime}}(y_{\pperp}) \rangle\!=\!\!\int^{\infty}_{-\infty}\!dz^-dw^-dz^{-\prime}dw^{-\prime}
\!\left\langle \!\frac{\partial^{i}\tilde{\rho}^{e}(z^-,x_{\pperp})}{\nabla^2}U^{ea}(z^-,x_{\pperp})\right.\nonumber\\
\left.\frac{\partial^{k}\tilde{\rho}^{f}(w^-,x_{\pperp})}{\nabla^2}U^{fc}(w^-,x_{\pperp})\frac{\partial^{i^{\prime}}\tilde{\rho}^{e^{\prime}}\!(z^{-\prime},y_{\pperp})}{\nabla^2}U^{e'a'}\!(z^{-\prime},y_{\pperp})\frac{\partial^{k^{\prime}}\tilde{\rho}^{f^{\prime}}\!(w^{-\prime},y_{\pperp})}{\nabla^2}U^{f'\!c'}\!(w^{-\prime},y_{\pperp}) \!\!\right\rangle\!.\label{MainCorrelator}
 \end{align}
This is an extended and more complicated version of \eqref{Correlator1}, with twice as many color sources depending on different longitudinal coordinates. The correlations between its different elements result in the following sum:
\begin{align}
\left\langle\tilde{\rho}^{i,e}_{x}U^{ea}_{x}\tilde{\rho}^{k,f}_{x}U^{fc}_{x}\tilde{\rho}^{i'\!,e'}_{y}U^{e'\!a'}_{y}\tilde{\rho}^{k'\!,f^{\prime}}_{y}U^{f'\!c'}_{y} \right\rangle\!=\!\left\langle \tilde{\rho}^{i,e}_{x}\tilde{\rho}^{k,f}_{x}\tilde{\rho}^{i'\!,e'}_{y}\tilde{\rho}^{k'\!,f'}_{y}\!\right\rangle\!\left\langle U^{ea}_{x}U^{fc}_{x}U^{e'\!a'}_{y}U^{f'\!c'}_{y} \!\right\rangle\!&\nonumber\\
+\!\left\langle \tilde{\rho}^{i,e}_{x}\tilde{\rho}^{k,f}_{x} \right\rangle\!\left\langle \tilde{\rho}^{i'\!,e'}_{y}\tilde{\rho}^{k'\!,f'}_{y} U^{ea}_{x}U^{fc}_{x}U^{e'\!a'}_{y}U^{f'\!c'}_{y}\!\right\rangle_{\hspace{-0.09cm}\text{\tiny c}}\!+\!\left\langle \tilde{\rho}^{i'\!,e'}_{y}\tilde{\rho}^{k'\!,f'}_{y} \!\right\rangle\!\left\langle \tilde{\rho}^{i,e}_{x}\tilde{\rho}^{k,f}_{x} U^{ea}_{x}U^{fc}_{x}U^{e'\!a'}_{y}U^{f'\!c'}_{y}\!\right\rangle_{\hspace{-0.09cm}\text{\tiny c}}\!&\nonumber\\
+\!\left\langle \tilde{\rho}^{i,e}_{x}\tilde{\rho}^{i'\!,e'}_{y} \!\right\rangle\!\left\langle \tilde{\rho}^{k,f}_{x}\tilde{\rho}^{k'\!,f'}_{y} U^{ea}_{x}U^{fc}_{x}U^{e'\!a'}_{y}U^{f'\!c'}_{y}\!\right\rangle_{\text{\hspace{-0.09cm}\tiny c}}\!+\!\left\langle \tilde{\rho}^{i,e}_{x}\tilde{\rho}^{k'\!,f'}_{y} \!\right\rangle\!\left\langle\tilde{\rho}^{k,f}_{x}\tilde{\rho}^{i'\!,e'}_{y} U^{ea}_{x}U^{fc}_{x}U^{e'\!a'}_{y}U^{f'\!c'}_{y}\!\right\rangle_{\hspace{-0.09cm}\text{\tiny c}}\!&\nonumber\\
+\!\left\langle \tilde{\rho}^{k,f}_{x}\tilde{\rho}^{i'\!,e'}_{y} \!\right\rangle\!\left\langle \tilde{\rho}^{i,e}_{x}\tilde{\rho}^{k'\!,f'}_{y} U^{ea}_{x}U^{fc}_{x}U^{e'\!a'}_{y}U^{f'\!c'}_{y}\!\right\rangle_{\hspace{-0.09cm}\text{\tiny c}}\!+\!\left\langle \tilde{\rho}^{k,f}_{x}\tilde{\rho}^{k'\!,f'}_{y} \!\right\rangle\!\left\langle \tilde{\rho}^{i,e}_{x}\tilde{\rho}^{i'\!,e'}_{y} U^{ea}_{x}U^{fc}_{x}U^{e'\!a'}_{y}U^{f'\!c'}_{y}\!\right\rangle_{\hspace{-0.09cm}\text{\tiny c}}\!&\nonumber\\
+\!\left\langle\tilde{\rho}^{i,e}_{x}U^{ea}_{x}\tilde{\rho}^{k,f}_{x}U^{fc}_{x}\tilde{\rho}^{i'\!,e'}_{y}U^{e'\!a'}_{y}\tilde{\rho}^{k'\!,f^{\prime}}_{y}U^{f'\!c'}_{y} \!\right\rangle_{\hspace{-0.09cm}\text{\tiny c}}&.\label{Decomposition}
\end{align}
The details of the above decomposition are explained in Appendix \ref{BigCorrelator}. For simplicity we momentarily adopted a shorthand notation that omits the longitudinal coordinate dependence and the differential operators $1/\nabla^2$, $\partial^{i}$. In the previous expression a major source of difficulty stands out: unlike the average featured in \eqref{Correlator1}, the one in \eqref{MainCorrelator} gets non-trivial contributions from correlators connecting external color sources with those arranged inside Wilson lines. We name these `connected' correlators, and indicate them as $\langle...\rangle_{\hspace{-0.02cm}\text{\tiny c}}\,$. Based on the diagrammatic rules derived in \cite{PhysRevC.79.025204} we are able to compute these contributions and express them in terms of the following function:
\begin{align}
&C^{ij;kl}_{ab;cd}(u_{\pperp},u'_{\pperp},v_{\pperp},v'_{\pperp})= \!\!\int^{\infty}_{-\infty}dz^{-}dz^{-\prime}dw^{-}dw^{-\prime}\left\langle \tilde{\rho}^{i,e}_{u}\tilde{\rho}^{j,f}_{u'} \right\rangle\!\left\langle \tilde{\rho}^{k,e'}_{v}\tilde{\rho}^{l,f'}_{v'} U^{ea}_{u}U^{fb}_{u'}U^{e'\!c}_{v}U^{f'\!d}_{v'}\right\rangle_{\hspace{-0.09cm}\text{\tiny c}} \nonumber\\[-0.75em]
&=g^2h^3(b_{\pperp})\partial^{i}_{u}\partial^{j}_{u'}L(u_{\pperp}-u'_{\pperp})\!\int^{\infty}_{-\infty}\!\!\!dz^-\!\!\int^{z^-}_{-\infty}\!\!\!dw^-\!\!\int^{w^-}_{-\infty}\!\!\!\!dw^{-\prime}\mu^2(z^-)\mu^2(w^-)\mu^2(w^{-\prime})\nonumber\\[-0.5em]
&\times\!C_{\text{adj}}^{(2)}(z^-,w^-;u_{\pperp},u'_{\pperp})\bigg(\left[\partial^{k}_v\!\left(L(v_{\pperp}\!-u'_{\pperp})\!-\!L(v_{\pperp}\!-u_{\pperp})\right)\!C^{(3)}_{\text{adj}}(w^-,w^{-\prime};u_{\pperp},u'_{\pperp},v_{\pperp})\right.\nonumber\\
&\times\!\partial^{l}_{v'}\!\left( f^{AeD}f^{CBe}L(v'_{\pperp}-u_{\pperp})\!+\!f^{ACe}f^{DBe}L(v'_{\pperp}-u'_{\pperp})+f^{ABe}f^{eCD}L(v'_{\pperp}-v_{\pperp})\right)\nonumber\\[-0.1em]
&\times Q^{ABCD}_{abcd}(w^{-\prime};u_{\pperp},u'_{\pperp},v_{\pperp},v'_{\pperp})\Big]+\scaleto{\begin{bmatrix} l&\longleftrightarrow &k \\[-0.5em] c&\longleftrightarrow &d \\[-0.5em] \scaleto{v_{\pperp}}{0.2cm}\!\!&\longleftrightarrow &\scaleto{v'_{\pperp}}{0.32cm}\end{bmatrix}}{1cm}\,\bigg),\label{ConnectedFunction}
\end{align}
where $b_{\pperp}\!=\!(x_{\pperp}\!+y_{\pperp})/2$ (detailed calculation in Appendix \ref{BigCorrelator}). Here we recover the notation used in the previous section for the adjoint dipole correlator and extend it to the case of three Wilson lines \cite{1126-6708-2007-06-040} as:
%Also calculated with our method as a check
\begin{align}
\langle U^{aa'}(w^-,x^{\scaleto{1}{4.5pt}}_{\pperp})U^{bb'}(w^-,x^{\scaleto{2}{4.5pt}}_{\pperp})U^{cc'}(w^-,x^{\scaleto{3}{4.5pt}}_{\pperp}) \rangle\nonumber\\
=\frac{1}{2N_c^2C_{\scaleto{F}{0.4em}}}\left( f^{abc}f^{a'b'c'}+\frac{N_c^2}{N_c^2-4}d^{abc}d^{a'b'c'}\right)\exp{\left\{ -g^2\frac{N_c}{4}\bar{\lambda}(w^-,b_{\pperp})\sum_{i>j}\Gamma(x^{i}_{\pperp}-x^{j}_{\pperp})\right\}}\nonumber\\
\equiv\frac{1}{2N_c^2C_{\scaleto{F}{0.4em}}}\left( f^{abc}f^{a'b'c'}+\frac{N_c^2}{N_c^2-4}d^{abc}d^{a'b'c'}\right)\!C^{(3)}_{\text{adj}}(w^-;x^{\scaleto{1}{4.5pt}}_{\pperp},x^{\scaleto{2}{4.5pt}}_{\pperp},x^{\scaleto{3}{4.5pt}}_{\pperp}).\label{Threepole}
\end{align}
The second longitudinal coordinate in the dependence of both $C^{(2)}_{\text{adj}}$ and $C^{(3)}_{\text{adj}}$ stands for the lower limit of the integral contained in their definition:
\begin{equation}
\ln\left(C^{(2),(3)}_{\text{adj}}(x^-,y^-)\right)\propto\int^{x^-}_{y^-}dz^-\lambda(z^-,b_{\pperp}).
\end{equation}
We also introduced the adjoint Wilson line quadrupole tensor:
\begin{equation}\label{Quadrupole}
Q^{aceg}_{bdfh}(w^-;x_{\pperp},x'_{\pperp},y_{\pperp},y'_{\pperp})=\left\langle U^{ab}(w^-,x_{\pperp})U^{cd}(w^-,x'_{\pperp})U^{ef}(w^-,y_{\pperp})U^{gh}(w^-,y'_{\pperp})\right\rangle\!.
\end{equation}
The fully connected term (last term of \eqref{Decomposition}) vanishes and thus we are able to write all connected contributions in terms of $C^{ij;kl}_{ab;cd}$. The remaining contributions result from the factorization of external and internal color sources (first term after the equal sign in \eqref{Decomposition}):
 \begin{align}
\int^{\infty}_{-\infty}dz^-dw^-dz^{-\prime}dw^{-\prime}&\left\langle \frac{\partial^{i}\tilde{\rho}^{e}(z^-,x_{\pperp})}{\nabla^2}\frac{\partial^{k}\tilde{\rho}^{f}(w^-,x_{\pperp})}{\nabla^2}\frac{\partial^{i^{\prime}}\!\tilde{\rho}^{e^{\prime}}\!(z^{-\prime},y_{\pperp})}{\nabla^2}\frac{\partial^{k^{\prime}}\!\tilde{\rho}^{f^{\prime}}\!(w^{-\prime},y_{\pperp})}{\nabla^2}\right\rangle\nonumber\\
&\times\!\left\langle U^{ea}(z^-,x_{\pperp})U^{fc}(w^-,x_{\pperp})U^{e'a'}\!(z^{-\prime},y_{\pperp})U^{f'\!c'}\!(w^{-\prime},y_{\pperp}) \right\rangle.\label{Disconnected1}
 \end{align}
This term can be further expanded by application of Wick's theorem, which tells us that the external source correlator breaks down into the following sum of pairwise contractions:
\begin{align}
\left\langle\tilde{\rho}^{i,e}_{x}\,\tilde{\rho}^{k,f}_{x}\,\tilde{\rho}^{i'\!,e'}_{y}\,\tilde{\rho}^{k'\!,f'}_{y}\right\rangle\!=\!\langle \tilde{\rho}^{i,e}_{x}\,\tilde{\rho}^{k,f}_{x}\rangle\langle \tilde{\rho}^{i'\!,e'}_{y}\tilde{\rho}^{k'\!,f'}_{y}\rangle\!+\!\langle \tilde{\rho}^{i,e}_{x}\,\tilde{\rho}^{i'\!,e'}_{y}\rangle\langle \tilde{\rho}^{k,f}_{x} \tilde{\rho}^{k'\!,f'}_{y}\rangle\!+\!\langle \tilde{\rho}^{i,e}_{x}\,\tilde{\rho}^{k'\!,f'}_{y}\rangle\langle \tilde{\rho}^{k,f}_{x}\tilde{\rho}^{i'\!,e'}_{y}\rangle.\label{Wick}
\end{align}
Following this decomposition, \eqref{Disconnected1} yields three terms that we can address in terms of the `disconnected' function, which we derive explicitly in the following lines:
\begin{align}
D^{ij;kl}_{ab;cd}&(u_{\pperp},u'_{\pperp},v_{\pperp},v'_{\pperp})\nonumber\\[-0.4em]
=&\!\int^{\infty}_{-\infty}\!dz^-dz^{-\prime}dw^-dw^{-\prime}\!\left\langle\!\frac{\partial^{i}\tilde{\rho}^{a'}\!(z^-,u_{\pperp})}{\nabla^2}\frac{\partial^{j}\tilde{\rho}^{b'}\!(z^{-\prime},u'_{\pperp})}{\nabla^2}\!\right\rangle\!\left\langle\!\frac{\partial^{k}\tilde{\rho}^{c'}\!(w^-,v_{\pperp})}{\nabla^2}\frac{\partial^{l}\tilde{\rho}^{d'}\!(w^{-\prime},v'_{\pperp})}{\nabla^2}\!\right\rangle\nonumber\\
&\times\!\left\langle\!U^{a'a}(z^-,u_{\pperp})U^{b'b}(z^{-\prime},u'_{\pperp})U^{c'c}(w^-,v_{\pperp})U^{d'd}(w^{-\prime},v'_{\pperp})\!\right\rangle\nonumber\\
=&\!\!\int^{\infty}_{-\infty}\!dz^-\!dz^{-\prime}\!dw^-\!dw^{-\prime} \delta^{a'b'}\lambda(z^-,b_{\pperp})\delta(z^-\!-z^{-\prime}\!)\partial^{i}_u\partial^{j}_{u'}L(u_{\pperp}\!-u'_{\pperp})\delta^{c'd'}\!\lambda(w^-,b_{\pperp})\nonumber\\
&\times\delta(w^{-}\!-w^{-\prime})\partial^{k}_{v}\partial^{l}_{v'}L(v_{\pperp}\!-v'_{\pperp})\!\left\langle\!U^{a'a}(z^-,u_{\pperp})U^{b'b}(z^{-\prime},u'_{\pperp})U^{c'c}(w^-,v_{\pperp})U^{d'd}(w^{-\prime},v'_{\pperp})\!\right\rangle\nonumber\\
=&\,T^{\,ij;kl}\!(u_{\pperp},u'_{\pperp},v_{\pperp},v'_{\pperp})\!\!\int^{\infty}_{-\infty}\!\!dz^-\!dw^{-}\!\lambda(z^-,b_{\pperp})\lambda(w^-,b_{\pperp})\nonumber\\
&\times\!\left\langle U^{a'a}(z^-,u_{\pperp})U^{a'b}(z^-,u'_{\pperp})U^{c'c}(w^-,v_{\pperp})U^{c'd}(w^-,v'_{\pperp})\right\rangle\!,\label{Discon}
\end{align}
where $b_{\pperp}\!\!=\!(x_{\pperp}\!+y_{\pperp})/2$. Note that both here and in the connected function \eqref{ConnectedFunction} we substituted the result of \eqref{2pointfunction2}, which implies that we adopt the same assumptions over $h(b_{\pperp})$ and $f(x_{\pperp}\!-y_{\pperp})$ as in the previous section. We also made use of the knowledge that eventually all the transverse positions that enter this expression will be either $x_{\pperp}$ or $y_{\pperp}$, which allows us to neglect the corrections to an expansion of $h((u_{\pperp}\!+u'_{\pperp})/2)$ around $h((x_{\pperp}\!+y_{\pperp})/2)$ (see Appendix \ref{GreenDerivatives} for details). This approximation was also taken in \eqref{Threepole}, allowing us to extract $\lambda(w^-,b_{\pperp})$ as a common factor of the sum.
Going back to \eqref{Discon}, note that we introduced the following function:
\begin{equation}
T^{\,ij;kl}(u_{\pperp},u'_{\pperp},v_{\pperp},v'_{\pperp})\!\equiv\! \partial^{i}_u\partial^{j}_{u'}L(u_{\pperp}\!-u'_{\pperp})\partial^{k}_{v}\partial^{l}_{v'}L(v_{\pperp}\!-v'_{\pperp}),
\end{equation}
where, as was also the case in the connected function \eqref{ConnectedFunction}, we encounter double derivatives of $L(x_{\pperp}\!-x'_{\pperp})$. From their symmetries and dimension, we can parameterize them as:
\begin{equation}\label{Derivatives}
\partial^{i}_{x}\partial^{j}_{y}L(r_{\pperp})=A(r_{\pperp})\delta^{ij}+B(r_{\pperp})\!\left( \frac{\delta^{ij}}{2}-\frac{r^{i}r^{j}}{r^2}\right)\!.
\end{equation}
In Appendix \ref{GreenDerivatives} we obtain expressions for the coefficients $A(r_{\pperp})$ and $B(r_{\pperp})$ in terms of $f(r_{\pperp})$ and provide an explicit calculation in the specific case of the MV model (where $f(r_{\pperp})\!=\!\delta^{(2)}(r_{\pperp})$). However, for now we prefer to stay in the most general case and leave them undetermined.

\begin{figure}
\centering
\includegraphics[width=0.82\textwidth]{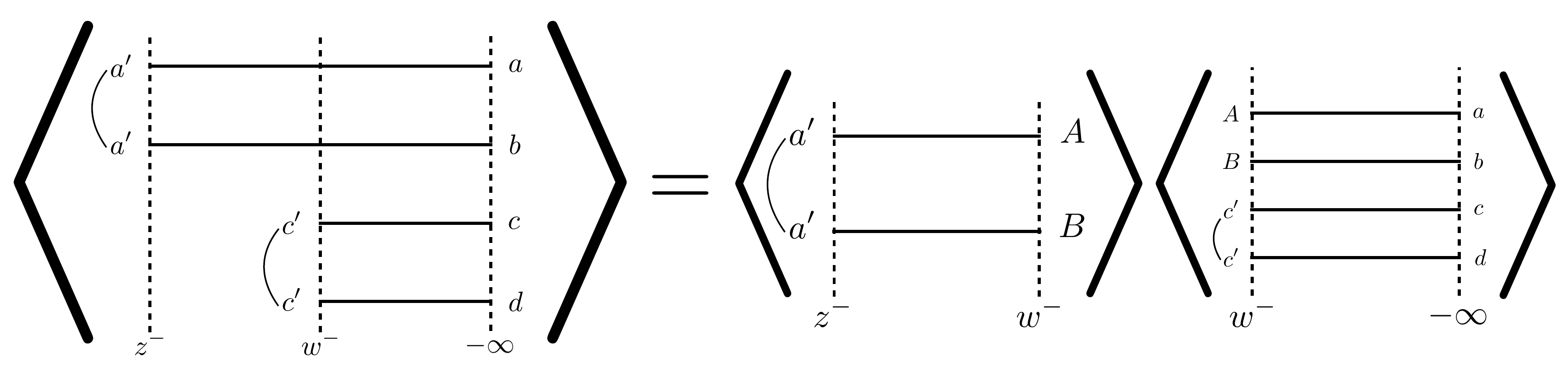}
\caption{Schematic representation of the correlator factorization performed in \eqref{FactorizationCorrelator}.}
\label{FigFactorization}
\end{figure}

In order to solve the integral present in $D^{ij;kl}_{ab;cd}$ we consider separately the region where $z^-\!>\!w^-$ and its complementary. Assuming a certain ordering in the integration variables allows us to factorize the Wilson line correlator by applying the locality in rapidity implied in \eqref{2point}. For instance, in the region $z^-\!>\!w^-$ (see \figref{FigFactorization}):
\begin{align}
\hspace{-0.24cm}\left\langle U^{a'a}(z^-,u_{\pperp})U^{a'b}(z^-,u'_{\pperp})U^{c'c}(w^-,v\right.\left.\!\!_{\pperp})U^{c'd}(w^-,v'_{\pperp})\right\rangle\!=\!\left\langle U^{a'A}(z^-,w^-;u_{\pperp})U^{a'B}(z^-,w^-;u'_{\pperp})\right\rangle \nonumber\\
\times\!\left\langle U^{Aa}(w^-,u_{\pperp})U^{Bb}(w^-,u'_{\pperp})U^{c'c}(w^-,v_{\pperp})U^{c'd}(w^-,v'_{\pperp})\right\rangle\nonumber\\
=C^{(2)}_{\text{adj}}(z^-,w^-;u_{\pperp},u'_{\pperp})\!\left\langle U^{Aa}(w^-,u_{\pperp})U^{Ab}(w^-,u'_{\pperp})U^{c'c}(w^-,v_{\pperp})U^{c'd}(w^-,v'_{\pperp})\right\rangle\nonumber\\
=C^{(2)}_{\text{adj}}(z^-,w^-;u_{\pperp},u'_{\pperp})Q^{AAc'c'}_{abcd}(w^-;u_{\pperp},u'_{\pperp},v_{\pperp},v'_{\pperp}).\label{FactorizationCorrelator}
\end{align}
Summing the contributions from each integration region $z^-\!>\!w^-$ and $w^-\!>\!z^-$ we get:
\begin{align}
D^{ij;kl}_{ab;cd}(u_{\pperp},u'_{\pperp},v_{\pperp},v'_{\pperp})=T^{\,ij;kl}\!(u_{\pperp},u'_{\pperp},v_{\pperp},v'_{\pperp})\!\!\int^{\infty}_{-\infty}\!\!dz^-\!\int^{z^-}_{-\infty}\!dw^{-}\!\lambda(z^-,b_{\pperp})\lambda(w^-,b_{\pperp})\nonumber\\
\times\!\left(C^{(2)}_{\text{adj}}(z^-,w^-;u_{\pperp},u'_{\pperp})+C^{(2)}_{\text{adj}}(z^-,w^-;v_{\pperp},v'_{\pperp})\right)\delta^{AB}\delta^{CD}Q^{ABCD}_{abcd}(w^-;u_{\pperp},u'_{\pperp},v_{\pperp},v'_{\pperp}).
\end{align}
Having defined these functions, we can rewrite our building block \eqref{MainCorrelator} as:
\begin{align}
\langle \alpha^{i\,a}(x_{\pperp}) \alpha^{k\,c}(x_{\pperp}) \alpha^{i^{\prime}a^{\prime}}(y_{\pperp})\alpha^{k^{\prime}c^{\prime}}(y_{\pperp}) \rangle= D^{ik;i'k'}_{ac;a'c'}(x_{\pperp},x_{\pperp},y_{\pperp},y_{\pperp})+D^{ii';kk'}_{aa';cc'}(x_{\pperp},y_{\pperp},x_{\pperp},y_{\pperp})\nonumber\\
+D^{ik';ki'}_{ac';ca'}(x_{\pperp},y_{\pperp},x_{\pperp},y_{\pperp})+C^{ii';kk'}_{aa';cc'}(x_{\pperp},y_{\pperp},x_{\pperp},y_{\pperp})+C^{ik';ki'}_{ac';ca'}(x_{\pperp},y_{\pperp},x_{\pperp},y_{\pperp})\nonumber\\
+C^{kk';ii'}_{cc';aa'}(x_{\pperp},y_{\pperp},x_{\pperp},y_{\pperp})+C^{ki';ik'}_{ca';ac'}(x_{\pperp},y_{\pperp},x_{\pperp},y_{\pperp}).\label{MainCorrelator2}
\end{align}
Note that, in addition to the fully connected correlator, also the first two partially connected terms of \eqref{Decomposition} vanish (see Appendix \ref{BigCorrelator}).

Remarkably, in both $D^{ij;kl}_{ab;cd}$ and $C^{ij;kl}_{ab;cd}$ we find different projections of the adjoint Wilson line quadrupole \eqref{Quadrupole}, which is a quite complex object. However, the fact that in our calculation we only deal with two transverse coordinates $x_{\pperp}$ and $y_{\pperp}$ yields great simplification in some instances. For example, the first term after the equal sign in \eqref{MainCorrelator2} corresponds to:
\begin{align}
D^{ik;i'k'}_{ac;a'c'}(x_{\pperp},x_{\pperp},y_{\pperp},y_{\pperp})\!=2\,T^{\,ik;i'k'}\!(x_{\pperp},x_{\pperp},y_{\pperp},y_{\pperp})\!\!\int^{\infty}_{-\infty}\!\!dz^-\!\int^{z^-}_{-\infty}\!dw^{-}\!\lambda(z^-,b_{\pperp})\lambda(w^-,b_{\pperp})\nonumber\\
Q^{AACC}_{aca'c'}(w^-;x_{\pperp},x_{\pperp},y_{\pperp},y_{\pperp}).\label{SimpleProjection}
\end{align}
In this case, the projection of the adjoint Wilson line quadrupole can be obtained in a straightforward way. Writing it explicitly:
\begin{equation}
Q^{AACC}_{aca'c'}(w^-;x_{\pperp},x_{\pperp},y_{\pperp},y_{\pperp})=\left\langle U^{Aa}(w^-,x_{\pperp})U^{Ac}(w^-,x_{\pperp})U^{Ca'}(w^-,y_{\pperp})U^{Cc'}(w^-,y_{\pperp})\right\rangle
\end{equation}
and expanding the first pair of adjoint Wilson lines in terms of fundamental Wilson lines as $U^{ab}\!=\!2\,\text{Tr}\left\{ U^{\dagger}t^{a}Ut^{b}\right\}$, we get:
\begin{align}
U^{Aa}U^{Ac}&=4\,U^{\dagger}_{ij}t^{A}_{jk}U_{kl}t^{a}_{li}U^{\dagger}_{i^{\prime}j^{\prime}}t^{A}_{j^{\prime}k^{\prime}}U_{k^{\prime}l^{\prime}}t^{c}_{l^{\prime}i^{\prime}}.\nonumber\\
\intertext{Now, applying the Fierz identity $t^{a}_{ij}t^{a}_{kl}\!=\!\frac{1}{2}(\delta_{il}\delta_{jk}\!-\!\frac{1}{N_c}\delta_{ij}\delta_{kl})$:}\nonumber\\[-2.5em]
&=2\left( \delta_{jk^{\prime}}\delta_{kj^{\prime}}-\frac{1}{N_c}\delta_{jk}\delta_{j^{\prime}k^{\prime}}\right)U^{\dagger}_{ij}U_{kl}t^{a}_{li}U^{\dagger}_{i^{\prime}j^{\prime}}U_{k^{\prime}l^{\prime}}t^{c}_{l^{\prime}i^{\prime}}\nonumber\\
&=2\left( U^{\dagger}_{ij}U_{kl}U^{\dagger}_{i^{\prime}k}U_{jl^{\prime}}-\frac{1}{N_c}U^{\dagger}_{ij}U_{jl}U^{\dagger}_{i^{\prime}j^{\prime}}U_{j^{\prime}l^{\prime}}\right)t^{a}_{li}t^{c}_{l^{\prime}i^{\prime}}\nonumber\\
&=2\left( \delta_{il^{\prime}}\delta_{i^{\prime}l}-\frac{1}{N_c}\delta_{il}\delta_{i^{\prime}l^{\prime}}\right)t^{a}_{li}t^{c}_{l^{\prime}i^{\prime}}=2\left(\text{Tr}\{t^{a}t^{c}\}-\frac{1}{N_c}\text{Tr}\{t^{a}\}\text{Tr}\{t^{c}\}\right)\!=\delta^{ac}.\label{Trick}
\end{align}
Therefore:
\begin{align}
D^{ik;i'k'}_{ac;a'c'}(x_{\pperp},x_{\pperp},y_{\pperp},y_{\pperp})\!=2\,T^{\,ik;i'k'}\!(x_{\pperp},x_{\pperp},y_{\pperp},y_{\pperp})\delta^{ac}\delta^{a^{\prime}c^{\prime}}\!\!\!\int^{\infty}_{-\infty}\!\!\!dz^-\!\!\int^{z^-}_{-\infty}\!\!\!\!dw^{-}\!\lambda(z^-,b_{\pperp})\lambda(w^-,b_{\pperp})\nonumber\\
=T^{\,ik;i'k'}\!(x_{\pperp},x_{\pperp},y_{\pperp},y_{\pperp})\delta^{ac}\delta^{a^{\prime}c^{\prime}}\bar{\lambda}^2(b_{\pperp})=\frac{1}{4}\delta^{ik}\delta^{i'k'}\!\!\left(\partial^2L(0_{\pperp})\right)^2\!\delta^{ac}\delta^{a^{\prime}c^{\prime}}\bar{\lambda}^2(b_{\pperp}).
\end{align}
In the two remaining disconnected terms the Wilson lines that share a color index depend on different transverse coordinates, which prevents the Fierz identity from simplifying the expression. In other words, while in \eqref{SimpleProjection} we have
\begin{equation}
\delta^{AB}\delta^{CD}Q^{ABCD}_{abcd}(w^-;x_{\pperp},x_{\pperp},y_{\pperp},y_{\pperp})=\delta^{ab}\delta^{cd},
\end{equation}
which corresponds to the trivial propagation of an eigenvector in color space, in the other two particular cases of $D^{ij;kl}_{ab;cd}$ we find
\begin{equation}\label{HardProjection}
\delta^{AC}\delta^{BD}Q^{ABCD}_{acbd}(w^-;x_{\pperp},x_{\pperp},y_{\pperp},y_{\pperp})
\end{equation}
instead, whose calculation requires expressing $\delta^{AC}\delta^{BD}$ in terms of the eigenvectors of $Q^{ABCD}_{acbd}$. This is a highly non-trivial problem that we analyze in depth in Appendix \ref{Kov}. Substituting the result of \eqref{HardProjection} and solving the double integrals, we obtain:
\begin{align}
\hspace{-1cm}D^{ij;kl}_{ab;cd}(x_{\pperp},y_{\pperp},x_{\pperp},y_{\pperp})\!=\!2\!\left(\!\delta^{ab}\delta^{cd}\!\left[\frac{N_c^2-4}{2N_c^2}f_{\scaleto{1}{4pt}}+\frac{2}{N_c^2}f_{\scaleto{2}{4pt}}+\frac{N_c+2}{4N_c}f_{\scaleto{3}{4pt}}+\frac{N_c-2}{4N_c}f_{\scaleto{4}{4pt}}\right]\right.\nonumber\\
+\delta^{ac}\delta^{bd}\!\left[\frac{1}{N_c^2-1}f_{\scaleto{5}{4pt}}-\frac{N_c+2}{2N_c(N_c+1)}f_{\scaleto{3}{4pt}}+\frac{N_c-2}{2N_c(N_c-1)}f_{\scaleto{4}{4pt}}\right]\nonumber\\
+\delta^{ad}\delta^{bc}\!\left[-\frac{N_c^2-4}{2N_c^2}f_{\scaleto{1}{4pt}}-\frac{2}{N_c^2}f_{\scaleto{2}{4pt}}+\frac{N_c+2}{4N_c}f_{\scaleto{3}{4pt}}+\frac{N_c-2}{4N_c}f_{\scaleto{4}{4pt}}\right]\nonumber\\
+d^{abm}d^{cdm}\!\left[-\frac{1}{N_c}f_{\scaleto{1}{4pt}}+\frac{1}{N_c}f_{\scaleto{2}{4pt}}+\frac{1}{4}f_{\scaleto{3}{4pt}}-\frac{1}{4}f_{\scaleto{4}{4pt}}\right]+d^{adm}d^{cbm}\!\left[ \frac{1}{N_c}f_{\scaleto{1}{4pt}}-\frac{1}{N_c}f_{\scaleto{2}{4pt}}+\frac{1}{4}f_{\scaleto{3}{4pt}}-\frac{1}{4}f_{\scaleto{4}{4pt}}\right]\nonumber\\
\left.+d^{acm}d^{bdm}\!\left[\frac{N_c}{N_c^2-4}f_{\scaleto{2}{4pt}}-\frac{N_c+4}{4(N_c+2)}f_{\scaleto{3}{4pt}}+\frac{N_c-4}{4(N_c-2)}f_{\scaleto{4}{4pt}}\right]\,\right)\!T^{\,ij;kl}\!(x_{\pperp},y_{\pperp},x_{\pperp},y_{\pperp}),\label{DisconResult}
\end{align}
where:
\begin{align}
    f_{\scaleto{1}{4pt}}=&\frac{2}{(N_cg^2\Gamma)^2}(1-C^{(2)}_{\text{adj}}(x_{\pperp},y_{\pperp}))^2\\
    f_{\scaleto{2}{4pt}} =&\frac{2}{N_cg^2\Gamma}\left( \frac{2}{N_cg^2\Gamma}(1-C^{(2)}_{\text{adj}}(x_{\pperp},y_{\pperp}))-\bar{\lambda}(b_{\pperp})C^{(2)}_{\text{adj}}(x_{\pperp},y_{\pperp})\right)\\
    f_{\scaleto{3}{4pt}} =& \left( \frac{4}{N_c(N_c+2)g^4\Gamma^2}(1-C^{(2)}_{\text{adj}}(x_{\pperp},y_{\pperp}))\right.\nonumber\\
   &\left.-\frac{2}{(N_c+2)(N_c+1)g^4\Gamma^2}(1-(C^{(2)}_{\text{adj}}(x_{\pperp},y_{\pperp}))^2\exp{\left\{-g^2\Gamma \bar{\lambda}(b_{\pperp}) \right\}})\right)\\
    f_{\scaleto{4}{4pt}} =& \left( \frac{4}{N_c(N_c-2)g^4\Gamma^2}(1-C^{(2)}_{\text{adj}}(x_{\pperp},y_{\pperp}))\right.\nonumber\\
    &\left.-\frac{2}{(N_c-2)(N_c-1)g^4\Gamma^2}(1-(C^{(2)}_{\text{adj}}(x_{\pperp},y_{\pperp}))^2\exp{\left\{g^2\Gamma \bar{\lambda}(b_{\pperp}) \right\}})\right)\\
    f_{\scaleto{5}{4pt}} =& \frac{2}{N_cg^2\Gamma}\left( \bar{\lambda}(b_{\pperp})-\frac{2}{N_cg^2\Gamma}(1-C^{(2)}_{\text{adj}}(x_{\pperp},y_{\pperp}))\right)\!.
\end{align}
As for the $C^{ij;kl}_{ab;cd}$ function, only one particular case enters \eqref{MainCorrelator2}:
\begin{align}
C^{ij;kl}_{ab;cd}(x_{\pperp}&,y_{\pperp},x_{\pperp},y_{\pperp})\!=\frac{g^2}{2}\partial^{i}_{x}\partial^{j}_{y}L(x_{\pperp}\!-\!y_{\pperp})h^3(b_{\pperp})\partial^{k}_{x}\Gamma(x_{\pperp}-y_{\pperp})\partial^{l}_{y}\Gamma(y_{\pperp}-x_{\pperp})\nonumber\\[-0.3em]
&\times\int^{\infty}_{-\infty}\!dz^-\!\!\int^{z^-}_{-\infty}\!\!dw^-\!\!\int^{w^-}_{-\infty}\!\!dw^{-\prime}\mu^2(z^-)\mu^2(w^-)\mu^2(w^{-\prime})C_{\text{adj}}^{(2)}(z^-,w^{-\prime};x_{\pperp},y_{\pperp})\nonumber\\
&\times f^{ACe}f^{BDe}Q^{ABCD}_{abcd}(w^{-\prime};x_{\pperp},y_{\pperp},x_{\pperp},y_{\pperp}).
\end{align}
Remarkably, the previous expression contains the propagation of the color vector $f^{ACe}f^{BDe}$ by the adjoint Wilson line quadrupole. In Appendix \ref{Kov} we show that it is actually an eigenvector of $Q^{ABCD}_{abcd}$, yielding the following straightforward result:
\begin{align}
f^{ACe}f^{BDe}Q^{ABCD}_{abcd}(w^{-\prime};x_{\pperp},y_{\pperp},x_{\pperp},y_{\pperp})=f^{ABe}f^{CDe}Q^{ABCD}_{acbd}(w^{-\prime};x_{\pperp},x_{\pperp},y_{\pperp},y_{\pperp})\nonumber\\
=f^{ace}f^{bde}C^{(2)}_{\text{adj}}(w^{-\prime};x_{\pperp},y_{\pperp}).
\end{align}
Substituting and solving the double integrals, we get:
\begin{align}
C^{ij;kl}_{ab;cd}(x_{\pperp},&\,y_{\pperp},x_{\pperp},y_{\pperp})\!=f^{ace}f^{bde}\partial^{i}_{x}\partial^{j}_{y}L(x_{\pperp}\!-y_{\pperp})\partial^{k}_{x}\Gamma(x_{\pperp}-y_{\pperp})\partial^{l}_{y}\Gamma(y_{\pperp}-x_{\pperp})\nonumber\\
&\times\!\left( \frac{4}{\Gamma ^3 g^4 N_c^3}-\left(\frac{\bar{\lambda}^2(b_{\pperp}) }{2\Gamma N_c}+\frac{4}{\Gamma ^3 g^4 N_c^3}+\frac{2 \bar{\lambda}(b_{\pperp})}{\Gamma ^2 g^2 N_c^2}\right)\!C_{\text{adj}}^{(2)}(x_{\pperp},y_{\pperp})\right)\!,
\end{align}
which concludes the calculation of the building block $\langle \alpha^{i,a}_{x} \alpha^{k,c}_{x}\alpha^{i^{\prime}\!,a^{\prime}}_{y}\!\alpha^{k^{\prime}\!,c^{\prime}}_{y} \rangle$. The final step consists in explicitly expanding the color contractions between these objects (one for each nucleus) and the transverse and color structure tensors defined earlier:
\begin{equation}
\langle \epsilon_{\scaleto{0}{3.8pt}}(x_{\pperp})\epsilon_{\scaleto{0}{3.8pt}}(y_{\pperp}) \rangle\!=\!{\cal A}^{ik;i'k'}_{jl;j'l'}{\cal F}^{ac;a'c'}_{bd;b'd'}\langle \alpha^{i,a}_{x} \alpha^{k,c}_{x}\alpha^{i^{\prime}\!,a^{\prime}}_{y}\alpha^{k^{\prime}\!,c^{\prime}}_{y} \rangle_{\scaleto{1}{4.2pt}}\langle \alpha^{j,b}_{x} \alpha^{l,d}_{x}\alpha^{j^{\prime}\!,b^{\prime}}_{y}\alpha^{l^{\prime}\!,d^{\prime}}_{y} \rangle_{\scaleto{2}{4.2pt}}.
\end{equation}
The product of the seven terms corresponding to each nucleus (\eqref{MainCorrelator2}) yields a total of 49 terms, which, by application of the symmetries of the tensors ${\cal A}^{ik;i'k'}_{jl;j'l'}$ and ${\cal F}^{ac;a'c'}_{bd;b'd'}$, can be reduced to:
\begin{align}
\langle \epsilon_{\scaleto{0}{3.8pt}}(x_{\pperp})\epsilon_{\scaleto{0}{3.8pt}}(y_{\pperp}) \rangle\!=\!\left[\frac{1}{2}D^{\,\;ik;i'k'}_{\scaleto{\!1}{4.25pt}\,ac;a'c'}(x_{\pperp},x_{\pperp},y_{\pperp},y_{\pperp})D^{\,\;jl;j'l'}_{\scaleto{\!2}{4.25pt}\,\,bd;b'd'}(x_{\pperp},x_{\pperp},y_{\pperp},y_{\pperp}){\cal A}^{ik;i'k'}_{jl;j'l'}{\cal F}^{ac;a'c'}_{bd;b'd'}\right.\nonumber\\[-0.5em]
+\left(D^{\,\;ik;i'k'}_{\scaleto{\!1}{4.25pt}\,ac;a'c'}(x_{\pperp},x_{\pperp},y_{\pperp},y_{\pperp})D^{\,\;jj';ll'}_{\scaleto{\!2}{4.25pt}\,\,bb';dd'}(x_{\pperp},y_{\pperp},x_{\pperp},y_{\pperp})\right.\nonumber\\
\left.+D^{\,\;ii';kk'}_{\scaleto{\!1}{4.25pt}\,aa';cc'}(x_{\pperp},y_{\pperp},x_{\pperp},y_{\pperp})D^{\,\;jj';ll'}_{\scaleto{\!2}{4.25pt}\,\,bb';dd'}(x_{\pperp},y_{\pperp},x_{\pperp},y_{\pperp})\right)\!\left[ {\cal A}^{ik;i'k'}_{jl;j'l'}{\cal F}^{ac;a'c'}_{bd;b'd'}+{\cal A}^{ik;i'k'}_{jl;l'j'}{\cal F}^{ac;a'c'}_{bd;d'b'} \right]\nonumber\\
+\left(D^{\,\;ik;i'k'}_{\scaleto{\!1}{4.25pt}\,ac;a'c'}(x_{\pperp},x_{\pperp},y_{\pperp},y_{\pperp})C^{\,\;jj';ll'}_{\scaleto{\!2}{4.25pt}\,\,bb';dd'}(x_{\pperp},y_{\pperp},x_{\pperp},y_{\pperp})\right.\nonumber\\
\left.+2\,D^{\,\;ii';kk'}_{\scaleto{\!1}{4.25pt}\,aa';cc'}(x_{\pperp},y_{\pperp},x_{\pperp},y_{\pperp})C^{\,\;jj';ll'}_{\scaleto{\!2}{4.25pt}\,\,bb';dd'}(x_{\pperp},y_{\pperp},x_{\pperp},y_{\pperp})\right.\nonumber\\
\left.+2\,C^{\,\;ii';kk'}_{\scaleto{\!1}{4.25pt}\,aa';cc'}(x_{\pperp},y_{\pperp},x_{\pperp},y_{\pperp})C^{\,\;jj';ll'}_{\scaleto{\!2}{4.25pt}\,\,bb';dd'}(x_{\pperp},y_{\pperp},x_{\pperp},y_{\pperp})\right)\nonumber\\[-0.3em]
\times\!\left[{\cal A}^{ik;i'k'}_{jl;j'l'}{\cal F}^{ac;a'c'}_{bd;b'd'}+{\cal A}^{ik;i'k'}_{lj;j'l'}{\cal F}^{ac;a'c'}_{db;b'd'}+{\cal A}^{ik;i'k'}_{jl;l'j'}{\cal F}^{ac;a'c'}_{bd;d'b'}+{\cal A}^{ik;i'k'}_{lj;l'j'}{\cal F}^{ac;a'c'}_{db;d'b'} \right]\bigg]+\left[1\leftrightarrow 2\right].\label{ResultExpansion}
\end{align}
It is worth mentioning that the terms resulting of the first contraction after the equal sign in \eqref{ResultExpansion} are identical to the product of the separate averages of $\epsilon_{\scaleto{0}{3.8pt}}(x_{\pperp})$ and $\epsilon_{\scaleto{0}{3.8pt}}(y_{\pperp})$:
\begin{align}
\hspace{-0.05cm}D^{ik;i'k'}_{ac;a'c'}(x_{\pperp},x_{\pperp},y_{\pperp},y_{\pperp})D^{jl;j'l'}_{bd;b'd'}(x_{\pperp},x_{\pperp},y_{\pperp},y_{\pperp}){\cal A}^{ik;i'k'}_{jl;j'l'}{\cal F}^{ac;a'c'}_{bd;b'd'}\nonumber\\=g^4(\partial^2L(0_{\pperp}))^4\!N_c^4C_{\scaleto{F}{0.4em}}^2\,\bar{\lambda}_{\scaleto{1}{4pt}}^2(b_{\pperp})\bar{\lambda}_{\scaleto{2}{4pt}}^2(b_{\pperp})\nonumber\\[-0.5em]
=\frac{1}{g^4}\alpha^4_s(4\pi\,\partial^2L(0_{\pperp}))^4\!N_c^4C_{\scaleto{F}{0.4em}}^2\,\bar{\lambda}_{\scaleto{1}{4pt}}^2(b_{\pperp})\bar{\lambda}_{\scaleto{2}{4pt}}^2(b_{\pperp})=\frac{C_{\scaleto{F}{0.4em}}^2}{g^4}\bar{Q}^4_{s\scaleto{1}{4pt}}\bar{Q}^4_{s\scaleto{2}{4pt}}(4\pi\,\partial^2L(0_{\pperp}))^4\approx\!\langle\epsilon_{\scaleto{0}{3.8pt}}(x_{\pperp})\rangle\langle\epsilon_{\scaleto{0}{3.8pt}}(y_{\pperp})\rangle,
\end{align}
where we approximated $h(x_{\pperp})$ and $h(y_{\pperp})$ with $h(b_{\pperp})$, as repeatedly done throughout the calculation. Therefore, the result of $\text{Cov}[\epsilon\,](\tau\!=\!0^+;x_{\pperp},y_{\pperp})\!=\!\langle \epsilon_{\scaleto{0}{3.8pt}}(x_{\pperp})\epsilon_{\scaleto{0}{3.8pt}}(y_{\pperp})\rangle\!-\!\langle \epsilon_{\scaleto{0}{3.8pt}}(x_{\pperp})\rangle\langle \epsilon_{\scaleto{0}{3.8pt}}(y_{\pperp})\rangle$ corresponds to the remaining terms. 
We use the Mathematica package FeynCalc \cite{MERTIG1991345, SHTABOVENKO2016432} to perform the contractions featured in \eqref{ResultExpansion}. After doing so we arrive at the main result of this work:
\begin{align}
\text{Cov}&[\epsilon\,](\tau\!=\!0^+;x_{\pperp},y_{\pperp})  \equiv \langle\epsilon_{\scaleto{0}{3.8pt}}(x_{\pperp})\epsilon_{\scaleto{0}{3.8pt}}(y_{\pperp})\rangle\!-\!\langle\epsilon_{\scaleto{0}{3.8pt}}(x_{\pperp})\rangle\langle\epsilon_{\scaleto{0}{3.8pt}}(y_{\pperp})\rangle= \nonumber \\
& \frac{\partial^{i}_{x}\Gamma\partial^{i}_{y}\Gamma(N_c^2-1)A(4A^2-B^2)}{16N_c^2\Gamma^5g^4}(p_{\scaleto{1}{4.2pt}}q_{\scaleto{2}{4.2pt}}+p_{\scaleto{2}{4.2pt}}q_{\scaleto{1}{4.2pt}}) \nonumber \\
&+\frac{(N_c^2-1)(16A^4+B^4)}{2N_c^2\Gamma^4g^4}p_{\scaleto{1}{4.2pt}}p_{\scaleto{2}{4.2pt}}+\frac{(\partial^{i}_{x}\Gamma\partial^{i}_{y}\Gamma)^2(N_c^2-1)A^2}{64N_c^2\Gamma^6g^4}q_{\scaleto{1}{4.2pt}}q_{\scaleto{2}{4.2pt}} \nonumber \\
& +\frac{(N_c^2-1)(4A^2+B^2)}{2N_c^2\Gamma^2g^4}\!\left(4\pi\,\partial^2L(0_{\pperp})\right)^2\!\left(\left[\bar{Q}_{s\scaleto{1}{4pt}}^4(Q_{s\scaleto{2}{4pt}}^2r^2-4+4e^{-\frac{Q_{s\scaleto{2}{4pt}}^2r^2}{4}})\right]+[1\leftrightarrow2]\right) \nonumber \\
& +\frac{(4A^2+B^2)^2}{g^4\Gamma^4N_c^2}\Bigg(\Bigg[ \frac{N_c^6+2 N_c^4-19 N_c^2+8}{(N_c^2-1)^2}-4\frac{N_c^6\!-3N_c^4\!-26N_c^2+16}{(N_c^2-1)(N_c^2-4)}e^{-\frac{Q_{s\scaleto{1}{4pt}}^2r^2}{4}} \nonumber \\
& +\frac{(N_c-1)(N_c+3)N_c^3}{(N_c+1)^2(N_c+2)^2}\!\left(\frac{N_c}{2}e^{-\frac{(N_c+1) r^2 Q_{s\scaleto{2}{4pt}}^2}{2 N_c}}\!+(N_c+2)-2(N_c+1)e^{-\frac{Q_{s\scaleto{2}{4pt}}^2r^2}{4}}\right)\!e^{-\frac{(N_c+1) r^2 Q_{s\scaleto{1}{4pt}}^2}{2 N_c}} \nonumber\\
& +\frac{(N_c+1)(N_c-3)N_c^3}{(N_c-1)^2(N_c-2)^2}\!\left(\frac{N_c}{2}e^{-\frac{(N_c-1) r^2 Q_{s\scaleto{2}{4pt}}^2}{2 N_c}}\!+(N_c-2)-2(N_c-1)e^{-\frac{Q_{s\scaleto{2}{4pt}}^2r^2}{4}}\right)\!e^{-\frac{(N_c-1) r^2 Q_{s\scaleto{1}{4pt}}^2}{2 N_c}} \nonumber \\
& +\frac{r^4}{2}Q_{s\scaleto{1}{4pt}}^2Q_{s\scaleto{2}{4pt}}^2-4r^2Q_{s\scaleto{1}{4pt}}^2\left(1-e^{-\frac{Q_{s\scaleto{2}{4pt}}^2r^2}{4}}\right) +4 \frac{(N_c^2 - 8) (N_c^2 - 1) (N_c^2 + 4)}{(N_c^2-4)^2} e^{-\frac{\left(Q_{s\scaleto{1}{4pt}}^2+Q_{s\scaleto{2}{4pt}}^2\right)r^2}{4}}\Bigg]\nonumber\\
&+[1\leftrightarrow2]\Bigg),\label{FullResult}
\end{align}
where the dependencies have been omitted for readability. The covariance of the full EMT is simply obtained from the previous expression as 
\begin{equation}
\text{Cov}[T^{\mu\nu}](0^+;x_{\pperp},y_{\pperp})\!=\!\text{Cov}[\epsilon\,](0^+;x_{\pperp},y_{\pperp})\!\times\!t^{\mu\nu}t^{\sigma\rho}.
\end{equation}
The factors $A(r_{\pperp})$ and $B(r_{\pperp})$ were introduced in \eqref{Derivatives}.  Explicit expressions for them in the general case are given in Appendix \ref{GreenDerivatives} and in Eqs. $\!$(\ref{ABMV1}), (\ref{ABMV2}) below for the specific case of the original MV model.  Also, in order to make our final result more compact we have defined:
\begin{align}
p_{\scaleto{1}{4pt},\scaleto{2}{4pt}}&\equiv e^{-\frac{Q_{s\scaleto{1}{4pt},\scaleto{2}{4pt}}^2r^2}{4}}(Q_{s\scaleto{1}{4pt},\scaleto{2}{4pt}}^2r^2+4)-4\\
q_{\scaleto{1}{4pt},\scaleto{2}{4pt}}&\equiv e^{-\frac{Q_{s\scaleto{1}{4pt},\scaleto{2}{4pt}}^2r^2}{4}}\left(Q_{s\scaleto{1}{4pt},\scaleto{2}{4pt}}^4r^4+8Q_{s\scaleto{1}{4pt},\scaleto{2}{4pt}}^2r^2+32\right)-32.
\end{align}
For simplicity, in the previous expressions we also defined the following momentum scale:
\begin{equation}\label{Saturation2}
\frac{r^2Q_s^2}{4}=g^2\frac{N_c}{2}\Gamma(r_{\pperp})\bar{\lambda}(b_{\pperp}),
\end{equation}
which is related to the one introduced in \secref{sec:t} by:
\begin{align}
Q_s^2(r_{\pperp},b_{\pperp})\overset{r\rightarrow 0}{=}\bar{Q}^2_s(b_{\pperp})\left(-4\pi\,\partial^2L(0_{\pperp})\right)\!.\label{satscales}
\end{align}
(See Appendix \ref{GreenDerivatives}). Since \eqref{FullResult} is somewhat lengthy, in the next sections we discuss a few simplifying limits in the context of the original MV model.

\subsection{$N_c$-expansion in the MV model} 
\label{sec:mv}

\begin{figure}[t]
\centering
\subfigure{\label{fig:2a}\includegraphics[width=0.47\textwidth]{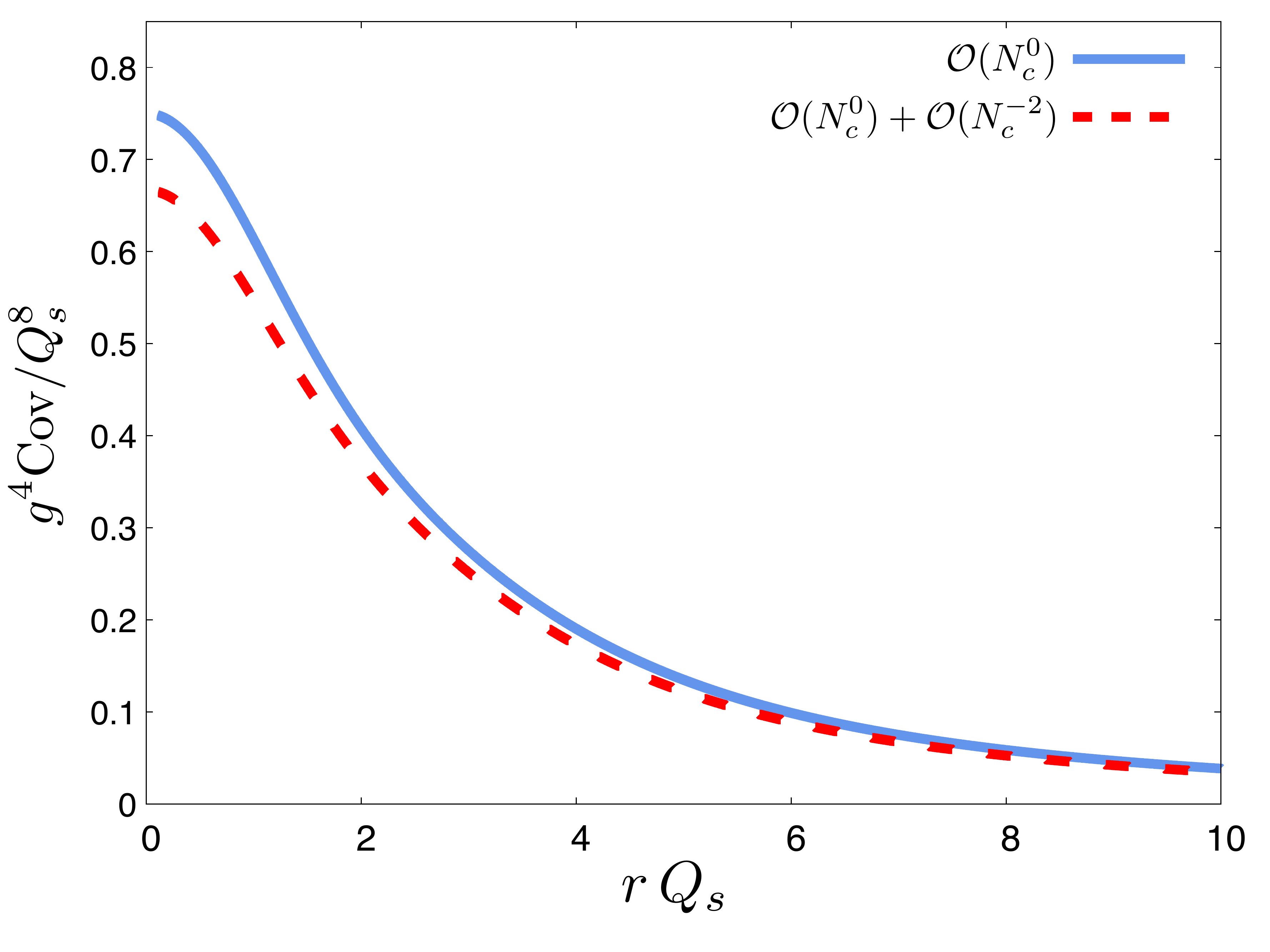}}
\subfigure{\label{fig:2a}\includegraphics[width=0.47\textwidth]{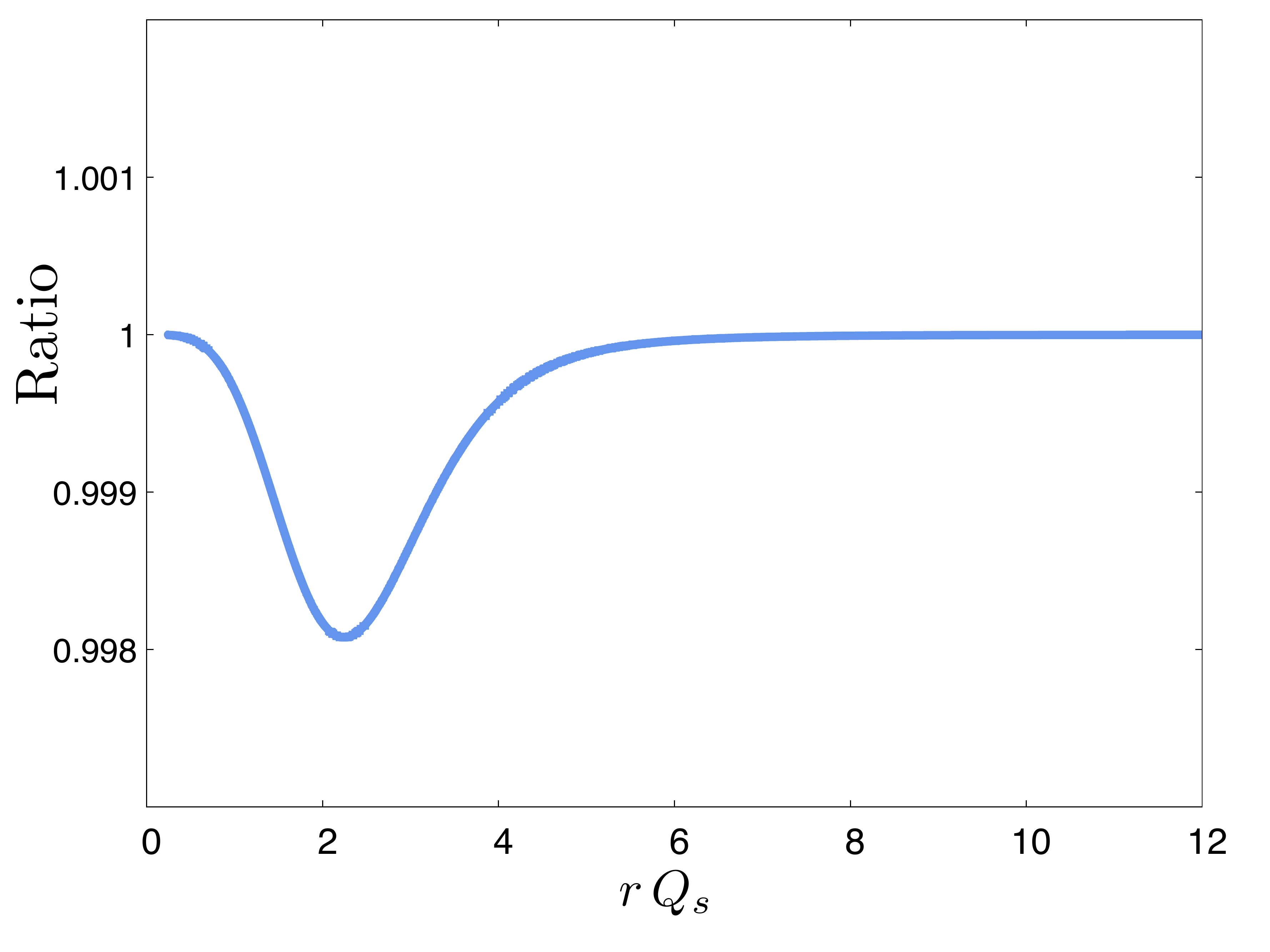}}
\caption{LEFT: Sum of the first two orders of the $N_c$-expansion of the energy density covariance against $rQ_s$ for $Q_{s\scaleto{1}{4pt}}\!=\!Q_{s\scaleto{2}{4pt}}$ and $N_c\!=\!3$. Blue full curve: $N_c^0$-order term. Red dashed curve: Sum of $N_c^0$- and $N_c^{-2}$-order terms. RIGHT: Ratio between the complete result and the sum of the first two orders of the $N_c$-expansion.}
\label{Plot2}
\end{figure}

Our generalization of the classical approach yields a few aspects of the previous calculation that had to be left undetermined. For instance, the function $f(r_{\pperp})$ featured in the two-point correlator of \eqref{2point} introduces some ambiguity in the computation of the double derivative of $L(r_{\pperp})$, which is left in terms of the unknown coefficients $A(r_{\pperp})$ and $B(r_{\pperp})$ (\eqref{Derivatives}). In the particular case of the MV model, where $f(r_{\pperp})$ is taken as a Dirac delta, we are able to compute them as:
\begin{align}
A(r_{\pperp})_{\scaleto{\text{MV}}{0.12cm}}&=-\frac{1}{2}G(r_{\pperp})=\frac{1}{4\pi}K_{\scaleto{0}{4.2pt}}(mr)\label{ABMV1}\\
B(r_{\pperp})_{\scaleto{\text{MV}}{0.12cm}}&=\frac{1}{4\pi},\label{ABMV2}
\end{align}
where $K_{\scaleto{0}{4.2pt}}$ is a modified Bessel function. The mass $m$ is an infrared scale that we introduce to regularize the divergent Green's function $G(r_{\pperp})$. For simplicity we choose $m$ to be the same mass scale introduced earlier in \eqref{Hierark}. %For a very small (although greater than 0) value of $m$, $A(r_{\pperp})_{\scaleto{\text{MV}}{0.12cm}}$ can be approximated as:
The leading behavior in the $m\!\rightarrow\!0$ limit is:
\begin{align}
A(r_{\pperp})_{\scaleto{\text{MV}}{0.12cm}}\approx\frac{1}{8\pi}\ln\!\left(\frac{4}{m^2r^2}\right)\!,
\end{align}
and $B_{\scaleto{\text{MV}}{0.12cm}}$, being a constant, yields a negligible correction to this logarithm. In the same limit, the leading behavior of $\Gamma(r_{\pperp})$ and the product of its derivatives corresponds to the following expressions:
\begin{align}
\Gamma(r_{\pperp})_{\scaleto{\text{MV}}{0.12cm}}&=\frac{1}{2\pi m^2}-\frac{r}{2\pi m}K_{\scaleto{1}{4.2pt}}(mr)\approx\frac{r^2}{8\pi}\ln\left(\frac{4}{m^2r^2}\right)\\
\left[\partial^{i}_{x}\Gamma\partial^{i}_{y}\Gamma\right]_{\scaleto{\text{MV}}{0.12cm}}&\approx-\frac{r^2}{16\pi^2}\ln\left(\frac{m^2r^2}{4}\right)^2,
\end{align}
and for the saturation scale:
\begin{align}
Q_s^2(r,b_{\pperp})_{\scaleto{\text{MV}}{0.12cm}}&\approx\bar{Q}^2_s(b_{\pperp})\ln\!\left(\frac{4}{m^2r^2}\right).
\end{align}
(See Appendix \ref{GreenDerivatives} for a detailed derivation of these expressions). These factors exhibit logarithmic divergences of different nature. While $\Gamma$ and $\partial^{i}_{x}\Gamma\partial^{i}_{y}\Gamma$ diverge only in the infrared limit $m\!\rightarrow\!0$, $A$ and $Q_s^2$ are divergent in both infrared and ultraviolet $r\!\rightarrow\!0$ limits. The latter case enters our solution explicitly through the terms multiplied by the factor $\partial^2L(0_{\pperp})\!\equiv\!-2\lim\limits_{r\rightarrow0}A(r_{\pperp})$. In the end those terms will be the only ones yielding a divergence, as the logarithms stemming from $A$, $\Gamma$ and $\partial^{i}_{x}\Gamma\partial^{i}_{y}\Gamma$ are exactly cancelled in \eqref{FullResult}. Therefore, the overall effect of taking the MV limit on the complete result of the energy density covariance only consists in replacing all $r$-depending coefficients (except the saturation scales) with constants. As this substitution does not yield a significant simplification to the final formula, instead of showing that result we prefer to display the first orders of its $N_c$-expansion.
%\begin{figure}
%	\centering
%	\includegraphics[width=0.48\textwidth]{Corrected/Corrections_good.pdf}
%	\caption{$N_c$-expansion of the covariance of energy density $\epsilon(x_{\pperp})$ against $r\!=\!|r_{\pperp}|$. We compare the leading order term (of order $N_c^0$, blue full curve), the first correction (of order $N_c^{-2}$, red dashed curve) and the next correction (of order $N_c^{-4}$, green dot-dashed curve).}
%	\label{Plot1}
%\end{figure}
Note that in these expressions we are not taking the complete, strict MV limit, which would imply $h(b_{\pperp})\!=\!1$; instead, we are only assuming locality in the transversal plane. The leading order of the expansion, of order $N^{0}_c$, reads:
\begin{align}
\left[\text{Cov}[\epsilon_{\scaleto{\text{\tiny MV}}{0.12cm}}](0^+;x_{\pperp},y_{\pperp})\right]_{N_c^0}=\left[\frac{1}{g^4\,r^8}e^{-\frac{r^2}{2} \left(Q_{s\scaleto{1}{4pt}}^2+Q_{s\scaleto{2}{4pt}}^2\right)}\!\left(16+32 e^{\frac{Q_{s\scaleto{1}{4pt}}^2 r^2}{2}}\right.\right.\nonumber\\
-64e^{\frac{Q_{s\scaleto{1}{4pt}}^2 r^2}{4}}\!-4 e^{\frac{r^2}{4} \left(2 Q_{s\scaleto{1}{4pt}}^2+Q_{s\scaleto{2}{4pt}}^2\right)}\!\left( Q_{s\scaleto{2}{4pt}}^4 r^4\!-2\left(4\pi\,\partial^2L(0_{\pperp})\right)^2\!\bar{Q}_{s\scaleto{1}{4pt}}^4 r^4+ 8\,Q_{s\scaleto{2}{4pt}}^2 r^2+48\right)\nonumber\\
+\frac{1}{8}e^{\frac{r^2}{4} \left(Q_{s\scaleto{1}{4pt}}^2+Q_{s\scaleto{2}{4pt}}^2\right)}\!\Big(Q_{s\scaleto{1}{4pt}}^4 Q_{s\scaleto{2}{4pt}}^4 r^8\!+(4\,Q_{s\scaleto{1}{4pt}}^2 Q_{s\scaleto{2}{4pt}}^2 r^6+128\,r^2)\!\left(Q_{s\scaleto{1}{4pt}}^2+Q_{s\scaleto{2}{4pt}}^2\right)\!+16\,r^4\!\left( Q_{s\scaleto{1}{4pt}}^2+ Q_{s\scaleto{2}{4pt}}^2\right)^2\!+1024\Big)\nonumber\\
\left.\left.+\,2 e^{\frac{r^2}{2} \left(Q_{s\scaleto{1}{4pt}}^2+Q_{s\scaleto{2}{4pt}}^2\right)}\!\left( \bar{Q}_{s\scaleto{1}{4pt}}^4r^4\!\left(Q_{s\scaleto{2}{4pt}}^2 r^2-4\right)\!\left(4\pi\,\partial^2L(0_{\pperp})\right)^2\!+40\right)\right)\right]\!+\left[1\leftrightarrow2\right]\!.\label{LargeN-NLO}
\end{align}
The next term, of order $N_c^{-2}$, reads:
\begin{align}
\left[\text{Cov}[\epsilon_{\scaleto{\text{\tiny MV}}{0.12cm}}](0^+;x_{\pperp},y_{\pperp})\right]_{N_c^{-2}}=\left[\frac{1}{N_c^2g^4\,r^8}e^{-\frac{r^2}{2} \left(Q_{s\scaleto{1}{4pt}}^2+Q_{s\scaleto{2}{4pt}}^2\right)}\bigg(2\left(Q_{s\scaleto{1}{4pt}}^2r^2+Q_{s\scaleto{2}{4pt}}^2r^2+8\right)^2\right.\nonumber\\
+4\,Q_{s\scaleto{1}{4pt}}^2r^2(8+Q_{s\scaleto{1}{4pt}}^2r^2)e^{\frac{Q_{s\scaleto{2}{4pt}}^2 r^2}{2}}\!-8(8+Q_{s\scaleto{1}{4pt}}^2r^2)(4+Q_{s\scaleto{1}{4pt}}^2r^2)e^{\frac{Q_{s\scaleto{2}{4pt}}^2 r^2}{4}}\nonumber\\
+4\, e^{\frac{r^2}{4} \left(2Q_{s\scaleto{1}{4pt}}^2+Q_{s\scaleto{2}{4pt}}^2\right)} \Big( Q_{s\scaleto{2}{4pt}}^4r^4-2(4\pi\,\partial^2L(0_{\pperp}))^2\bar{Q}_{s\scaleto{1}{4pt}}^4r^4+8Q_{s\scaleto{2}{4pt}}^2r^2+16 Q_{s\scaleto{1}{4pt}}^2r^2\Big)\nonumber\\
-\frac{1}{8}e^{\frac{r^2}{4} \left(Q_{s\scaleto{1}{4pt}}^2+Q_{s\scaleto{2}{4pt}}^2\right)}\!\Big(Q_{s\scaleto{1}{4pt}}^4 Q_{s\scaleto{2}{4pt}}^4 r^8+(4 Q_{s\scaleto{1}{4pt}}^2 Q_{s\scaleto{2}{4pt}}^2 r^6+128r^2)\!\left(Q_{s\scaleto{1}{4pt}}^2+Q_{s\scaleto{2}{4pt}}^2\right)+16\,r^4\!\left(Q_{s\scaleto{1}{4pt}}^2+ Q_{s\scaleto{2}{4pt}}^2\right)^2\!-1024\Big)\nonumber\\
-2\, e^{\frac{r^2}{2} \left(Q_{s\scaleto{1}{4pt}}^2+Q_{s\scaleto{2}{4pt}}^2\right)}\!\Big(\bar{Q}_{s\scaleto{1}{4pt}}^4r^4(Q_{s\scaleto{2}{4pt}}^2r^2-4)(4\pi\,\partial^2L(0_{\pperp}))^2+32Q_{s\scaleto{1}{4pt}}^2r^2-4\,Q^2_{s\scaleto{1}{4pt}}Q^2_{s\scaleto{2}{4pt}}r^4\Big)\bigg)\!\bigg]\!+\left[1\leftrightarrow 2\right].\label{LargeN-NNLO}
\end{align}
In order to give a general idea of the magnitude and analytical features of our solution, on \figref{Plot2} we draw these functions in the GBW model as a function of the dimensionless product $rQ_s$ for $Q_{s\scaleto{1}{4pt}}\!=\!Q_{s\scaleto{2}{4pt}}$. Note that in this limit, as we are ignoring all logarithmic factors, we also have $Q_s\!=\!\bar{Q}_s$.

The $N_c^{-2}$ term yields a small but noticeable negative correction (see red dashed curve of \figref{Plot2}). As the next terms are negligible, the first two orders of the $N_c$-expansion provide a neat approximation to the complete result (see right plot of \figref{Plot2}).
Comparing this curve with the $N_c^0$-order term we notice that the large-$N_c$ limit yields a 12.5\% error in the $r\!\rightarrow\!0$ limit, which is a reasonable approximation. In the $rQ_s\gg1$ limit our result vanishes following a power-law behavior. The leading term of this limit results from a combination of terms included in the first two orders of the $N_c$-expansion presented above, \eqref{LargeN-NLO} and \eqref{LargeN-NNLO}:
\begin{equation}
\lim_{r Q_s\gg1}\text{Cov}[\epsilon_{\scaleto{\text{\tiny MV}}{0.12cm}}](0^+;x_{\pperp},y_{\pperp})=\frac{2 \left(N_c^2-1\right)\left(4\pi\,\partial^2L(0_{\pperp})\right)^2\!\left(\bar{Q}_{s\scaleto{1}{4pt}}^4Q_{s\scaleto{2}{4pt}}^2+\bar{Q}_{s\scaleto{2}{4pt}}^4Q_{s\scaleto{1}{4pt}}^2\right)}{g^4N_c^2 r^2}.
\end{equation}
Note that this power-law tail is a non-trivial feature of our general result (shown in \eqref{FullResult}) that is also displayed in the particular case of the MV model.
Normalizing the previous result with a single one-point correlator we obtain the following expression:
\begin{align}
\lim_{r Q_s\gg1}\left(\frac{\text{Cov}[\epsilon](0^+;x_{\pperp},y_{\pperp})}{\langle \epsilon_{\scaleto{0}{3.8pt}}(x_{\pperp})\rangle}\right)_{\!\!\scaleto{\text{\tiny MV}}{0.12cm}}\!\!&=\!\frac{4}{g^2N_cr^2}\left(\frac{\bar{Q}_{s\scaleto{1}{4pt}}^2Q_{s\scaleto{2}{4pt}}^2}{\bar{Q}_{s\scaleto{2}{4pt}}^2}+\frac{\bar{Q}_{s\scaleto{2}{4pt}}^2Q_{s\scaleto{1}{4pt}}^2}{\bar{Q}_{s\scaleto{1}{4pt}}^2}\right).
\end{align}
In the opposite limit, $r\!\rightarrow\!0$, the covariance tends to:
\begin{align}
\lim_{r\rightarrow0}\text{Cov}[\epsilon_{\scaleto{\text{\tiny MV}}{0.12cm}}](0^+;x_{\pperp},y_{\pperp})=\frac{C_{\scaleto{F}{0.4em}}}{2N_cg^4}\left(Q_{s\scaleto{1}{4pt}}^4Q_{s\scaleto{2}{4pt}}^4+\left(4\pi\partial^2L(0_{\pperp})\right)^2\!\left(\bar{Q}_{s\scaleto{1}{4pt}}^4Q_{s\scaleto{2}{4pt}}^4+\bar{Q}_{s\scaleto{2}{4pt}}^4Q_{s\scaleto{1}{4pt}}^4\right)\right)\nonumber\\
=\frac{3C_{\scaleto{F}{0.4em}}}{2N_cg^4}\left(4\pi\partial^2L(0_{\pperp})\right)^4\bar{Q}_{s\scaleto{1}{4pt}}^4\bar{Q}_{s\scaleto{2}{4pt}}^4,
\end{align}
%which, using the physical value $N_c\!=\!3$, yields $\frac{2}{3}Q_{s\,\scaleto{1}{4pt}}^4Q_{s\,\scaleto{2}{4pt}}^4$ in the GBW model.
and the normalized covariance:
\begin{align}
\lim_{r\rightarrow 0}\left(\frac{\text{Cov}[\epsilon](0^+;x_{\pperp},y_{\pperp})}{\langle \epsilon_{\scaleto{0}{3.8pt}}(x_{\pperp})\rangle\langle \epsilon_{\scaleto{0}{3.8pt}}(y_{\pperp})\rangle}\right)_{\!\!\scaleto{\text{\tiny MV}}{0.12cm}}\!\!&=\!\frac{Q_{s\scaleto{1}{4pt}}^4Q_{s\scaleto{2}{4pt}}^4+\left(4\pi\partial^2L(0_{\pperp})\right)^2\!\left(\bar{Q}_{s\scaleto{1}{4pt}}^4Q_{s\scaleto{2}{4pt}}^4+\bar{Q}_{s\scaleto{2}{4pt}}^4Q_{s\scaleto{1}{4pt}}^4\right)}{\left(4\pi\partial^2L(0_{\pperp})\right)^4\bar{Q}_{s\scaleto{1}{4pt}}^4\bar{Q}_{s\scaleto{2}{4pt}}^4(N_c^2-1)}=\frac{3}{N^2_c-1}.
\end{align}
In both expressions we applied \eqref{satscales} in the last step.

\subsection{The Glasma Graph approximation} 
\label{sec:gg}
An alternative approach to this calculation is proposed in \cite{PhysRevD.97.034034} under the Glasma Graph approximation, whereby it is assumed that the four-point correlation functions of the WW fields characterizing the single nucleus solution of the classical Yang-Mills equations of motion can be factorized into products of two-point correlation functions such that:
\begin{align}
\langle \alpha^{i,a}(x_{\pperp}) \alpha^{k,c}(x_{\pperp}) \alpha^{i^{\prime}\!,a^{\prime}}(y_{\pperp})\alpha^{k^{\prime}\!,c^{\prime}}(y_{\pperp}) \rangle_{\scaleto{\text{GG}}{0.12cm}}=\,&\langle\alpha^{i,a}(x_{\pperp}) \alpha^{k,c}(x_{\pperp})\rangle\langle \alpha^{i^{\prime}\!,a^{\prime}}(y_{\pperp})\alpha^{k^{\prime}\!,c^{\prime}}(y_{\pperp}) \rangle\nonumber\\
+ &\langle \alpha^{i,a}(x_{\pperp})\alpha^{i^{\prime}\!,a^{\prime}}(y_{\pperp})\rangle\langle \alpha^{k,c}(x_{\pperp})\alpha^{k^{\prime}\!,c^{\prime}}(y_{\pperp})\rangle\nonumber\\
+ &\langle\alpha^{i,a}(x_{\pperp})\alpha^{k^{\prime}\!,c^{\prime}}(y_{\pperp})\rangle\langle\alpha^{k,c}(x_{\pperp})\alpha^{i^{\prime}\!,a^{\prime}}(y_{\pperp})\rangle.\label{GGallin}
\end{align}
\begin{figure}
\centering
\subfigure{\label{fig:2a}\includegraphics[width=0.47\textwidth]{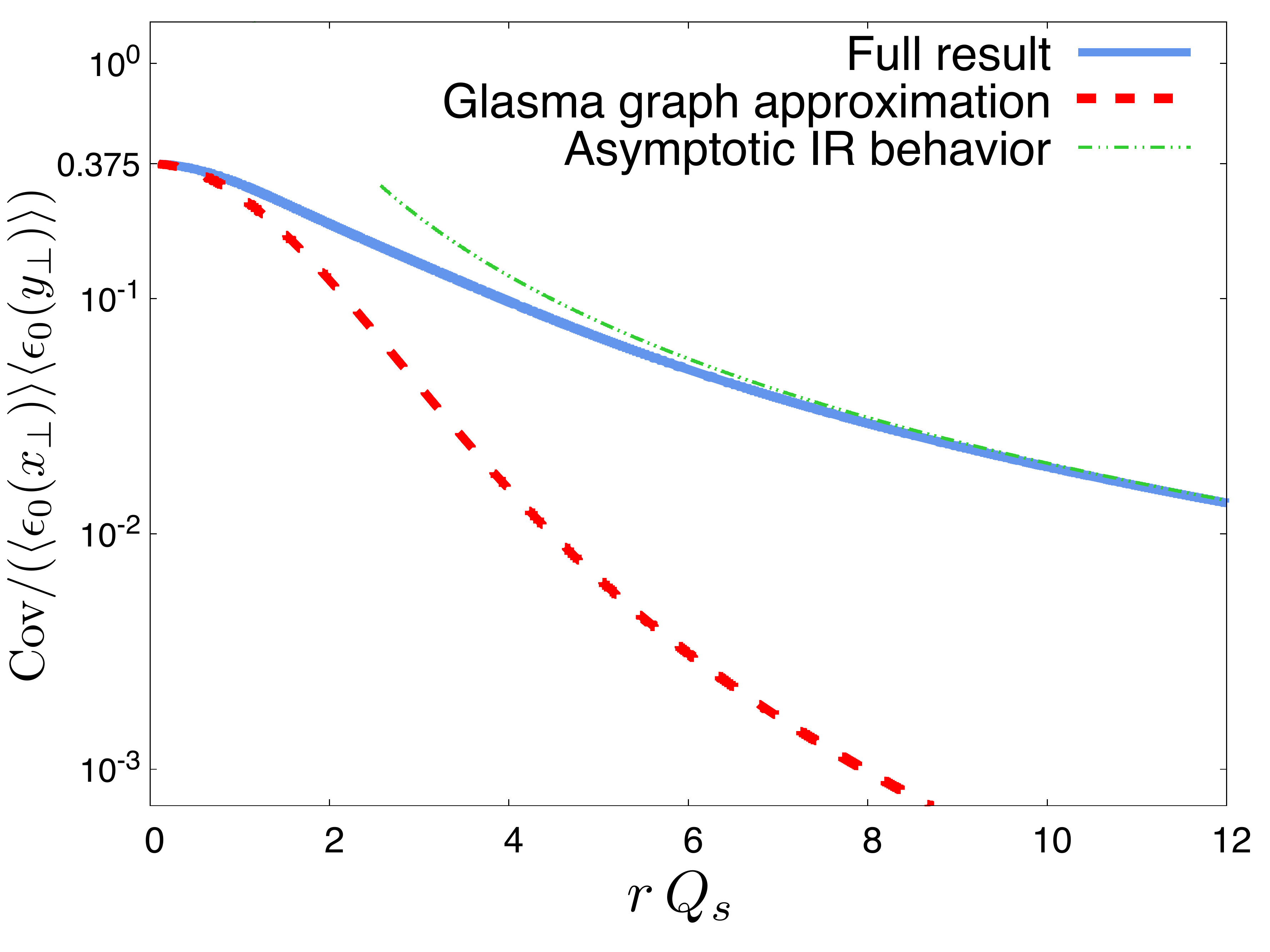}}
\subfigure{\label{fig:2a}\includegraphics[width=0.47\textwidth]{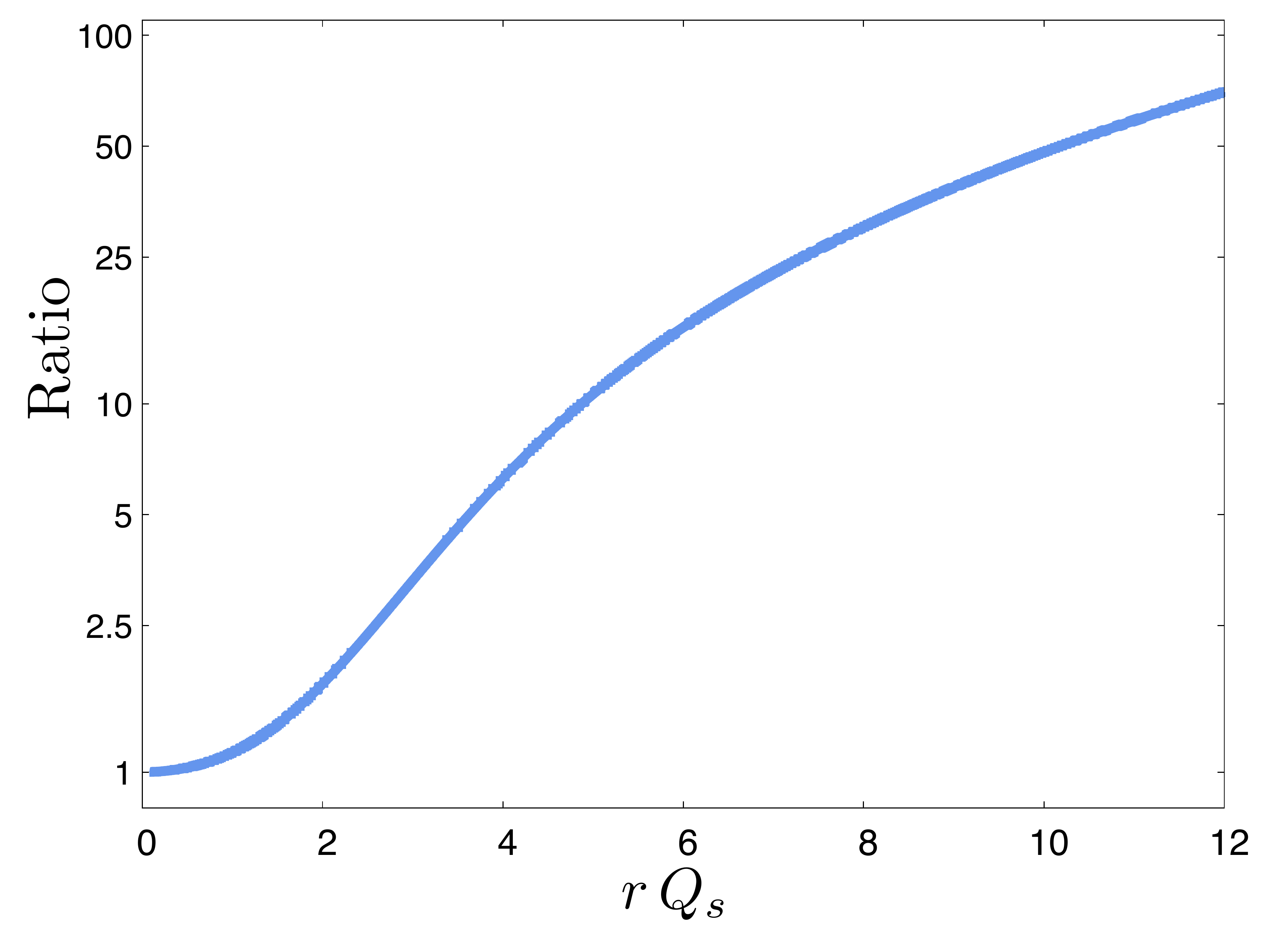}}
\caption{LEFT: Comparison of the normalized covariance of energy density $\epsilon_{\scaleto{0}{3.8pt}}$ against $r\,Q_s$ for $Q_{s\scaleto{1}{4pt}}\!=\!Q_{s\scaleto{2}{4pt}}$, $N_c\!=\!3$ in the exact analytical approach (blue full curve) and the Glasma Graph approximation (red dashed curve). As a visual aid we also indicate the asymptotic behavior in the IR limit, which is $16/[(N_c^2-1)r^2Q_s^2]$ (green dot-dashed curve). RIGHT: Ratio of exact analytical result to the Glasma Graph result.}
\label{Plot0}
\end{figure}This Wick theorem-like decomposition is equivalent to assuming that the WW fields obey Gaussian statistics. This is not generally correct, since the dynamics relating the Gaussianly distributed color sources and the generated gluon fields (encoded in the Yang-Mills equations) are non-linear. However, as we shall see, the Glasma Graph method remains a good approximation in the limit of small transverse separations $r\!\to\!0$. In that limit the dynamics linearize and reduce to two-gluon exchanges, effectively mapping a Gaussian distribution (for the color sources) onto another one (for the WW fields).
%The Glasma Graph approximation results in a factorization of double parton distributions into all possible products of single parton distributions, which yields great simplification in the context of the calculation of di-hadron correlators \cite{PhysRevD.97.034034}. In the same spirit, \eqref{GGallin} proposes a similar approach to the calculation of the EMT two-point correlator, which is expressed in terms of the building block needed for the calculation of the one-point EMT correlator, namely the two-point correlator of WW fields defined in \eqref{Correlator1}.

We compare the normalized covariance from our result (in the strict MV model) with the one computed according to the decomposition defined in \eqref{GGallin}. As can be seen in \figref{Plot0}, although both results agree exactly in the UV limit $r\!\rightarrow\!0$, in the rest of the spectrum our computation yields a harder curve. Another remarkable difference is that, while our result for the normalized covariance shows a slowly vanishing behavior in the infrared limit, the Glasma Graph approximation yields a much steeper tail:
\begin{equation}\label{GGcov}
\lim_{rQ_s\gg1}\left(\frac{\text{Cov}[\epsilon](0^+;x_{\pperp},y_{\pperp})}{\langle\epsilon_{\scaleto{0}{3.8pt}}(x_{\pperp})\rangle\langle\epsilon_{\scaleto{0}{3.8pt}}(y_{\pperp})\rangle}\right)_{\!\!\scaleto{\text{GG}}{0.3em}}\!=\frac{16(\bar{Q}_{s\scaleto{1}{4pt}}^4+\bar{Q}_{s\scaleto{2}{4pt}}^4)}{r^4(N_c^2-1)\bar{Q}_{s\scaleto{1}{4pt}}^4\bar{Q}_{s\scaleto{2}{4pt}}^4(4\pi\partial^2L(0_{\pperp}))^2}.
\end{equation}
The $\sim\!1/r^4$ decreasing behavior displayed by \eqref{GGcov} is in clear contrast with the $\sim\!1/r^2$ asymptotic behavior of our result. This potentially implies much different results and physical interpretations for any observable built from this quantity.

\section{Discussion and Outlook} 
\label{sec:end}
In this paper we provided an analytical expression for the covariance of the EMT characterizing the Glasma state produced in the early stages of an ultra-relativistic heavy ion collision. We performed this calculation in a classical framework based on the CGC effective theory, which we introduced by outlining the solution to the Yang-Mills equations for two nuclei at $\tau\!=\!0^+$.  In our approach we assumed a non-local two-point correlator of color source densities with an explicit impact parameter dependence. These modifications were introduced as small but non-negligible deviations from the original MV model. In this framework we obtained a remarkably lengthy, but still simple, formula whose general behavior we analyzed in the GBW model. We find that the first order of the $N_c$-expansion (of order $N_c^0$) yields a good approximation of the complete result (especially at large correlation distances), and that only the first correction to this term (of order $N_c^{-2}$) yields a non-negligible contribution. Finally, we compare our result with a recent calculation performed in the Glasma Graph approximation \cite{PhysRevD.97.034034}. In this work they assume a Gaussian-like decomposition of correlators at the level of the gluon fields rather than the color source densities, thus neglecting many contributions to the medium average. We find that this approximation quickly becomes unsatisfactory as we move out of the UV ($r\!\rightarrow\!0$) limit. In fact, the most striking difference emerges in the IR ($rQ_s\!\gg\!1$) limit, in which our result for the normalized covariance vanishes following a inverse square law whereas the Glasma Graph approximation yields a much more rapidly decreasing inverse fourth power. Arguably, these relatively long-range correlations could have an impact on any observable based on the integration of this quantity in the transverse plane, such as the mean square eccentricity fluctuations.

The results presented here provide a first step towards a first-principles computation of the initial conditions of the hydrodynamical expansion of QGP. The subsequent $\tau$-evolution up to thermalization time, as well as the calculation of observables relevant to QGP phenomenology, will be analyzed in a forthcoming publication.
%Specifically, we suggest that our results could be straightforwardly applied to the calculation of eccentricity fluctuations through the expressions presented in \cite{BLAIZOT2014166}.
%ALSO THIS LAST SENTENCE, OUT

\appendix

\section{Operations involving the 2-D Laplacian Green's function}\label{GreenDerivatives}

Throughout the computation of the covariance of $T^{\mu\nu\!}_{\scaleto{0}{3.8pt}}$ we encounter several non-trivial calculations involving the Green's function for the 2-dimensional Laplace operator $G(x_{\pperp}\!-y_{\pperp})$. For instance, when computing the correlator of two gluon fields (\eqref{Correlator1}), we find:
\begin{align}
\frac{1}{\nabla^2_x}\frac{1}{\nabla^2_y}\left(h(b_{\pperp})f(x_{\pperp}\!-y_{\pperp})\right)&=\!\!\int dz^2_{\pperp}du^2_{\pperp}G(z_{\pperp}\!-x_{\pperp})G(u_{\pperp}\!-y_{\pperp})h\!\left(\frac{z_{\pperp}+u_{\pperp}}{2}\right)\!f(z_{\pperp}\!-u_{\pperp}).\label{Green1}
\end{align}
This expression includes two undetermined functions, $h(b_{\pperp})$ and $f(x_{\pperp}\!-y_{\pperp})$, introduced in the two-point correlator (\eqref{2point}) in order to generalize the MV model. However, we do not take these functions as completely general. For $h(b_{\pperp})$, in addition to overall good analytical properties, we assume a slowly varying behavior over lengths of the order of a length scale $1/m$ or smaller (as proposed in \cite{PhysRevC.92.064912}):
\begin{align}
|h(b_{\pperp})|\gg m^{-1}|\partial^{i}h(b_{\pperp})|\gg m^{-2}|\partial^{i}\partial^{j}h(b_{\pperp})|\gg ...\label{MHierarchy}
\end{align}
where we take $m$ as the infrared regulator. We require that:
\begin{align}
\frac{1}{Q_s}\ll \frac{1}{m} \ll R_A,
\end{align}
where $R_A$ is the nuclear radius. Thus, the interaction distances of interest in our calculation obey $r=|x_{\pperp}\!-y_{\pperp}|\ll m^{-1}$. This requirement, as well as the assumed behavior for $h(b_{\pperp})$, yield a significant simplification to \eqref{Green1}. To see this, we expand $h\left((z_{\pperp}\!+u_{\pperp})/2\right)$ around $b_{\pperp}\!=\!(x_{\pperp}\!+y_{\pperp})/2$:
\begin{align}
h(b'_{\pperp})=h(b_{\pperp})+(b'_{\pperp}-b_{\pperp})^{i}\partial^{i}h(b_{\pperp})+...
\end{align}
where $b'_{\pperp}\!\!=\!(z_{\pperp}\!+u_{\pperp})/2$. Cutting the expansion at first order, \eqref{Green1} yields the following terms:
\begin{align}
h(b_{\pperp})\!\!\int\!d^2 z_{\pperp} d^2 u_{\pperp} G(z_{\pperp}-x_{\pperp})G(u_{\pperp}-y_{\pperp})f(z_{\pperp}-u_{\pperp})\nonumber\\
+\,\partial^{i}h(b_{\pperp})\!\!\int\!d^2 z_{\pperp} d^2 u_{\pperp} G(z_{\pperp}-x_{\pperp})G(u_{\pperp}-y_{\pperp})(b'_{\pperp}-b_{\pperp})^{i}f(z_{\pperp}-u_{\pperp}).\label{GreenFirst}
\end{align}
First, we focus on the leading order term:
\begin{align}
h(b_{\pperp})\!\!\int\!d^2 z_{\pperp} d^2 u_{\pperp} G(z_{\pperp}-x_{\pperp})G(u_{\pperp}-y_{\pperp})f(z_{\pperp}-u_{\pperp})\equiv h(b_{\pperp})L(x_{\pperp}-y_{\pperp}).\label{Green0}
\end{align}
In order to further transform $L(x_{\pperp}\!-y_{\pperp})$ we go to momentum space. The Green's function $G(x_{\pperp}\!-y_{\pperp})$ admits a simple Fourier representation:
\begin{align}
G(x_{\pperp}-y_{\pperp})=-\!\!\int\! \frac{d^2k_{\pperp}}{(2\pi)^2}\frac{e^{ik_{\pperp}\!\cdot(x_{\pperp}-y_{\pperp})}}{k^2},\label{Green}
\end{align}
which we substitute in $L(x_{\pperp}\!-y_{\pperp})$, yielding:
\begin{align}
L(x_{\pperp}-y_{\pperp})&=\!\!\int\!\frac{d^2z_{\pperp}}{(2\pi)^2}\frac{d^2u_{\pperp}}{(2\pi)^2}\frac{d^2k_{\pperp}}{k^2}\frac{d^2q_{\pperp}}{q^2}e^{ik_{\pperp}\!\cdot(z_{\pperp}-x_{\pperp})}e^{iq_{\pperp}\!\cdot(u_{\pperp}-y_{\pperp})}f(z_{\pperp}-u_{\pperp})\nonumber\\
&=\!\!\int\!\frac{d^2q_{\pperp}}{(2\pi)^2}\hat{f}(q_{\pperp})\frac{e^{iq_{\pperp}\!\cdot r_{\pperp}}}{q^4}.\label{Lready}
\end{align}
In the last step we introduced the inverse Fourier transform of $f$, defined as:
\begin{align}
\hat{f}(q_{\pperp})=\!\!\int\! d^2w_{\pperp}e^{-iq_{\pperp}\!\cdot w_{\pperp}}f(w_{\pperp}).
\end{align}
Now we turn to the linear term of the expansion (second term of \eqref{GreenFirst}), which we want to compare with $h(b_{\pperp})L(r_{\pperp})$. By performing a simple variable change, it can be written as:
\begin{align}
\frac{1}{4}\partial^{i}h(b_{\pperp})\!\!\int\! d^2 v_{\pperp} d^2 w_{\pperp} G\left(\frac{v_{\pperp}+w_{\pperp}}{2}-\frac{r_{\pperp}}{2}\right)G\left(\frac{v_{\pperp}-w_{\pperp}}{2}+\frac{r_{\pperp}}{2}\right)\frac{(v_{\pperp})^{i}}{2}f(w_{\pperp}),\label{Green2}
\end{align}
where $v_{\pperp}\!=\!z_{\pperp}\!+u_{\pperp}$ and $w_{\pperp}\!=\!z_{\pperp}\!-u_{\pperp}$. Substituting \eqref{Green} and performing some transformations, we get to:
\begin{align}
\frac{1}{8}\partial^{i}h(b_{\pperp})\!\!\int\!\frac{d^2 v_{\pperp}}{(2\pi)^4} \frac{d^2k_{\pperp}}{k^{2}}\frac{d^2q_{\pperp}}{q^{2}}e^{i(k_{\pperp}+q_{\pperp})\cdot v_{\pperp}}e^{-i(k_{\pperp}-q_{\pperp})\cdot r_{\pperp}}(v_{\pperp})^{i}\hat{f}(q_{\pperp}-k_{\pperp}).\label{Green3}
\end{align}
The integration in $v_{\pperp}$ yields a distribution derivative of the Dirac delta function:
\begin{align}
\int d^2 v_{\pperp}e^{i(k_{\pperp}+q_{\pperp})\cdot v_{\pperp}}(v)^{i}=-i(2\pi)^2\partial^{i}\delta(k_{\pperp}+q_{\pperp}).
\end{align}
Substituting this result in \eqref{Green3} and integrating by parts, we finally obtain:
\begin{align}
\frac{1}{2}r^{i}_{\pperp}\partial^{i}h(b_{\pperp})\!\!\int\! \frac{d^2q_{\pperp}}{(2\pi)^2}\hat{f}(q_{\pperp})\frac{e^{iq_{\pperp}\!\cdot r_{\pperp}}}{q^4}=\frac{1}{2}r^{i}_{\pperp}\partial^{i}h(b_{\pperp})L(r_{\pperp}),
\end{align}
and thus, \eqref{GreenFirst} yields:
\begin{align}
\left(h(b_{\pperp})+\frac{1}{2}r^{i}_{\pperp}\partial^{i}h(b_{\pperp})\right)L(r_{\pperp})\approx h(b_{\pperp})L(r_{\pperp}).
\end{align}
%Vector notation here?
Here we applied the fact that $r^{i}_{\pperp}\partial^{i}h(b_{\pperp})\leq |r_{\pperp}| |\vec{\partial} h(b_{\pperp})| \ll m^{-1}|\vec{\partial} h(b_{\pperp})| \ll |h(b_{\pperp})|$. We will take this expression as a good approximation of \eqref{Green1}. The next step in the calculation of the two gluon field correlator \eqref{Correlator1} is the computation of the double derivative:
\begin{align}
\partial^{i}_{x}\partial^{j}_{y} \left(h(b_{\pperp})L(r_{\pperp})\right)&= (\partial^{i}_x\partial^{j}_yh)L+(\partial^{j}_yh)(\partial^{i}_xL)+(\partial^{i}_xh)(\partial^{j}_yL)+h(\partial^{i}_x\partial^{j}_yL)\nonumber\\
&\approx h(b_{\pperp})\partial^{i}_x\partial^{j}_yL(r_{\pperp}).
\end{align}
The reasoning behind the last approximate equality follows from the dimension of $L(r_{\pperp})$, its IR behavior, and the fact that we imposed an infrared cut-off mass scale $m$. In order to be able to discuss $L(r_{\pperp})$ in the infrared region we need to assume a certain behavior of $\hat{f}(q_{\pperp})$ in this regime. We assume $\hat{f}_{\text{IR}}\!\sim\!1$, just like in the MV model, as we do not expect other possible choices of models to differ in that regime. Then, we can safely assume that $L\!\propto\!m^{-2}$, which makes the term $(\partial^{i}_x\partial^{j}_yh)L$ suppressed with respect to $\partial^{i}_x\partial^{j}_yL$ (a dimensionless object). Also, this takes us to $\partial^{i}L\!\propto\!m^{-1}$, making the terms of the form $(\partial^{j}h)(\partial^{i}L)$ negligible as well. Thus, we are left with the following double derivative:
%At a later stage, the calculation of the covariance requires the explicit computation of the following double derivative:
\begin{equation}\label{Derivatives0}
\partial^{i}_{x}\partial^{j}_{y}L(r_{\pperp})=\!\int\!\frac{d^2q_{\pperp}}{(2\pi)^2}\hat{f}(q_{\pperp})\frac{e^{iq_{\pperp}\!\cdot r_{\pperp}}}{q^4}q^{i}q^{j}.
\end{equation}
From its symmetries and dimension, the previous expression can be parameterized as:
\begin{equation}\label{DerivativesMV0}
\partial^{i}_{x}\partial^{j}_{y}L(r_{\pperp})=A(r_{\pperp})\delta^{ij}+B(r_{\pperp})\!\left( \frac{\delta^{ij}}{2}-\frac{r^{i}r^{j}}{r^2}\right)\!.
\end{equation}
A priori, this decomposition is not possible when $r\!\rightarrow\!0$. However, as it is a symmetric object in $i$, $j$, we can make a different parameterization in this limit:
\begin{equation}\label{Derivatives02}
\lim_{r\rightarrow0}\partial^{i}_{x}\partial^{j}_{y}L(r_{\pperp})=C\delta^{ij},
\end{equation}
that we can relate to:
\begin{equation}
\partial^{i}_{x}\partial^{j}_{x}L(r_{\pperp})=\frac{\partial}{\partial x^{i}}\left( \frac{\partial}{\partial y^{j}}\frac{\partial y^{j}}{\partial r^{j}}\frac{\partial r^{j}}{\partial x^{j}} \right)L(r_{\pperp})=-\partial^{i}_{x}\partial^{j}_{y}L(r_{\pperp}).
\end{equation}
Now, taking the limit $r\rightarrow 0$:
\begin{equation}
\lim_{r\rightarrow0}\partial^{i}_{x}\partial^{j}_{x}L(r_{\pperp})=-\lim_{r\rightarrow0}\partial^{i}_{x}\partial^{j}_{y}L(r_{\pperp})=-C\delta^{ij}
\end{equation}
and contracting with $\delta^{ij}$:
\begin{equation}
\delta^{ij}\lim_{r\rightarrow0}\partial^{i}_{x}\partial^{j}_{x}L(r_{\pperp})=-2C\equiv \partial^2_{\pperp}L(0_{\pperp}),
\end{equation}
we have $C=-\frac{1}{2}\partial^2_{\pperp}L(0_{\pperp})$, which is the notation we use in the body of the article (the same as in \cite{PhysRevC.79.024909}). We can express these coefficients in terms of $\hat{f}(q_{\pperp})$ by computing the following projections of \eqref{Derivatives0} and \eqref{Derivatives02}:
\begin{align}
A(r_{\pperp})&=\!\frac{1}{2}\delta^{ij}\partial^{i}_{x}\partial^{j}_{y}L(r_{\pperp})=\!\frac{1}{2}\!\int\!\frac{d^2q_{\pperp}}{(2\pi)^2}\hat{f}(q_{\pperp})\frac{e^{iq_{\pperp}\!\cdot r_{\pperp}}}{q^2}\\
B(r_{\pperp})&=\!2\!\left( \frac{\delta^{ij}}{2}-\frac{r^{i}r^{j}}{r^2}\right)\!\partial^{i}_{x}\partial^{j}_{y}L(r_{\pperp})=\!\int\!\frac{d^2q_{\pperp}}{(2\pi)^2}\hat{f}(q_{\pperp})\frac{e^{iq_{\pperp}\!\cdot r_{\pperp}}}{q^4}q^{i}q^{j}\left( \delta^{ij}-2\frac{r^{i}r^{j}}{r^2}\right)\nonumber\\
&=\!\int\!\frac{d^2q_{\pperp}}{(2\pi)^2}\hat{f}(q_{\pperp})\frac{e^{iq_{\pperp}\!\cdot r_{\pperp}}}{q^4}\left( q^2-2\frac{q^{i}r^{i}q^{j}r^{j}}{r^2}\right)\nonumber\\
&=\!\int\!\frac{d^2q_{\pperp}}{(2\pi)^2}\hat{f}(q_{\pperp})\frac{e^{iq\,r\cos\theta}}{q^2}\left( 1-2\cos^2\theta\right)=-\int\!\frac{d^2q_{\pperp}}{(2\pi)^2}\hat{f}(q_{\pperp})\frac{e^{iq\,r\cos\theta}}{q^2}\cos(2\theta)\\
C&=\frac{1}{2}\delta^{ij}\lim_{r\rightarrow0}\partial^{i}_x\partial^{j}_yL(r_{\pperp})=\frac{1}{2}\int\frac{d^2q_{\pperp}}{(2\pi)^2}\hat{f}(q_{\pperp})\frac{1}{q^2}.
\end{align}
Note that, as $\lim\limits_{r\rightarrow0}A(r_{\pperp})\!=\!C$ and $\lim\limits_{r\rightarrow0}B(r_{\pperp})\!=\!0$, this parameterization of $\partial^{i}_{x}\partial^{j}_{y}L(r_{\pperp})$ is continuous in $r$. We can relate $C$ to the factor $\Gamma$, defined as:
\begin{align}
\Gamma(x_{\pperp}-y_{\pperp})=2(L(0_{\pperp})-L(x_{\pperp}-y_{\pperp}))=2\!\int\frac{d^2q_{\pperp}}{(2\pi)^2}\frac{\hat{f}(q_{\pperp})}{q^4}(1-e^{iq_{\pperp}\!\cdot r_{\pperp}}),
\end{align}
by taking the limit $r\!\rightarrow\!0$:
\begin{align}
\lim_{r\rightarrow0}\Gamma(x_{\pperp}-y_{\pperp})=2\!\int\frac{d^2q_{\pperp}}{(2\pi)^2}\frac{\hat{f}(q_{\pperp})}{q^4}\left(-i(q_{\pperp}\!\cdot r_{\pperp})+\frac{1}{2}(q_{\pperp}\!\cdot r_{\pperp})^2\right)&=\frac{r^2}{2}\!\int\frac{d^2q_{\pperp}}{(2\pi)^2}\frac{\hat{f}(q_{\pperp})}{q^2}\nonumber\\
&=r^2C=r^2\!\left(-\frac{1}{2}\partial^2_{\pperp}L(0_{\pperp})\right)\!,
\end{align}
where we assumed that $\hat{f}(q_{\pperp})\!=\!\hat{f}(|q_{\pperp}|)$.

\subsubsection*{The MV model}
In the specific case where $f(z_{\pperp}\!-w_{\pperp})\!=\!\delta^{2}(z_{\pperp}\!-w_{\pperp})$, i.e.$\;$the MV model, we have $\hat{f}(q_{\pperp})\!=\!1$ and thus we can explicitly compute our coefficients:
\begin{align}
A(r_{\pperp})_{\scaleto{\text{MV}}{0.12cm}}&=-\frac{1}{2}G(r_{\pperp})\\
B(r_{\pperp})_{\scaleto{\text{MV}}{0.12cm}}&=\!-\!\int^{\infty}_0\int^{2\pi}_0\!\frac{dq\,d\theta}{(2\pi)^2}\frac{e^{iq\,r\cos\theta}}{q}\cos(2\theta)=\!\frac{1}{2\pi}\!\int^{\infty}_0\frac{dq}{q}J_{\scaleto{2}{4.2pt}}(q\,r)=\frac{1}{4\pi}\\
C_{\scaleto{\text{MV}}{0.12cm}}&=\frac{1}{4\pi}\int\frac{dq}{q}=-\frac{1}{2}\lim_{r\rightarrow0} G(r_{\pperp}).
\end{align}
Both $A(r_{\pperp})_{\scaleto{\text{MV}}{0.12cm}}$ and $C_{\scaleto{\text{MV}}{0.12cm}}$ yield an infrared logarithmic divergence, which we deal with by introducing a regularizing mass in the Fourier representation of $G(r_{\pperp})$:
\begin{equation}
G(r_{\pperp})=\!-\!\int \frac{d^2q_{\pperp}}{(2\pi)^2}\frac{e^{iq_{\pperp}\!\cdot r_{\pperp}}}{q^2+m^2}=-\frac{1}{2\pi}K_{\scaleto{0}{4.2pt}}(m\,r),
\end{equation}
where $K_{\scaleto{0}{4.2pt}}$ is a modified Bessel function. For simplicity we choose $m$ to be the same mass scale introduced earlier in \eqref{MHierarchy} (although it could be an unrelated infrared scale). %For very small (although greater than 0) values of $m$ we get to:
In our calculation we will keep only the leading behavior in the $m\!\rightarrow\!0$ limit, which is:
\begin{align}
A(r_{\pperp})_{\scaleto{\text{MV}}{0.12cm}}\approx-\frac{1}{4\pi}\left( \ln{\left(\frac{mr}{2}\right)}+\gamma\right)\approx-\frac{1}{4\pi} \ln{\left(\frac{mr}{2}\right)},
\end{align}
(where $\gamma$ is the Euler constant) and thus:
\begin{align}
\partial^{i}_{x}\partial^{j}_{y}L(r_{\pperp})_{\scaleto{\text{MV}}{0.12cm}}\approx\frac{1}{4\pi}\left[-\delta^{ij}\ln{\left(\frac{mr}{2}\right)}+\left( \frac{\delta^{ij}}{2}-\frac{r^{i}r^{j}}{r_{\pperp}^2}\right)\right].\label{GRes1}
\end{align}
The coefficient $C_{\scaleto{\text{MV}}{0.12cm}}$ corresponds to the UV limit of the previous expression ($r\!\rightarrow\!0$):
\begin{align}
\lim_{r\rightarrow0}\partial^{i}_{x}\partial^{j}_{y}L(r_{\pperp})_{\scaleto{\text{MV}}{0.12cm}}=C_{\scaleto{\text{MV}}{0.12cm}}\delta^{ij}\approx\frac{\delta^{ij}}{4\pi}\lim_{r\rightarrow0}\left[\ln{\left(\frac{2}{mr}\right)}\right],
\end{align}
and thus:
\begin{align}
\partial^2_{\pperp}L(0_{\pperp})_{\scaleto{\text{MV}}{0.12cm}}=\frac{1}{4\pi}\lim_{r\rightarrow0}\left[\ln{\left(\frac{m^2r^2}{4}\right)}\right],
\end{align}
which also exhibits a logarithmic divergence. As for $\Gamma$, we have:
\begin{align}
\Gamma(x_{\pperp}-y_{\pperp})_{\scaleto{\text{MV}}{0.12cm}}=2\!\int\frac{d^2q_{\pperp}}{(2\pi)^2}\frac{1}{(q^2+m^2)^2}(1-e^{iq_{\pperp}\!\cdot r_{\pperp}})=\frac{1}{2\pi m^2}-\frac{r}{2\pi m}K_{\scaleto{1}{4.2pt}}(mr).
\end{align}
The leading behavior of the previous expression in the $m\!\rightarrow\!0$ yields:
\begin{align}
\Gamma(x_{\pperp}-y_{\pperp})_{\scaleto{\text{MV}}{0.12cm}}\approx-\frac{r^2}{8\pi}\left(\log\left(\frac{m^2r^2}{4}\right)+2\gamma-1\right)\approx\frac{r^2}{8\pi}\log\left(\frac{4}{m^2r^2}\right).
\end{align}

\section{Correlators of $n$ Wilson lines and $m$ external color sources}\label{BigCorrelator}
\begin{figure}
\centering
\includegraphics[width=0.85\textwidth]{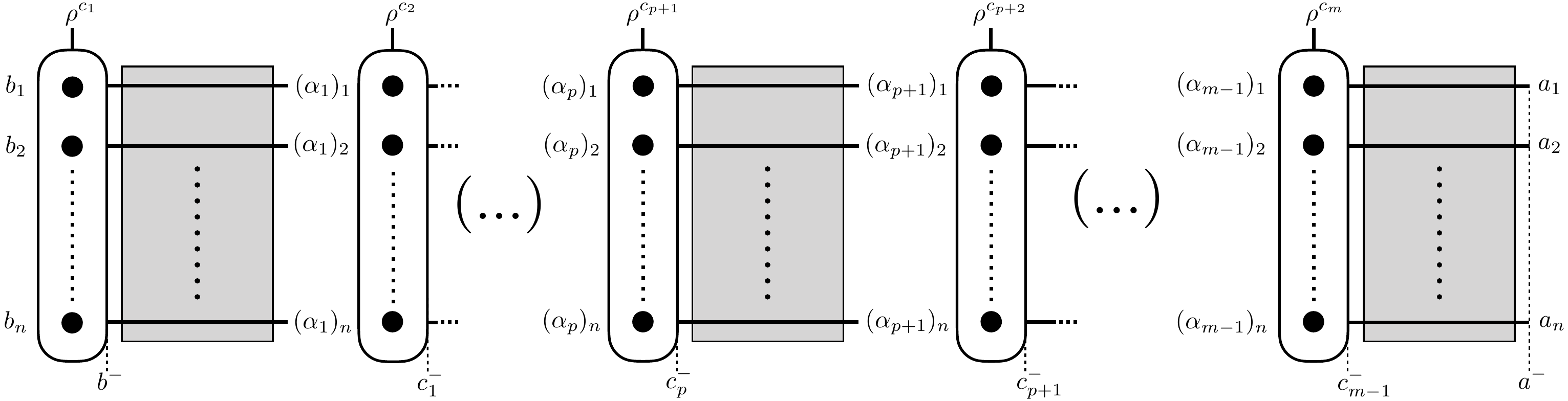}
\caption{Schematic representation of the connected correlator $H^{m,n}$ (see \eqref{fformula1}) for the particular case featured in our calculation. The oblong shapes represent the sum of all possible contractions between an external source and $n$ Wilson lines, whose correlator is represented as a dark square (see \figref{H1n}).}
\label{Hmn}
\end{figure}

In this appendix we expand on the calculation of the correlator featured in \eqref{MainCorrelator}:
\begin{align}
\left\langle \frac{\partial^{i}\tilde{\rho}^{e}(z^-,x_{\pperp})}{\nabla^2}U^{ea}(z^-,x_{\pperp})\frac{\partial^{k}\tilde{\rho}^{f}(w^-,x_{\pperp})}{\nabla^2}U^{fc}(w^-,x_{\pperp})\frac{\partial^{i^{\prime}}\tilde{\rho}^{e^{\prime}}(z^{-\prime},y_{\pperp})}{\nabla^2}U^{e'\!a'}(z^{-\prime},y_{\pperp})\right.\nonumber\\
\left.\times\frac{\partial^{k^{\prime}}\tilde{\rho}^{f^{\prime}}(w^{-\prime},y_{\pperp})}{\nabla^2}U^{f'\!c'}(w^{-\prime},y_{\pperp}) \right\rangle,\label{fformula0}
\end{align}
which we perform by application of the techniques derived in \cite{PhysRevC.79.025204}. In one of the appendices of said work they analyze the general case of the correlator of $n$ Wilson lines and $m$ color charge densities\footnote{\eqref{fformula} is derived for the cases where $m$ is even. In \cite{PhysRevC.79.025204} the odd $m$ formula is also provided.}:
\begin{align}
F^{m,n}(b^-,a^-)\equiv G^{m}H^{0,n}\nonumber\\
+\sum_{i,j,i<j}G^{m-2}_{(1,...,i-1,\{i\},i+1,...,j-1,\{j\},j+1,...,m)}H^{2,n}_{(\{1,...,i-1\},i,\{i+1,...,j-1\},j,\{j+1,...,m\})}\nonumber\\
+\sum_{i,j,k,l,i<j<k<l}G^{m-4}_{(1,...i-1,\{i\},i+1,...,j-1,\{j\},j+1,...,k-1,\{k\},k+1,...,l-1,\{l\},l+1,...,m)}\nonumber\\
\times H^{4,n}_{(\{1,...,i-1\},i,\{i+1,...,j-1\},j,\{j+1,...,k-1\},k,\{k+1,...,l-1\},l,\{l+1,...,m\})}\nonumber\\
+...+\sum_{i,j,i<j}G^{2}_{(\{1,...,i-1\},i,\{i+1,...,j-1\},j,\{j+1,...,m\})}H^{2,n}_{(1,...,i-1,\{i\},i+1,...,j-1,\{j\},j+1,...,m)}+H^{m,n},\label{fformula}
\end{align}
where
\begin{equation}\label{AppG}
G^{m-1}_{(1,...,j-1,\{j\},j+1,...,m)}\equiv\langle \rho_{\scaleto{1}{4pt}}...\rho_{\scaleto{j-1}{5.2pt}}\rho_{\scaleto{j+1}{5.2pt}}...\rho_{\scaleto{m}{3pt}}\rangle
\end{equation}
is the correlator of $m\!-\!1$ color charge densities. In the notation adopted here, the indices corresponding to sources that are `missing' from the correlators are indicated by brackets $\{...\}$. We also have:
\begin{equation}\label{AppH}
H^{j,n}_{(\{1,...,J_{1}-1\},J_{1},\{J_{1}+1,...,J_{2}-1\},J_{2},\{J_{2}+1,...\}...\{J_{j}-1\},J_{j},\{J_{j+1},...,m\})}\equiv\langle \rho_{\scaleto{J_{1}}{4.8pt}}\rho_{\scaleto{J_{2}}{4.8pt}}...\rho_{\scaleto{J_{j}}{5.6pt}}U_{\scaleto{1}{4.2pt}}...U_{\scaleto{n}{3.2pt}}\rangle_{\hspace{-0.02cm}\text{\tiny c}},
\end{equation}
which is the `connected' correlator of $n$ Wilson lines with $j$ insertions of external sources at the positions $J_{\scaleto{1}{4.2pt}}$, $J_{\scaleto{2}{4.2pt}}$, ... $J_{\scaleto{j}{5.6pt}}$ (with $J_{\scaleto{1}{4.2pt}}<J_{\scaleto{2}{4.2pt}}<...<J_{\scaleto{j}{5.6pt}}$). This is a special kind of correlator that does not include contractions between color sources outside the Wilson lines. Therefore, when computing it, any of these external sources can only be linked to those arranged inside Wilson lines. This object can be factorized as:
\begin{align}
H^{m,n}(b^-,a^-|\{b\},\{a\})= H^{1,n}(b^-,c_{\scaleto{1}{4.2pt}}^-|\{b\},\{\alpha_{\scaleto{1}{4.2pt}}\})\!\left[ \prod^{m-2}_{p=1} H^{1,n}(c^-_p,c^-_{p+1}|\{\alpha_p\},\{\alpha_{p+1}\})\right]\nonumber\\
\times H^{1,n}(c^-_{m-1},a^-|\{\alpha_{m-1}\},\{a\}),\label{fformula1}
\end{align}
where $H^{1,n}$ is the basic building block of the connected correlators, having only one external source being linked to those inside the $n$ Wilson lines (see \figref{Hmn}). Applying our generalized version of the MV model (embodied in the two-point correlator \eqref{2point}), it yields the following expression:
\begin{align}
H^{1,n}(b^-,a^-|\{b\},\{a\})\equiv g\sum^{n}_{j=1}\mu^2(y^-)F^{n}(b^-,y^-|\{b\}\{\beta\})|_{\beta_{j}=d}F^{n}(y^-,a^-|\{\beta\}\{a\})|_{\beta_{j}=d'}\nonumber\\
\times\!\!\int dz_{\pperp}G(z_{\pperp}-x_{j\pperp})f(z_{\pperp}-y_{\pperp})h\left( \frac{z_{\pperp}+y_{\pperp}}{2}\right)\!f^{cdd'},
\end{align}
However, a fundamental difference between our calculation and the one featured in \cite{PhysRevC.79.025204} is that in our case the external sources are affected by the differential operators $1/\nabla^2$ and $\partial^{i}$. This aspect can be comprised in a redefinition of $H^{1,n}$ as:
\begin{align}
H^{1,n}(b^-,a^-|\{b\},\{a\})^{i}\!\equiv g\sum^{n}_{j=1}\mu^2(y^-)F^{n}(b^-,y^-|\{b\}\{\beta\})|_{\beta_{j}=d}F^{n}(y^-,a^-|\{\beta\}\{a\})|_{\beta_{j}=d'}\nonumber\\
\times\partial^{i}_y\!\!\int dz_{\pperp}dw_{\pperp}G(z_{\pperp}-x_{j\pperp})G(w_{\pperp}-y_{\pperp})f(z_{\pperp}-w_{\pperp})h\left( \frac{z_{\pperp}+w_{\pperp}}{2}\right)\!f^{cdd'},\label{Noapp}
\end{align}
where $F^{n}$ denotes the correlator of $n$ Wilson lines. Note that in these formulas the bracketed indices represent a set of $n$ color indices (not indices from `missing' sources, as in \eqref{AppG} and \eqref{AppH}). By application of the approximations outlined in Appendix \ref{GreenDerivatives}, the previous expression can be rewritten as:
\begin{align}
H^{1,n}(b^-,a^-|\{b\},\{a\})^{i}\!\approx g\sum^{n}_{j=1}\mu^2(y^-)F^{n}(b^-,y^-|\{b\}\{\beta\})|_{\beta_{j}=d}F^{n}(y^-,a^-|\{\beta\}\{a\})|_{\beta_{j}=d'}\nonumber\\
\times h\left(b_{\pperp}\right)\partial^{i}_y L(x_{j\pperp}-y_{\pperp})f^{cdd'}\nonumber\\
=g\lambda(y^-,b_{\pperp})\sum^{n}_{j=1}\partial^{i}_y L(x_{j\pperp}-y_{\pperp})f^{cdd'}F^{n}(b^-,y^-|\{b\}\{\beta\})|_{\beta_{j}=d}F^{n}(y^-,a^-|\{\beta\}\{a\})|_{\beta_{j}=d'},
\end{align}
where $b_{\pperp}\!=\!(x_{\pperp}\!+y_{\pperp})/2$. In this step we have made use of the knowledge that all the transverse positions that enter our calculation are either $x_{\pperp}$ or $y_{\pperp}$. Thus, when expanding $h((z_{\pperp}\!+w_{\pperp})/2)$ around $h(b_{\pperp})$ in \eqref{Noapp}, the linear term of the expansion yields a correction proportional to a product of the form $(x'_{\pperp}\!-b_{\pperp})^{i}\partial^{i}h(b_{\pperp})$. Whether $x'_{\pperp}\!=x_{\pperp}$ or $x'_{\pperp}\!=y_{\pperp}$, this term is suppressed with respect to $h(b_{\pperp})$ according to the assumptions detailed in Appendix \ref{GreenDerivatives}.

Another difference between our calculation and the one performed in the aforementioned paper is that in the latter the insertion of external sources is assumed to take place at a longitudinal position $y^-$ that satisfies $b^-\!<\!y^-\!<\!a^-$. However, in our particular case the longitudinal coordinate on which the external color source $\tilde{\rho}^{a}$ depends is the same as the one of the Wilson line that it is attached to, yielding the following simplification of the previous expression (see \figref{H1n}):
\begin{equation}\label{fformula2}
H^{1,n}(b^-,a^-|\{b\},\{a\})^{i}=g\lambda(b^-,b_{\pperp})\sum^{n}_{j=1}\partial^{i}_y L(x_{j\pperp}-y_{\pperp})f^{c\,b_jb'}F^{n}(b^-,a^-|\{\beta\}\{a\})|_{\beta_j=b'}.
\end{equation}
Having defined all the basic pieces of the calculation of a correlator with $m$ external sources and $n$ Wilson lines, we can go back to our particular case. According to the notation used in \cite{PhysRevC.79.025204}, the correlator in \eqref{fformula0} corresponds to $F^{m,n}$ with $m\!=\!4$, $n\!=\!4$. By direct application of \eqref{fformula}:
\begin{figure}
\centering
\includegraphics[width=1\textwidth]{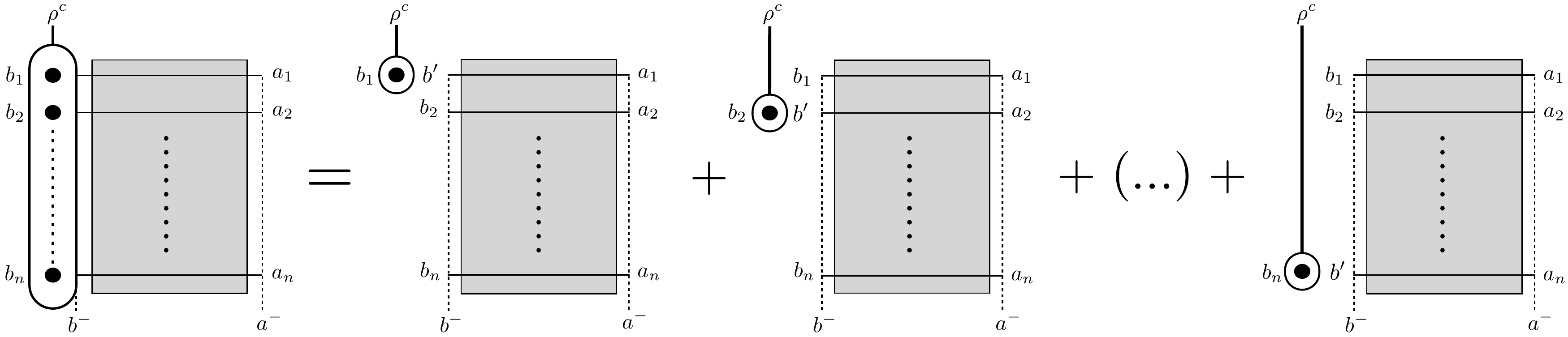}
\caption{Schematic representation of \eqref{fformula2}. The circles in the right hand of the equation represent couplings of the external source $\rho^{c}$ to Wilson lines inside the correlator (dark square). Each of these couplings multiplies the correlator by a $g\lambda(b^-,b_{\pperp})\,f^{cb_{j}b'}\partial_y L(x_{j\pperp}-y_{\pperp})$ factor.}
\label{H1n}
\end{figure}
\begin{align}
\left\langle\tilde{\rho}^{i,e}_{x}U^{ea}_{x}\tilde{\rho}^{k,f}_{x}U^{fc}_{x}\tilde{\rho}^{i'\!,e'}_{y}U^{e'\!a'}_{y}\tilde{\rho}^{k'\!,f^{\prime}}_{y}U^{f'\!c'}_{y} \right\rangle\!=\!\left\langle \tilde{\rho}^{i,e}_{x}\tilde{\rho}^{k,f}_{x}\tilde{\rho}^{i'\!,e'}_{y}\tilde{\rho}^{k'\!,f'}_{y}\!\right\rangle\!\left\langle U^{ea}_{x}U^{fc}_{x}U^{e'\!a'}_{y}U^{f'\!c'}_{y} \!\right\rangle\nonumber\\
+\!\left\langle \tilde{\rho}^{i,e}_{x}\tilde{\rho}^{k,f}_{x} \right\rangle\!\left\langle \tilde{\rho}^{i'\!,e'}_{y}\tilde{\rho}^{k'\!,f'}_{y} U^{ea}_{x}U^{fc}_{x}U^{e'\!a'}_{y}U^{f'\!c'}_{y}\!\right\rangle_{\hspace{-0.09cm}\text{\tiny c}}\!+\!\left\langle \tilde{\rho}^{i'\!,e'}_{y}\tilde{\rho}^{k'\!,f'}_{y} \!\right\rangle\!\left\langle \tilde{\rho}^{i,e}_{x}\tilde{\rho}^{k,f}_{x} U^{ea}_{x}U^{fc}_{x}U^{e'\!a'}_{y}U^{f'\!c'}_{y}\!\right\rangle_{\hspace{-0.09cm}\text{\tiny c}}\nonumber\\
+\!\left\langle \tilde{\rho}^{i,e}_{x}\tilde{\rho}^{i'\!,e'}_{y} \!\right\rangle\!\left\langle \tilde{\rho}^{k,f}_{x}\tilde{\rho}^{k'\!,f'}_{y} U^{ea}_{x}U^{fc}_{x}U^{e'\!a'}_{y}U^{f'\!c'}_{y}\!\right\rangle_{\text{\hspace{-0.09cm}\tiny c}}\!+\!\left\langle \tilde{\rho}^{i,e}_{x}\tilde{\rho}^{k'\!,f'}_{y} \!\right\rangle\!\left\langle\tilde{\rho}^{k,f}_{x}\tilde{\rho}^{i'\!,e'}_{y} U^{ea}_{x}U^{fc}_{x}U^{e'\!a'}_{y}U^{f'\!c'}_{y}\!\right\rangle_{\hspace{-0.09cm}\text{\tiny c}}\nonumber\\
+\!\left\langle \tilde{\rho}^{k,f}_{x}\tilde{\rho}^{i'\!,e'}_{y} \!\right\rangle\!\left\langle \tilde{\rho}^{i,e}_{x}\tilde{\rho}^{k'\!,f'}_{y} U^{ea}_{x}U^{fc}_{x}U^{e'\!a'}_{y}U^{f'\!c'}_{y}\!\right\rangle_{\hspace{-0.09cm}\text{\tiny c}}\!+\!\left\langle \tilde{\rho}^{k,f}_{x}\tilde{\rho}^{k'\!,f'}_{y} \!\right\rangle\!\left\langle \tilde{\rho}^{i,e}_{x}\tilde{\rho}^{i'\!,e'}_{y} U^{ea}_{x}U^{fc}_{x}U^{e'\!a'}_{y}U^{f'\!c'}_{y}\!\right\rangle_{\hspace{-0.09cm}\text{\tiny c}}\nonumber\\
+\!\left\langle\tilde{\rho}^{i,e}_{x}U^{ea}_{x}\tilde{\rho}^{k,f}_{x}U^{fc}_{x}\tilde{\rho}^{i'\!,e'}_{y}U^{e'\!a'}_{y}\tilde{\rho}^{k'\!,f^{\prime}}_{y}U^{f'\!c'}_{y} \!\right\rangle_{\hspace{-0.09cm}\text{\tiny c}}.\label{decomp1}
\end{align}
For simplicity we momentarily adopted a shorthand notation that omits the longitudinal coordinate dependence and the differential operators $1/\nabla^2$, $\partial^{i}$. However, it should be kept in mind that the external sources and Wilson lines that share an index depend on the same longitudinal coordinate. The first term after the equal sign, which corresponds to a complete factorization of external sources and Wilson lines, can be further expanded by application of the Wick's theorem, which tells us that the factor involving the external sources breaks down into the following sum of pairwise contractions:
\begin{align}
\left\langle\tilde{\rho}^{i,e}_{x}\,\tilde{\rho}^{k,f}_{x}\,\tilde{\rho}^{i'\!,e'}_{y}\,\tilde{\rho}^{k'\!,f'}_{y}\right\rangle\!=\!\langle \tilde{\rho}^{i,e}_{x}\,\tilde{\rho}^{k,f}_{x}\rangle\langle \tilde{\rho}^{i'\!,e'}_{y}\,\tilde{\rho}^{k'\!,f'}_{y}\rangle+\langle \tilde{\rho}^{i,e}_{x} \tilde{\rho}^{i'\!,e'}_{y}\rangle\langle \tilde{\rho}^{k,f}_{x} \tilde{\rho}^{k'\!,f'}_{y}\rangle+\langle \tilde{\rho}^{i,e}_{x} \tilde{\rho}^{k'\!,f'}_{y}\rangle\langle \tilde{\rho}^{k,f}_{x}\,\tilde{\rho}^{i'\!,e'}_{y}\rangle.
\end{align}
The three terms resulting from this expansion are addressed on section \ref{sec:tt}, where we derive the `disconnected' function:
\begin{align}
D^{ij;kl}_{ab;cd}(u_{\pperp},u'_{\pperp},v_{\pperp},v'_{\pperp})=T^{\,ij;kl}\!(u_{\pperp},u'_{\pperp},v_{\pperp},v'_{\pperp})\!\!\int^{\infty}_{-\infty}\!\!dz^-\!\int^{z^-}_{-\infty}\!dw^{-}\!\lambda(z^-,b_{\pperp})\lambda(w^-,b_{\pperp})\nonumber\\
\times\!\left(C^{(2)}_{\text{adj}}(z^-,w^-;u_{\pperp},u'_{\pperp})+C^{(2)}_{\text{adj}}(z^-,w^-;v_{\pperp},v'_{\pperp})\right)\delta^{AB}\delta^{CD}Q^{ABCD}_{abcd}(w^-;u_{\pperp},u'_{\pperp},v_{\pperp},v'_{\pperp}),
\end{align}
with $b_{\pperp}\!=\!(x_{\pperp}\!+y_{\pperp})/2$. Again, to get to this expression we anticipate that in our particular case the general transverse positions $u_{\pperp}$, $u'_{\pperp}$, $v_{\pperp}$, and $v'_{\pperp}$ will take the values $x_{\pperp}$ or $y_{\pperp}$. Here we also introduce the following function:
\begin{equation}
T^{\,ij;kl}(x_{\pperp},x'_{\pperp},y_{\pperp},y'_{\pperp})\equiv \partial^{i}_x\partial^{j}_{x'}L(x_{\pperp}-x'_{\pperp})\partial^{k}_{y}\partial^{l}_{y'}L(y_{\pperp}-y'_{\pperp}).
\end{equation}
The disconnected function is used to express part of the outcome of \eqref{MainCorrelator}, which consists in the integration of the correlator \eqref{fformula0} in the longitudinal direction. Specifically:
\begin{align}
\int^{\infty}_{-\infty}dz^-dw^-dz^{-\prime}dw^{-\prime}\left\langle \frac{\partial^{i}\tilde{\rho}^{e}(z^-,x_{\pperp})}{\nabla^2}\frac{\partial^{k}\tilde{\rho}^{f}(w^-,x_{\pperp})}{\nabla^2}\frac{\partial^{i^{\prime}}\tilde{\rho}^{e^{\prime}}(z^{-\prime},y_{\pperp})}{\nabla^2}\frac{\partial^{k^{\prime}}\tilde{\rho}^{f^{\prime}}(w^{-\prime},y_{\pperp})}{\nabla^2}\right\rangle\nonumber\\
\times\!\left\langle U^{ea}(z^-,x_{\pperp})U^{fc}(w^-,x_{\pperp})U^{e'a'}(z^{-\prime},y_{\pperp})U^{f'\!c'}(w^{-\prime},y_{\pperp}) \right\rangle\nonumber\\
=D^{ik;i'k'}_{ac;a'c'}(x_{\pperp},x_{\pperp},y_{\pperp},y_{\pperp})+D^{ii';kk'}_{aa';cc'}(x_{\pperp},y_{\pperp},x_{\pperp},y_{\pperp})+D^{ik';ki'}_{ac';ca'}(x_{\pperp},y_{\pperp},x_{\pperp},y_{\pperp}).
 \end{align}
%\begin{align*}
%\int^{\infty}_{-\infty}dz^-dw^-dz^{-\prime}dw^{-\prime}\left\langle \tilde{\rho}^{e}_{x_{\pperp}}\tilde{\rho}^{f}_{x_{\pperp}}\tilde{\rho}^{e'}_{y_{\pperp}}\tilde{\rho}^{f'}_{y_{\pperp}}\right\rangle\!\left\langle U^{ea}_{x_{\pperp}}U^{fc}_{x_{\pperp}}U^{e'\!a'}_{y_{\pperp}}U^{f'\!c'}_{y_{\pperp}} \right\rangle=D^{ik;i'k'}_{ac;a'c'}(x_{\pperp},x_{\pperp},y_{\pperp},y_{\pperp})&+D^{ii';kk'}_{aa';cc'}(x_{\pperp},y_{\pperp},x_{\pperp},y_{\pperp})\\
%&+D^{ik';ki'}_{ac';ca'}(x_{\pperp},y_{\pperp},x_{\pperp},y_{\pperp}).
%\end{align*}
All remaining terms of \eqref{decomp1} contain the connected correlator $\langle... \rangle_{\hspace{-0.02cm}\text{\tiny c}\hspace{0.05cm}}$, which can be computed by application of formulas \eqref{fformula1} and \eqref{fformula2}. However, our case of interest is somewhat more general than the one covered in these equations, as in our correlator each Wilson line depends on a different longitudinal coordinate. Even though this may seem a source of extra difficulty, it actually yields great simplification. For example, let us take what seems to be the most dreadful term of our calculation,  namely the fully connected version of the correlator (last term in \eqref{decomp1}):
\begin{align}
\int^{\infty}_{-\infty}dz^-dw^-dz^{-\prime}dw^{-\prime}&\left\langle\frac{\partial^{i}\tilde{\rho}^{e}(z^-,x_{\pperp})}{\nabla^2}\frac{\partial^{k}\tilde{\rho}^{f}(w^-,x_{\pperp})}{\nabla^2}\frac{\partial^{i^{\prime}}\!\tilde{\rho}^{e^{\prime}}\!(z^{-\prime},y_{\pperp})}{\nabla^2}\frac{\partial^{k^{\prime}}\!\tilde{\rho}^{f^{\prime}}\!(w^{-\prime},y_{\pperp})}{\nabla^2}\right.\nonumber\\
&\hspace{0.15cm}\left.\times U^{ea}(z^-,x_{\pperp})U^{fc}(w^-,x_{\pperp})U^{e'a'}(z^{-\prime},y_{\pperp})U^{f'\!c'}(w^{-\prime},y_{\pperp}) \right\rangle_{\hspace{-0.08cm}\text{c}}.\label{fullcon}
\end{align}
As we have four different longitudinal coordinates, in order to compute \eqref{fullcon} we need to consider all regions of the integration space\footnote{Namely the regions where $z^-\!>\!z^{-\prime}\!>\!w^-\!>\!w^{-\prime}$, $z^-\!>\!z^{-\prime}\!>\!w^{-\prime}\!>\!w^-$, etcetera. As is also the case for a single point in a 1-dimensional integral or a line in a 2-dimensional one, the regions where two or more of the coordinates have the same values (for example $z^-\!=\!z^{-\prime}\!>\!w^-\!>\!w^{-\prime}$) yield a negligible contribution. Therefore, we must always consider a certain ordering in our integration variables. The same logic was applied when splitting the double integral featured in the disconnected function $D^{ij;kl}_{ab;cd}$.}. For example, applying \eqref{fformula1} in the region where $z^-\!>\!w^-\!>\!z^{-\prime}\!>\!w^{-\prime}$ we have:
\begin{align}
\hspace{-0.2cm}H^{4,4}(z^-,-\infty |e,f,e',f'\,;\,a,c,a',c')\!=\!H^{1,1}(z^-,w^-|e\,\,;\,\alpha_{\scaleto{1}{4.2pt}})^{i}H^{1,2}(w^-,z^{-\prime}|\alpha_{\scaleto{1}{4.2pt}},f\,\,;\,\alpha_{\scaleto{2}{4.2pt}}, \beta_{\scaleto{1}{4.2pt}})^{k}\nonumber\\
\times H^{1,3}(z^{-\prime},w^{-\prime}|\alpha_{\scaleto{2}{4.2pt}},\beta_{\scaleto{1}{4.2pt}},e'\,\,;\,\alpha_{\scaleto{3}{4.2pt}}, \beta_{\scaleto{2}{4.2pt}},\gamma_{\scaleto{1}{4.2pt}})^{i'}\nonumber\\
\times H^{1,4}(w^{-\prime},-\infty |\alpha_{\scaleto{3}{4.2pt}},\beta_{\scaleto{2}{4.2pt}},\gamma_{\scaleto{1}{4.2pt}},f'\,\,;\,a, c,a',c')^{k'},
\end{align}
where, according to \eqref{fformula2}, the first factor reads:
\begin{equation}
H^{1,1}(z^-,w^-|e\,\,;\,\alpha_{\scaleto{1}{4.2pt}})^{i}=g\,\lambda(z^-,b_{\pperp})\partial^{i}_{x}L(0_{\pperp})f^{ee\alpha}\left\langle U^{\alpha\alpha_1}(z^-,w^-;x_{\pperp})\right\rangle=0,
\end{equation}
which vanishes due to the antisymmetric property of the $SU(N_c)$ structure constants.
%It would also vanish because of the derivative BUT in general we prefer to keep this terms because we will need them later to build the Gamma functions
As we have the same contribution from every region of the integration space, \eqref{fullcon} yields 0. In order to address the remaining six terms we define the `connected' function:
\begin{align}
C^{ij;kl}_{ab;cd}(u_{\pperp},u'_{\pperp},v_{\pperp},v'_{\pperp})=\int^{\infty}_{-\infty}dz^-dz^{-\prime}dw^-dw^{-\prime}\left\langle \frac{\partial^{i}\tilde{\rho}^{a'}(z^-,u_{\pperp})}{\nabla^2}\frac{\partial^{j}\tilde{\rho}^{b'}(z^{-\prime},u'_{\pperp})}{\nabla^2} \right\rangle\nonumber\\
\times\!\left\langle \frac{\partial^{k}\tilde{\rho}^{c'}(w^-,v_{\pperp})}{\nabla^2}\frac{\partial^{l}\tilde{\rho}^{d'}(w^{-\prime},v'_{\pperp})}{\nabla^2} U^{a'a}(z^-,u_{\pperp})U^{b'b}(z^{-\prime},u'_{\pperp})U^{c'c}(w^-,v_{\pperp})U^{d'd}(w^{-\prime},v'_{\pperp})\right\rangle_{\hspace{-0.08cm}\text{c}}\nonumber\\
=\partial^{i}_{u}\partial^{j}_{u'}L(u_{\pperp}-u'_{\pperp})\int^{\infty}_{-\infty}dz^-dw^-dw^{-\prime}\lambda(z^-,b_{\pperp})\nonumber\\
\times\!\left\langle  \frac{\partial^{k}\tilde{\rho}^{c'}(w^-,v_{\pperp})}{\nabla^2}\frac{\partial^{l}\tilde{\rho}^{d'}(w^{-\prime},v'_{\pperp})}{\nabla^2} U^{a'a}(z^-,u_{\pperp})U^{a'b}(z^-,u'_{\pperp})U^{c'c}(w^-,v_{\pperp})U^{d'd}(w^{-\prime},v'_{\pperp})\right\rangle_{\hspace{-0.08cm}\text{c}},\label{connected1}
\end{align}
\begin{figure}
\centering
\includegraphics[width=0.48\textwidth]{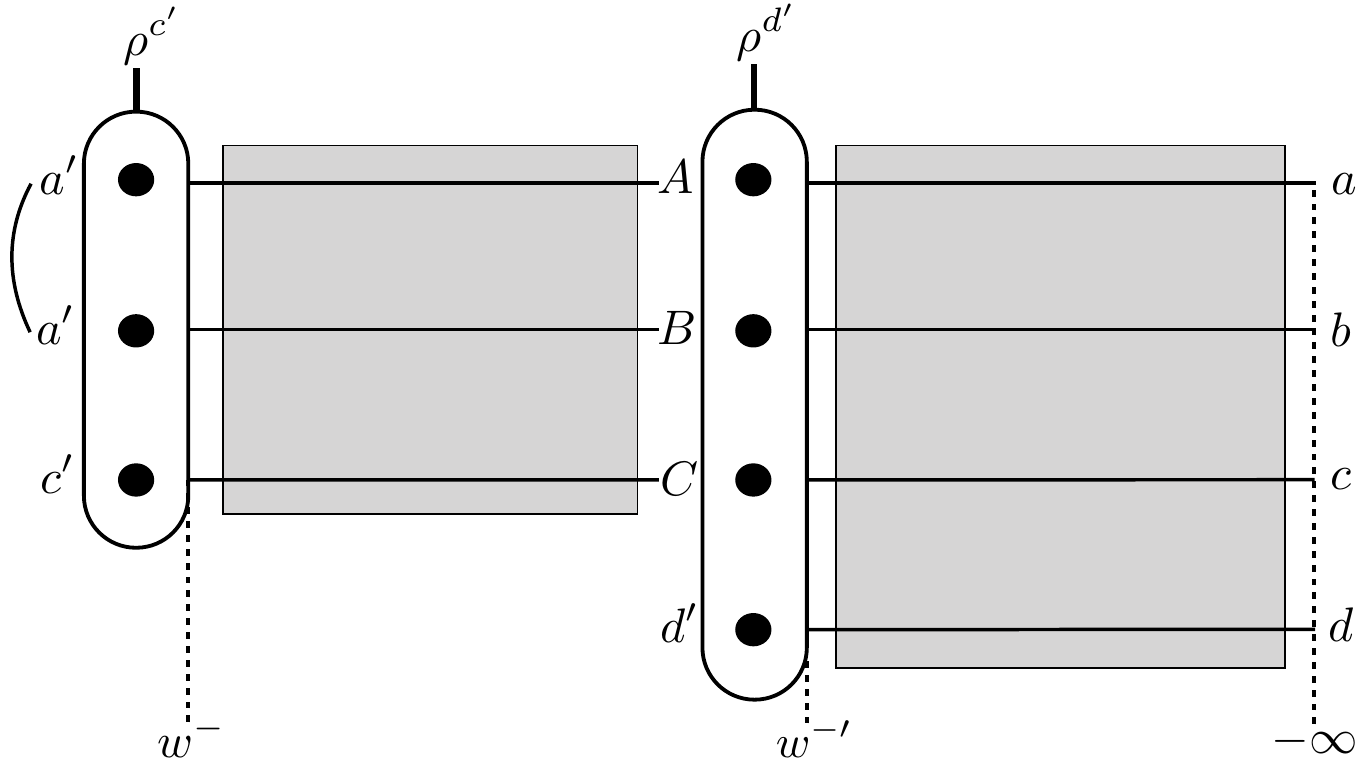}
\caption{Schematic representation of the connected correlator $H^{2,4}$ factorized in \eqref{connected2}.}
\label{H24}
\end{figure}where $b_{\pperp}\!=\!(x_{\pperp}\!+y_{\pperp})/2$. The only nonvanishing contributions to this integral come from the regions where $z^-\!>\!w^-\!>\!w^{-\prime}$ and $z^-\!>\!w^{-\prime}\!>\!w^-$, as in the other cases \eqref{fformula2} introduces a vanishing $H^{1,1}$ factor. For the $z^-\!>\!w^-\!>\!w^{-\prime}$ region we have (see \figref{H24}):
\begin{align}
&\partial^{i}_{u}\partial^{j}_{u'}L(u_{\pperp}-u'_{\pperp})\!\int^{\infty}_{-\infty}dz^-\!\int^{z^-}_{-\infty}\!dw^-\!\int^{w^-}_{-\infty}\!dw^{-\prime}\lambda(z^-,b_{\pperp})C_{\text{adj}}^{(2)}(z^-,w^-;u_{\pperp},u'_{\pperp})\nonumber\\
&\times\!\left\langle \frac{\partial^{k}\tilde{\rho}^{c'}(w^-,v_{\pperp})}{\nabla^2} U^{a'A}(w^-,w^{-\prime};u_{\pperp})U^{a'B}(w^-,w^{-\prime};u'_{\pperp})U^{c'C}(w^-,w^{-\prime};v_{\pperp})\right\rangle_{\hspace{-0.08cm}\text{c}}\nonumber\\
&\times\!\left\langle \frac{\partial^{l}\tilde{\rho}^{d'}(w^{-\prime},v'_{\pperp})}{\nabla^2} U^{Aa}(w^{-\prime},u_{\pperp})U^{Bb}(w^{-\prime},u'_{\pperp})U^{Cc}(w^{-\prime},v_{\pperp})U^{d'd}(w^{-\prime},v'_{\pperp})\right\rangle_{\hspace{-0.08cm}\text{c}},\label{connected2}
\end{align}
where we have applied the longitudinal locality of \eqref{2point} to factorize the correlator the same way we do in \eqref{FactorizationCorrelator}. We focus on the first connected correlator, which contains one external source and three Wilson lines, thus corresponding to:
\begin{align}
H^{1,3}(w^-,w^{-\prime}|a',a',c'\,\,;\,A,B,C)^{k}/(g\,\lambda(w^-,b_{\pperp}))\nonumber\\
=\partial^{k}_vL(v_{\pperp}\!-u_{\pperp})f^{c'a'\alpha}\!\left\langle U^{\alpha A}(w^-,w^{-\prime};u_{\pperp})U^{a' B}(w^-,w^{-\prime};u'_{\pperp})U^{c' C}(w^-,w^{-\prime};v_{\pperp})\right\rangle\nonumber\\
+\partial^{k}_vL(v_{\pperp}\!-u'_{\pperp})f^{c'a'\alpha}\!\left\langle U^{a' A}(w^-,w^{-\prime};u_{\pperp})U^{\alpha B}(w^-,w^{-\prime};u'_{\pperp})U^{c' C}(w^-,w^{-\prime};v_{\pperp})\right\rangle+0.
\end{align}
Substituting the result for the three-point adjoint correlator from \cite{1126-6708-2007-06-040}:
%I calculated it with our method
\begin{align}
\langle U^{aa'}(w^-,x^{\scaleto{1}{4.5pt}}_{\pperp})U^{bb'}(w^-,x^{\scaleto{2}{4.5pt}}_{\pperp})U^{cc'}(w^-,x^{\scaleto{3}{4.5pt}}_{\pperp}) \rangle\nonumber\\
=\frac{1}{2N_c^2C_{\scaleto{F}{0.4em}}}\left( f^{abc}f^{a'b'c'}+\frac{N_c^2}{N_c^2-4}d^{abc}d^{a'b'c'}\right)\exp{\left\{ -g^2\frac{N_c}{4}\bar{\lambda}(w^-,b_{\pperp})\sum_{i>j}\Gamma(x^{i}_{\pperp}\!-x^{j}_{\pperp})\right\}}\nonumber\\
\equiv\frac{1}{2N_c^2C_{\scaleto{F}{0.4em}}}\left( f^{abc}f^{a'b'c'}+\frac{N_c^2}{N_c^2-4}d^{abc}d^{a'b'c'}\right)\!C^{(3)}_{\text{adj}}(w^-;x^{\scaleto{1}{4.5pt}}_{\pperp},x^{\scaleto{2}{4.5pt}}_{\pperp},x^{\scaleto{3}{4.5pt}}_{\pperp}),
\end{align}
(where, again, we approximated the function $h$ as $h((x_{\pperp}\!+y_{\pperp})/2)$), we get to:
\begin{align}
H^{1,3}(w^-,w^{-\prime}|a',a',c'\,\,;\,A,B,C)^{k}=gf^{ABC}C^{(3)}_{\text{adj}}(w^-,w^{-\prime};u_{\pperp},u'_{\pperp},v_{\pperp})\nonumber\\
\times\lambda(w^-,b_{\pperp})\partial^{k}_v\!\left(L(v_{\pperp}-u'_{\pperp})-L(v_{\pperp}-u_{\pperp})\right).
\end{align}
Here, the color factor (the one that cancels $(2N_c^2C_{\scaleto{F}{0.4em}})^{-1}$) comes from the trace of the product of two structure constants. The remaining correlator, which contains an external source and four adjoint Wilson lines, yields:
\begin{align}
\hspace{-0.35cm}H^{1,4}(w^{-\prime}|\{A,B,C,d'\},\{a,b,c,d\})^{l}/(g\,\lambda(w^{-\prime},b_{\pperp}))\nonumber\\
=\partial^{l}_{v'}L(v'_{\pperp}\!-u_{\pperp})f^{d'A\alpha}\!\left\langle U^{\alpha a}(w^{-\prime},u_{\pperp})U^{Bb}(w^{-\prime},u'_{\pperp})U^{Cc}(w^{-\prime},v_{\pperp})U^{d'd}(w^{-\prime},v'_{\pperp})\right\rangle\nonumber\\
+\partial^{l}_{v'}L(v'_{\pperp}\!-u'_{\pperp})f^{d'B\alpha}\!\left\langle U^{A a}(w^{-\prime},u_{\pperp})U^{\alpha b}(w^{-\prime},u'_{\pperp})U^{Cc}(w^{-\prime},v_{\pperp})U^{d'd}(w^{-\prime},v'_{\pperp})\right\rangle\nonumber\\
+\partial^{l}_{v'}L(v'_{\pperp}\!-v_{\pperp})f^{d'C\alpha}\!\left\langle U^{A a}(w^{-\prime},u_{\pperp})U^{Bb}(w^{-\prime},u'_{\pperp})U^{\alpha c}(w^{-\prime},v_{\pperp})U^{d'd}(w^{-\prime},v'_{\pperp})\right\rangle+0.
\end{align}
Substituting this expression and summing the contribution from the $z^-\!>\!w^{-\prime}\!>\!w^-$ region the `connected' function finally yields:
\begin{align}
&\hspace{-0.75cm}C^{ij;kl}_{ab;cd}(u_{\pperp},u'_{\pperp},v_{\pperp},v'_{\pperp})\nonumber\\[-0.6em]
=&\,g^2h^3(b_{\pperp})\partial^{i}_{u}\partial^{j}_{u'}L(u_{\pperp}-u'_{\pperp})\!\int^{\infty}_{-\infty}\!\!\!dz^-\!\!\int^{z^-}_{-\infty}\!\!\!dw^-\!\!\int^{w^-}_{-\infty}\!\!\!\!dw^{-\prime}\mu^2(z^-)\mu^2(w^-)\mu^2(w^{-\prime})\nonumber\\
&\times\!C_{\text{adj}}^{(2)}(z^-,w^-;u_{\pperp},u'_{\pperp})\!\left(\left[\partial^{k}_v\!\left(L(v_{\pperp}\!-u'_{\pperp})\!-\!L(v_{\pperp}\!-u_{\pperp})\right)\!C^{(3)}_{\text{adj}}(w^-,w^{-\prime};u_{\pperp},u'_{\pperp},v_{\pperp})\right.\right.\nonumber\\
&\times\!\partial^{l}_{v'}\!\left( f^{AeD}f^{CBe}L(v'_{\pperp}-u_{\pperp})\!+\!f^{ACe}f^{DBe}L(v'_{\pperp}-u'_{\pperp})+f^{ABe}f^{eCD}L(v'_{\pperp}-v_{\pperp})\right)\nonumber\\
&\left.\left.\times Q^{ABCD}_{abcd}(w^{-\prime};u_{\pperp},u'_{\pperp},v_{\pperp},v'_{\pperp})\right]+\scaleto{\begin{bmatrix} l&\longleftrightarrow &k \\[-0.5em] c&\longleftrightarrow &d \\[-0.5em] \scaleto{v_{\pperp}}{0.2cm}\!\!&\longleftrightarrow &\scaleto{v'_{\pperp}}{0.32cm}\end{bmatrix}}{1cm}\right).\label{fformula3.1}
\end{align}
Now we can rewrite \eqref{MainCorrelator} in terms of $D^{ij;kl}_{ab;cd}$ and $C^{ij;kl}_{ab;cd}$ as:
\begin{align}
\langle \alpha^{i,a}(x_{\pperp}) \alpha^{k,c}(x_{\pperp}) \alpha^{i^{\prime}\!,a^{\prime}}(y_{\pperp})\alpha^{k^{\prime}\!,c^{\prime}}(y_{\pperp}) \rangle=D^{ik;i'k'}_{ac;a'c'}(x_{\pperp},x_{\pperp},y_{\pperp},y_{\pperp})+D^{ii';kk'}_{aa';cc'}(x_{\pperp},y_{\pperp},x_{\pperp},y_{\pperp})\nonumber\\
+D^{ik';ki'}_{ac';ca'}(x_{\pperp},y_{\pperp},x_{\pperp},y_{\pperp})+C^{ik;i'k'}_{ac;a'c'}(x_{\pperp},x_{\pperp},y_{\pperp},y_{\pperp})+C^{i'k';ik}_{a'c';ac}(y_{\pperp},y_{\pperp},x_{\pperp},x_{\pperp})\nonumber\\
+C^{ii';kk'}_{aa';cc'}(x_{\pperp},y_{\pperp},x_{\pperp},y_{\pperp})+C^{ik';ki'}_{ac';ca'}(x_{\pperp},y_{\pperp},x_{\pperp},y_{\pperp})+C^{ki';ik'}_{ca';ac'}(x_{\pperp},y_{\pperp},x_{\pperp},y_{\pperp})\nonumber\\+C^{kk';ii'}_{cc';aa'}(x_{\pperp},y_{\pperp},x_{\pperp},y_{\pperp}).\label{fformula4}
\end{align}
The fact that in our particular case the Wilson lines depend on only two transverse coordinates ($x_{\pperp}$ and $y_{\pperp}$) yields a significant simplification in the final expression. For example, by taking $u_{\pperp}\!=u_{\pperp}'$ in \eqref{fformula3.1} we can see that the first two connected terms of \eqref{fformula4} yield 0.
%Another way of seeing this is doing the explicit calculation, which gives us a vanishing H^{1,1} factor
The next four terms take the following form:
\begin{align}
C^{ij;kl}_{ab;cd}&(x_{\pperp},y_{\pperp},x_{\pperp},y_{\pperp})\nonumber\\[-0.6em]
=&\,2g^2h^3(b_{\pperp})\partial^{i}_{x}\partial^{j}_{y}L(x_{\pperp}-y_{\pperp})\!\int^{\infty}_{-\infty}\!\!\!dz^-\!\!\int^{z^-}_{-\infty}\!\!\!dw^-\!\!\int^{w^-}_{-\infty}\!\!\!\!dw^{-\prime}\mu^2(z^-)\mu^2(w^-)\mu^2(w^{-\prime})\nonumber\\[-0.3em]
&\times\!C_{\text{adj}}^{(2)}(z^-,w^-;x_{\pperp},y_{\pperp})\partial^{k}_{x}\!\left(L(x_{\pperp}\!-x_{\pperp})\!-\!L(x_{\pperp}\!-y_{\pperp})\right)\!C^{(3)}_{\text{adj}}(w^-,w^{-\prime};x_{\pperp},y_{\pperp},x_{\pperp})\nonumber\\
&\times\!\partial^{l}_{y}\!\left(L(y_{\pperp}\!-y_{\pperp})\!-\!L(y_{\pperp}\!-x_{\pperp})\right)f^{ACe}f^{BDe}Q^{ABCD}_{abcd}(w^{-\prime};x_{\pperp},y_{\pperp},x_{\pperp},y_{\pperp}),
\end{align}
where we applied the Jacobi identity of SU($N_c$). The previous expression contains a trivial projection of the adjoint Wilson line quadrupole:
\begin{align}
f^{ACe}f^{BDe}Q^{ABCD}_{abcd}(w^{-\prime};x_{\pperp},y_{\pperp},x_{\pperp},y_{\pperp})=f^{ABe}f^{CDe}Q^{ABCD}_{acbd}(w^{-\prime};x_{\pperp},x_{\pperp},y_{\pperp},y_{\pperp})\nonumber\\
=f^{ace}f^{bde}C^{(2)}_{\text{adj}}(w^{-\prime};x_{\pperp},y_{\pperp}).
\end{align}
(See Appendix \ref{Kov} for the detailed computation). Also, in the chosen limit the adjoint Wilson line tripole yields:
\begin{align}
C^{(3)}_{\text{adj}}(w^-,w^{-\prime};x_{\pperp},y_{\pperp},x_{\pperp})=C^{(2)}_{\text{adj}}(w^-,w^{-\prime};x_{\pperp},y_{\pperp}).
\end{align}
Both correlators tend to the dipole function in the case of only two different transverse coordinates. We are left with a product of three dipole functions that combine as:
\begin{align}
C_{\text{adj}}^{(2)}(z^-,w^-;x_{\pperp},y_{\pperp})C^{(2)}_{\text{adj}}(w^-,w^{-\prime};x_{\pperp},y_{\pperp})C^{(2)}_{\text{adj}}(w^{-\prime};x_{\pperp},y_{\pperp})\nonumber\\
=\exp{\left\{ -g^2\frac{N_c}{2}\Gamma(x_{\pperp}-y_{\pperp})h(b_{\pperp})\!\left(\bar{\mu}^2(z^-,w^-)+\bar{\mu}^2(w^-,w^{-\prime})+\bar{\mu}^2(w^{-\prime})\right)\right\}}\nonumber\\
=\exp{\left\{ -g^2\frac{N_c}{2}\Gamma(x_{\pperp}-y_{\pperp})h(b_{\pperp})\!\left( \int^{z^-}_{w^-}\!\!du^-\mu^2(u^-)\!+\!\int^{w^-}_{w^{-\prime}}\!\!du^-\mu^2(u^-)\!+\!\int^{w^{-\prime}}_{-\infty}\!\!du^-\mu^2(u^-)\right)\right\}}\nonumber\\
=C_{\text{adj}}^{(2)}(z^-;x_{\pperp},y_{\pperp}),
\end{align}
and therefore:
\begin{align}
C^{ij;kl}_{ab;cd}(x_{\pperp},y_{\pperp},x_{\pperp},y_{\pperp})\!=\frac{g^2}{2}f^{ace}f^{bde}h^3(b_{\pperp})\partial^{i}_{x}\partial^{j}_{y}L(x_{\pperp}\!-y_{\pperp})\partial^{k}_{x}\Gamma(x_{\pperp}-y_{\pperp})\partial^{l}_{y}\Gamma(y_{\pperp}-x_{\pperp})\nonumber\\
\times\!\int^{\infty}_{-\infty}\!dz^-\!\!\int^{z^-}_{-\infty}\!\!dw^-\!\!\int^{w^-}_{-\infty}\!\!dw^{-\prime}\mu^2(z^-)\mu^2(w^-)\mu^2(w'^-)C_{\text{adj}}^{(2)}(z^-;x_{\pperp},y_{\pperp}).
\end{align}
Solving the double integral, we obtain:
\begin{align}
C^{ij;kl}_{ab;cd}(x_{\pperp},y_{\pperp},x_{\pperp},y_{\pperp})\!=f^{ace}f^{bde}\partial^{i}_{x}\partial^{j}_{y}L(x_{\pperp}\!-y_{\pperp})\partial^{k}_{x}\Gamma(x_{\pperp}-y_{\pperp})\partial^{l}_{y}\Gamma(x_{\pperp}-y_{\pperp})\nonumber\\
\times\!\left( \frac{4}{\Gamma ^3 g^4 N_c^3}-\left(\frac{\bar{\lambda}^2(b_{\pperp}) }{2\Gamma N_c}+\frac{4}{\Gamma ^3 g^4 N_c^3}+\frac{2 \bar{\lambda}(b_{\pperp})}{\Gamma ^2 g^2 N_c^2}\right)\!C_{\text{adj}}^{(2)}(x_{\pperp},y_{\pperp})\right)\!.
\end{align}

\section{The correlator of four Wilson lines in the adjoint representation}\label{Kov}
\subsection{Reexponentiation method}\label{LengthyWilson}
%Some citations needed in this paragraph
Before addressing the problem of the adjoint Wilson line quadrupole we will briefly describe and apply a general method for the computation of Wilson line correlators. This technique, first applied in \cite{PhysRevD.64.114002}, is based in the discretization of the $x^-$ direction into $n$ layers of length $\Delta x^-$. Due to the properties of path ordered exponentials, this leads to the factorization of the Wilson line into a product of $n$ independent contributions from each zone:
\begin{equation}
U(x^-,x_{\pperp})_{ij}\equiv U^{(n)}_{ij}\equiv (U^{n}(x^{-}_{n},x_{\pperp})U^{n-1}(x^{-}_{n-1},x_{\pperp})...U^{1}(x^{-}_{1},x_{\pperp}))_{ij},
\end{equation}
assuming that $\Delta x^-$ is equal to or shorter than the correlation length of the gluon field fluctuations. This assumption is trivially satisfied in the MV model (and also in our generalized version), where interactions are local in rapidity, allowing us to take the limit $\Delta x^-\!\rightarrow\!0$. As a first step we expand one of these $n$ factors to order $g^2$:
\begin{align}
U(x^-,x_{\pperp})_{ij}\approx\!\left(\!\delta_{ik}+ig\tilde{A}^{+a}(x_n^-,x_{\pperp})t^{a}_{ik}\Delta x^-\!\!-\!g^{2}\frac{C_{\scaleto{F}{0.4em}}}{2} \lambda(x_n^-,b_{\pperp})L(0_{\pperp})\Delta x^-\delta_{ik}\!\right)\!U^{(n-1)}_{kj},\label{WLexp}
\end{align}
%Talk about the b_{\pperp}, the approximation, all that stuff?
where 
\begin{align}
\tilde{A}^{+a}(x^-,x_{\pperp})=-\frac{\tilde{\rho}^{\,a}(x^-,x_{\pperp})}{\nabla^2}
\end{align}
is the only non-trivial component of the gluon field expressed in the covariant gauge (defined by the condition $\partial_{\mu}\tilde{A}^{\mu}\!=\!0$). Note that in the $g^2$-order term of \eqref{WLexp} we already applied the two-point correlator, whose discretized version reads:
\begin{equation}
\langle \tilde{A}^{+a}(x^-,x_{\pperp})\tilde{A}^{+b}(y^-,y_{\pperp})\rangle = \lambda(x^-,b_{\pperp})\delta^{ab}L(x_{\pperp}-y_{\pperp})\frac{\delta_{x^-y^-}}{\Delta x^-}.
\end{equation}
We iterate this process $n-1$ more times neglecting terms of order $(\Delta x^-)^2$ or higher. Then, we rearrange the resulting terms in the form of the first orders of an expanded exponential. The last step is the reexponentiation, where we assume that the neglected terms complete the expansion. As an example, let us use this technique to calculate the well-known dipole function:
\begin{align}
\left\langle \text{Tr}\left\{ U(x_{\pperp})U^{\dagger}(y_{\pperp})\right\} \right\rangle\hspace{11.42cm}\nonumber\\
\hspace{0.2cm}\approx\!\left\langle U^{(n-1)}_{kl}(x_{\pperp})U^{(n-1)\dagger}_{lj}(y_{\pperp})\!\left( \delta_{ik}+ig\tilde{A}^{+a}(x_n^-,x_{\pperp})t^{a}_{ik}\Delta x^--g^{2}\frac{C_{\scaleto{F}{0.4em}}}{2}  \lambda(x_n^-,b_{\pperp})L(0_{\pperp})\Delta x^-\delta_{ik}\right)\right.\nonumber\\
\left.\times\!\left( \delta_{ji}-ig\tilde{A}^{+b}(x_n^-,y_{\pperp})t^{b}_{ji}\Delta x^-\!-\!g^{2}\frac{C_{\scaleto{F}{0.4em}}}{2}  \lambda(x_n^-,b_{\pperp})L(0_{\pperp})\Delta x^-\delta_{ji}\right)\right\rangle\nonumber\\
= \left\langle \text{Tr}\left\{ U(x_{\pperp})U^{\dagger}(y_{\pperp}) \right\} \right\rangle^{(n-1)}\!\!\left( 1-\frac{g^2}{2}C_{\scaleto{F}{0.4em}}\Delta x^-\lambda(x_n^-,b_{\pperp})\Gamma(x_{\pperp}-y_{\pperp})\right)\!.
\end{align}
In the last step we have made use of the locality in rapidity of the MV model to factorize the correlator of the remaining Wilson line slices $\left\langle\text{Tr}\left\{ U(x_{\pperp})U^{\dagger}(y_{\pperp}) \right\} \right\rangle^{(n-1)}$.
By iterating the process, we arrive at:
\begin{align}
\left\langle \text{Tr}\left\{ U(x_{\pperp})U^{\dagger}(y_{\pperp})\right\} \right\rangle\!&\approx\!\left( 1-\frac{g^2}{2}C_{\scaleto{F}{0.4em}}\,\Gamma(x_{\pperp}-y_{\pperp})h(b_{\pperp})\sum^{n}_{i=1}\Delta x^-\mu^2(x_i^-)\right)\nonumber\\
&=\!\left( 1-\frac{g^2}{2}C_{\scaleto{F}{0.4em}}\,\Gamma(x_{\pperp}-y_{\pperp})\bar{\lambda}(x^-,b_{\pperp})\right).
\end{align}
Lastly, we assume that the neglected higher order terms add up to an exponential expression, which reads:
\begin{equation}
\left\langle \text{Tr}\left\{ U(x_{\pperp})U^{\dagger}(y_{\pperp})\right\} \right\rangle \!=\! \exp{\left\{-\frac{g^2}{2}C_{\scaleto{F}{0.4em}}\,\Gamma(x_{\pperp}-y_{\pperp})\bar{\lambda}(x^-,b_{\pperp})\right\}}.
\end{equation}

\subsection{Diagonalization method}\label{Kovner}
One important shortcoming of the technique described above lies in the fact that there is no unique way in which we can arrange the terms resulting from expanding the Wilson lines. This step becomes more problematic as we increase the number of Wilson lines in the correlator. Nevertheless, we can reduce the inherent arbitrariness of the reexponentiation process by formulating the method as a diagonalization problem. This allows us to systematically account for all incoming and outgoing states of the interaction embodied in the medium average $\langle...\rangle$. In the next subsection we will make use of this technique to obtain the behavior of the following adjoint Wilson line quadrupole:
\begin{equation}
\left\langle U^{Aa}(x_{\pperp})U^{Bb}(x_{\pperp})U^{Cc}(y_{\pperp})U^{Dd}(y_{\pperp}) \right\rangle\!.
\end{equation}
under different color projections. However, to illustrate the method we will first reproduce the more general result obtained in \cite{PhysRevD.64.114002} for three different transverse coordinates:
\begin{equation}\label{adjointquad}
\left\langle U^{ab}(z_{\pperp})U^{cd}(z_{\pperp})U^{ef}(x_{\pperp})U^{gh}(y_{\pperp}) \right\rangle\!.
\end{equation}
%I leave the indices here like this, so it is easier to compare with the Kovner paper. Is it ok?
First, we need to expand the adjoint Wilson lines in a longitudinal position $x^-_n$. For the sake of simplicity in the following calculations we will momentarily adopt a shorthand notation similar to the one used in \cite{PhysRevD.64.114002}. We absorb the $g\Delta x^-$ factor in the definition of our fields:
\begin{equation}
g\tilde{A}^{+a}(x^-,x_{\pperp})\Delta x^-\equiv \tilde{A}^{+a}(x^-,x_{\pperp}),
\end{equation}
which yields the following two-point function:
\begin{equation}
\langle \tilde{A}^{+a}(x^-,x_{\pperp})\tilde{A}^{+b}(y^-,y_{\pperp})\rangle = \delta_{x^-y^-}\delta^{ab} B_{xy}(x^-,b_{\pperp}),
\end{equation}
where, due to the discretization of the rapidity range, the Kronecker delta $\delta_{x^-y^-}$ now takes the place of the Dirac delta. For simplicity we also introduced:
\begin{equation}
B_{xy}(x^-,b_{\pperp})\!\equiv\!g^2\Delta x^-\lambda(x^-,b_{\pperp})L(x_{\pperp}-y_{\pperp}).
\end{equation}
Using this notation the expansion to order $g^2$ of the adjoint Wilson line looks like:
\begin{equation}
U^{ab}(x^-,x_{\pperp})=(U^{ab_1})^{(n-1)}\left( \delta^{b_1 b}\left( 1-\frac{N_c}{2}B_x(x_{n}^-,b_{\pperp})\right) - \tilde{A}^{g}(x_{n}^-,x_{\pperp})f^{b_1 g b}\right)\!.
\end{equation}
Performing this expansion for every Wilson line in \eqref{adjointquad} and neglecting terms of order $(\Delta x^-)^2$ or higher we get:
\begin{align}
\left\langle U^{ab}(z_{\pperp})U^{cd}(z_{\pperp})U^{ef}(x_{\pperp})U^{gh}(y_{\pperp}) \right\rangle=\left\langle U^{aa'}(z_{\pperp})U^{cc'}(z_{\pperp})U^{ee'}(x_{\pperp})U^{gg'}(y_{\pperp}) \right\rangle^{(n-1)}\nonumber\\
\times\!\left( \delta^{a'b}\delta^{c'd}\delta^{e'f}\delta^{g'h}\left(1-\frac{N_c}{2}\left( 2B_z + B_x + B_y\right)\right)+\delta^{a'b}\delta^{c'd}f^{e'mf}f^{g'mh}B_{xy} \right.\nonumber\\
\left. + \delta^{a'b}\delta^{e'f}f^{c'md}f^{g'mh}B_{zy}+\delta^{a'b}\delta^{g'h}f^{e'mf}f^{c'md}B_{zx}+\delta^{e'f}\delta^{c'd}f^{a'mb}f^{g'mh}B_{zy}\right.\nonumber\\
\left.+\delta^{g'h}\delta^{c'd}f^{e'mf}f^{a'mb}B_{zx}+\delta^{e'f}\delta^{g'h}f^{a'mb}f^{c'md}B_{z}\right)\!.\label{adjointquad1}
\end{align}
We express the previous lines as a matrix equation: $U^{aceg}_{bdfh}\!=\!(U^{aceg}_{a'c'e'g'})^{(n-1)} T^{a'c'e'g'}_{bdfh}$, for which we introduce the following color vector basis:
\begin{align}
u_1&=\delta^{ea} \delta^{gc} \;  \;\hspace{0.76cm} u_2=\delta^{ca} \delta^{ge} \;  \; \hspace{0.73cm}u_3=\delta^{ga} \delta^{ec}\nonumber\\
w_1&=d^{eam} d^{gcm} \;  \; \hspace{0.18cm}w_2=d^{cam} d^{gem} \;  \; \hspace{0.14cm}w_3=d^{gam} d^{ecm}\nonumber\\
z_1&=d^{eam} f^{gcm} \;  \;\hspace{0.26cm} z_2=d^{cam} f^{gem} \;  \;\hspace{0.21cm} z_3=d^{gam} f^{ecm}.\label{colorbss}
\end{align}
It can be shown via color algebra arguments that this ensemble covers the entirety of possible interactions embodied in $T^{a'c'e'g'}_{bdfh}$ (see \cite{Dittner:1971fy})\footnote{In \cite{Dittner:1971fy}, the author mentions only 8 such tensors, but that is because he is dealing with $SU(3)$, and there exists a relation between the $SU(N_c)$ generators, valid only for $N_c=3$, which reduces the number of independent rank 4 tensors from 9 to 8 in that case.}. The last three ($z_1$, $z_2$, $z_3$) form a basis that does not mix with the rest of the vectors in the Gaussian model we are considering. Thus, if we expressed $T^{a'c'e'h'}_{bdfh}$ in this 9-dimensional space it would look like a block diagonal matrix with a $6\times 6$ part corresponding to the vectors $u_i$, $w_i$ and a $3 \times 3$ sector corresponding to the $z_i$ set. In our specific calculation, the vectors that we are interested in propagating live in the 6-dimensional space defined by the first two sets, so it will be enough to consider $T^{a'c'e'h'}_{bdfh}$ in the basis formed by $u_i$ and $w_i$. To build this matrix we propagate these six vectors using \eqref{adjointquad1}:
\begin{align}
T^{a'c'e'h'}_{bdfh}\delta^{e'a'} \delta^{g'c'}=\,&\delta^{fb} \delta^{hd}\left( 1 -\frac{N_c}{2}(2B_z+B_x+B_y - 2B_{zx} -2B_{zy}) \right)\nonumber\\
&+f^{bfm}f^{dhm}\left( B_z + B_{xy} - B_{zx} - B_{zy} \right)\nonumber\\
=\,&\delta^{fb} \delta^{hd}\left( 1 -g^2\frac{N_c}{2}\Delta x^- \lambda(x^-_n,b_{\pperp})(\Gamma(z_{\pperp}-x_{\pperp})+\Gamma(z_{\pperp}-y_{\pperp})) \right)\nonumber\\
&+f^{bfm}f^{dhm}\frac{g^2}{2}\Delta x^- \lambda(x^-_n,b_{\pperp})\left( \Gamma(z_{\pperp}-x_{\pperp})+\Gamma(z_{\pperp}-y_{\pperp})-\Gamma(x_{\pperp}-y_{\pperp}) \right)\!.
\end{align}
The SU($N_c$) factor $f^{bfm}f^{dhm}$, as well as the ones resulting from permutations of its indices, can be expressed in terms of our basis vectors by means of the following identity:
\begin{align}
f^{abm}f^{cdm}=\frac{2}{N_c}(\delta^{ac}\delta^{bd}-\delta^{ad}\delta^{bc})+d^{ace}d^{bde}-d^{ade}d^{bce},
\end{align}
and thus the propagation of $u_1$ reads:
\begin{align}
Tu_1=&\,u_1\left( 1 -g^2\frac{N_c}{2}\Delta x^- \lambda(x^-_n,b_{\pperp})(\Gamma(z_{\pperp}-x_{\pperp})+\Gamma(z_{\pperp}-y_{\pperp})) \right)\nonumber\\
&\hspace{-0.3cm}+\frac{g^2}{2}\Delta x^- \lambda(x^-_n,b_{\pperp})\!\left(\frac{2}{N_c}(u_2-u_3)+w_2-w_3\right)\!\left( \Gamma(z_{\pperp}\!-x_{\pperp})\!+\Gamma(z_{\pperp}\!-y_{\pperp})\!-\Gamma(x_{\pperp}\!-y_{\pperp}) \right)\!.
\end{align}
Repeating this process for the remaining vectors, we obtain:
\begin{align}
Tu_2=&\,u_2\!\left(\!1 -g^2\frac{N_c}{2}\Delta x^- \lambda(x^-_n,b_{\pperp})\Gamma(x_{\pperp}-y_{\pperp})\!\right)\\
Tu_3=&\,u_3\left( 1 -g^2\frac{N_c}{2}\Delta x^- \lambda(x^-_n,b_{\pperp})(\Gamma(z_{\pperp}-x_{\pperp})+\Gamma(z_{\pperp}-y_{\pperp})) \right)\nonumber\\
&\hspace{-0.2cm}+\frac{g^2}{2}\Delta x^- \lambda(x^-_n,b_{\pperp})\!\left(\frac{2}{N_c}(u_2-u_1)+w_2-w_1\right)\!\left( \Gamma(z_{\pperp}\!-x_{\pperp})\!+\!\Gamma(z_{\pperp}\!-y_{\pperp})\!-\!\Gamma(x_{\pperp}\!-y_{\pperp}) \right)\\
Tw_1=&\,w_1\left( 1-g^2\frac{N_c}{8}\Delta x^- \lambda(x^-_n,b_{\pperp})(\Gamma(x_{\pperp}-y_{\pperp})+3\Gamma(z_{\pperp}-x_{\pperp})+3\Gamma(z_{\pperp}-y_{\pperp}))\right)\nonumber\\
&+\frac{g^2}{2}\Delta x^- \lambda(x^-_n,b_{\pperp})\left(\left( \frac{2}{N_c}-\frac{N_c}{4}\right)\left(w_2-w_3\right)+\left(\frac{4}{N_c^2}-1\right)(u_2-u_3)\right)\nonumber\\
&\times\!\left( \Gamma(x_{\pperp}-y_{\pperp})-\Gamma(z_{\pperp}-x_{\pperp})-\Gamma(z_{\pperp}-y_{\pperp})\right)\\
Tw_2=&\,w_2\left( 1-g^2\frac{N_c}{4}\Delta x^- \lambda(x^-_n,b_{\pperp})\left( \Gamma(x_{\pperp}-y_{\pperp})+\Gamma(z_{\pperp}-x_{\pperp})+\Gamma(z_{\pperp}-y_{\pperp})\right)\right)\\
Tw_3=&\,w_3\left( 1-g^2\frac{N_c}{8}\Delta x^- \lambda(x^-_n,b_{\pperp})(\Gamma(x_{\pperp}-y_{\pperp})+3\Gamma(z_{\pperp}-x_{\pperp})+3\Gamma(z_{\pperp}-y_{\pperp}))\right)\nonumber\\
&+\frac{g^2}{2}\Delta x^- \lambda(x^-_n,b_{\pperp})\left(\left( \frac{2}{N_c}-\frac{N_c}{4}\right)\left(w_2-w_1\right)+\left(\frac{4}{N_c^2}-1\right)(u_2-u_1)\right)\nonumber\\
&\times\!\left( \Gamma(x_{\pperp}-y_{\pperp})-\Gamma(z_{\pperp}-x_{\pperp})-\Gamma(z_{\pperp}-y_{\pperp})\right)\!.
\end{align}
From the previous projections we can write \eqref{adjointquad1} in the following form:
\begin{equation}
U^{aceg}_{bdfh}=(U^{aceg}_{a'c'e'g'})^{(n-1)}T^{a'c'e'g'}_{bdfh}(x_n^-)=(U^{aceg}_{a'c'e'g'})^{(n-1)}(\mathbb{1}+M(x_n^-))^{a'c'e'g'}_{bdfh}\!,
\end{equation}
where the $M^{a'c'e'g'}_{bdfh}\!$ matrix is of order 1 in $\Delta x^-$. The next step of the method consists in iterating the expansion of the Wilson lines $n-1$ times. Doing this (and neglecting terms of order $(\Delta x^-)^2$ or higher), we get:
\begin{equation}\label{eqtodiag}
U^{aceg}_{bdfh}=\mathbb{1}+\sum^{n}_{i=1}M^{a'c'e'g'}_{bdfh}(x_i^-)=\mathbb{1}+\!\int^{x^-}\!\!dz^{\prime-}M^{a'c'e'g'}_{bdfh}(z^{\prime-})=\mathbb{1}+\bar{M}(x^-).
\end{equation}
It is worth reminding that we are omitting some of the dependencies of $\bar{M}$ for simplicity; this tensor also depends on the transverse coordinates, $\bar{M}(x^-;z_{\pperp},x_{\pperp},y_{\pperp})$.
%At this point it is trivial to match our notation with the one used by Kovner et al. in their paper.
In order to reproduce the notation of \cite{PhysRevD.64.114002}, we introduce the following functions:
\begin{align}
R_{a}&=-\frac{g^2}{2}\bar{\lambda}(x^-,b_{\pperp})\left(\Gamma(z_{\pperp}-x_{\pperp})-\Gamma(z_{\pperp}-y_{\pperp})\right)\\
R_{b}&=-\frac{g^2}{2}\bar{\lambda}(x^-,b_{\pperp})\left(\Gamma(x_{\pperp}-y_{\pperp})\right)\\
R_{d}&=R_b-R_a,
\end{align}
and thus we obtain the following expression for $\bar{M}$ (hereby correcting typos in the matrix given in \cite{PhysRevD.64.114002}):
\begin{equation}
\setlength{\arraycolsep}{1\arraycolsep}% this change is local
\begin{bmatrix}
    N_cR_a & 0 & -\frac{2}{N_c}R_d & 0 & 0 & R_d\left(\frac{4}{N_c^2}-1\right) \\[3.2pt]
    \frac{2}{N_c}R_d & N_cR_b & \frac{2}{N_c}R_d & -R_d\left(\frac{4}{N_c^2}-1\right) & 0 & -R_d\left(\frac{4}{N_c^2}-1\right) \\[3.2pt]
    -\frac{2}{N_c}R_d & 0 & N_cR_a & R_d\left(\frac{4}{N_c^2}-1\right) & 0 & 0 \\[3.2pt]
    0 & 0 & -R_d & \frac{N_c}{4}(3R_a+R_b) & 0 & R_d\left(\frac{2}{N_c}-\frac{N_c}{4}\right) \\[3.2pt]
    R_d & 0 & R_d & -R_d\left(\frac{2}{N_c}-\frac{N_c}{4}\right) & \frac{N_c}{2}(R_a+R_b) & -R_d\left(\frac{2}{N_c}-\frac{N_c}{4}\right) \\[3.2pt]
    -R_d & 0 & 0 & R_d\left(\frac{2}{N_c}-\frac{N_c}{4}\right) & 0 & \frac{N_c}{4}(3R_a+R_b) \\[3.2pt]
\end{bmatrix}\!,
\end{equation}
which we diagonalize using Mathematica:
\begin{equation}
\bar{M}_d\!=\!
\setlength{\arraycolsep}{1\arraycolsep}% this change is local
\begin{bmatrix}
    N_cR_a & 0 & 0 & 0 & 0 & 0 \\[3.2pt]
    0 & N_cR_b & 0 & 0 & 0 & 0 \\[3.2pt]
    0 & 0 & \frac{1}{2}(R_a+R_b)N_c & 0 & 0 & 0 \\[3.2pt]
    0 & 0 & 0 & \frac{1}{2}(R_a+R_b)N_c & 0 & 0 \\[3.2pt]
    0 & 0 & 0 & 0 & N_cR_a\!-\!R_d & 0 \\[3.2pt]
    0 & 0 & 0 & 0 & 0 & N_cR_a\!+\!R_d \\[3.2pt]
\end{bmatrix}\!.
\end{equation}
The final step is the reexponentiation of \eqref{eqtodiag}, which is straightforward for a diagonal matrix:
\begin{equation}\label{adjointquad2}
U^{aceg}_{bdfh}\,\dot{=}\,(\mathbb{1}+\bar{M}_d)^{aceg}_{bdfh}\;\longrightarrow\; U^{aceg}_{bdfh}\,\dot{=}\,(e^{\bar{M}_d})^{aceg}_{bdfh}.
\end{equation}
Here the dot stresses that in order to use this result we need to work in the basis defined by the eigenvectors of $\bar{M}$, which in the $(u_1,u_2,u_3,w_1,w_2,w_3)$ basis looks like:
\begin{equation*}
t_1=\begin{pmatrix} \frac{N_c^2-4}{2N_c}  \\[0.5em]  0  \\[0.5em] -\frac{N_c^2-4}{2N_c} \\[0.5em] -1 \\[0.5em] 0 \\[0.5em] 1\end{pmatrix}, \; t_2=\begin{pmatrix} 0  \\[0.5em]  1  \\[0.5em] 0 \\[0.5em] 0 \\[0.5em] 0 \\[0.5em] 0\end{pmatrix}, \; t_3=\begin{pmatrix} -\frac{2}{N_c}  \\[0.5em]  0  \\[0.5em] \frac{2}{N_c} \\[0.5em] -1 \\[0.5em] 0 \\[0.5em] 1\end{pmatrix},
\end{equation*}
\begin{equation}\label{eigenquad}
\hspace{1.5cm}t_4=\begin{pmatrix} 0  \\[0.5em]  0  \\[0.5em] 0 \\[0.5em] 0 \\[0.5em] 1 \\[0.5em] 0\end{pmatrix}, \; t_5=\begin{pmatrix} \frac{2+N_c}{N_c}  \\[0.5em]   -\frac{2}{N_c}\frac{2+N_c}{N_c+1}  \\[0.5em] \frac{2+N_c}{N_c} \\[0.5em] 1 \\[0.5em] -\frac{N_c+4}{N_c+2} \\[0.5em] 1\end{pmatrix}, \; t_6=\begin{pmatrix} \frac{2-N_c}{N_c}  \\[0.5em]  \frac{2}{N_c}\frac{2-N_c}{N_c-1} \\[0.5em] \frac{2-N_c}{N_c} \\[0.5em] 1 \\[0.5em] -\frac{N_c-4}{N_c-2} \\[0.5em] 1\end{pmatrix}.
\end{equation}
Remarkably, we have $t_2\!=\!u_2\!=\!\delta^{ca}\delta^{ge}$, $t_3\!=\!-f^{can}f^{gen}$, and $t_4\!=\!w_1\!=\!d^{ean}d^{gcn}$.

\subsection{Projections of the quadrupole}\label{projKovner}
Let us now go back to our particular case:
\begin{align}
\left\langle U^{Aa}(x_{\pperp})U^{Bb}(x_{\pperp})U^{Cc}(y_{\pperp})U^{Dd}(y_{\pperp}) \right\rangle,
\end{align}
which can be obtained from the quadrupole studied in the previous subsection by setting $z_{\pperp}\!\equiv\! x_{\pperp}$ and $x_{\pperp}\!=\!y_{\pperp}\!\equiv\! y_{\pperp}$. This simplifies the above result, as $R_b\!=\!0$ and $R_d\!=\!-R_a$. In this limit, $\bar{M}_d$ adopts the following form:
\begin{equation}
\bar{M}_d\!=\!
\setlength{\arraycolsep}{1\arraycolsep}% this change is local
\begin{bmatrix}
    N_cR_a & 0 & 0 & 0 & 0 & 0 \\[3.2pt]
    0 & 0 & 0 & 0 & 0 & 0 \\[3.2pt]
    0 & 0 & \frac{1}{2}N_cR_a & 0 & 0 & 0 \\[3.2pt]
    0 & 0 & 0 & \frac{1}{2}N_cR_a & 0 & 0 \\[3.2pt]
    0 & 0 & 0 & 0 & (N_c+1)R_a & 0 \\[3.2pt]
    0 & 0 & 0 & 0 & 0 & (N_c-1)R_a \\[3.2pt]
\end{bmatrix}\!.
\end{equation}
As part of the calculation of $\langle T^{\mu\nu}(x_{\pperp})T^{\sigma\rho}(y_{\pperp})\rangle$, we need to calculate the following projections of the adjoint Wilson line quadrupole:
\begin{align}
f^{ABe}f^{DCe}\left\langle U^{Aa}(x_{\pperp})U^{Bb}(x_{\pperp})U^{Cc}(y_{\pperp})U^{Dd}(y_{\pperp})\right\rangle\label{adjointquad5}\\
\delta^{AC}\delta^{BD}\left\langle U^{Aa}(x_{\pperp})U^{Bb}(x_{\pperp})U^{Cc}(y_{\pperp})U^{Dd}(y_{\pperp}) \right\rangle\!.\label{adjointquad4}
\end{align}
The first of them corresponds to the propagation of the eigenvector $t_3$, which is straightforward to compute:
\begin{align}
(e^{\bar{M}_d})^{ABCD}_{abcd}(t_3)^{ABCD} &= (t_3)^{abcd}\exp{\left\{ \frac{1}{2}N_cR_a\right\}} \nonumber\\
& =f^{abn}f^{dcn}\exp{\left\{-g^2\frac{N_c}{2}\Gamma(x_{\pperp}\!-y_{\pperp})\bar{\lambda}(x^-,b_{\pperp})\right\}}\!.
\end{align}
The case of \eqref{adjointquad4} corresponds to the propagation of $u_1$, which is not an eigenvector and thus requires that we express it in terms of the $t_i$ set:
\begin{equation}
u_1=\frac{1}{N_c}t_1+\frac{1}{N_c^2-1}t_2-\frac{1}{N_c}t_3+\frac{N_c}{N_c^2-4}t_4+\frac{1}{4}t_5-\frac{1}{4}t_6,
\end{equation}
and then:
\begin{align}
(e^{\bar{M}_d})^{ABCD}_{abcd}(u_1)^{ABCD} = \frac{1}{N_c} (t_1)^{abcd}\exp{\left\{ -g^2N_c\Gamma(x_{\pperp}-y_{\pperp})\bar{\lambda}\right\}}+ \frac{1}{N_c^2-1}(t_2)^{abcd}\nonumber\\
-\frac{1}{N_c}(t_3)^{abcd}\exp{\left\{ -g^2\frac{N_c}{2}\Gamma(x_{\pperp}-y_{\pperp})\bar{\lambda}\right\}} +\frac{N_c}{N_c^2-4}(t_4)^{abcd}\exp{\left\{ -g^2\frac{N_c}{2}\Gamma(x_{\pperp}-y_{\pperp})\bar{\lambda}\right\}}\nonumber\\
+\frac{1}{4}(t_5)^{abcd}\exp{\left\{-g^2(N_c+1)\Gamma(x_{\pperp}-y_{\pperp})\bar{\lambda}\right\}}-\frac{1}{4}(t_6)^{abcd}\exp{\left\{-g^2(N_c-1)\Gamma(x_{\pperp}-y_{\pperp})\bar{\lambda}\right\}},
\end{align}
where we omitted the dependencies of $\bar{\lambda}$ for simplicity. Expanding the eigenvectors in terms of our original basis \eqref{colorbss} we obtain:
\begin{align}
=\delta^{ac}\delta^{bd}\!\left(\frac{N_c^2-4}{2N_c^2}e^{-g^2N_c\Gamma\bar{\lambda}}+\frac{2}{N_c^2}e^{-g^2\frac{N_c}{2}\Gamma\bar{\lambda}}+\frac{N_c+2}{4N_c}e^{-g^2(N_c+1)\Gamma\bar{\lambda}}+\frac{N_c-2}{4N_c}e^{-g^2(N_c-1)\Gamma\bar{\lambda}} \right)\nonumber\\
+\delta^{ab}\delta^{cd}\!\left( \frac{1}{N_c^2-1}-\frac{N_c+2}{2N_c(N_c+1)}e^{-g^2(N_c+1)\Gamma\bar{\lambda}}+\frac{N_c-2}{2N_c(N_c-1)}e^{-g^2(N_c-1)\Gamma\bar{\lambda}}\right)\nonumber\\
+\delta^{ad}\delta^{bc}\!\left( -\frac{N_c^2-4}{2N_c^2}e^{-g^2N_c\Gamma\bar{\lambda}}-\frac{2}{N_c^2}e^{-g^2\frac{N_c}{2}\Gamma\bar{\lambda}}+\frac{N_c+2}{4N_c}e^{-g^2(N_c+1)\Gamma\bar{\lambda}}+\frac{N_c-2}{4N_c}e^{-g^2(N_c-1)\Gamma\bar{\lambda}}\right)\nonumber\\
+d^{acn}d^{bdn}\!\left(-\frac{1}{N_c}e^{-g^2N_c\Gamma\bar{\lambda}}+\frac{1}{N_c}e^{-g^2\frac{N_c}{2}\Gamma\bar{\lambda}}+\frac{1}{4}e^{-g^2(N_c+1)\Gamma\bar{\lambda}}-\frac{1}{4}e^{-g^2(N_c-1)\Gamma\bar{\lambda}} \right)\nonumber\\
+d^{abn}d^{cdn}\!\left(\frac{N_c}{N_c^2-4}e^{-g^2\frac{N_c}{2}\Gamma\bar{\lambda}}-\frac{N_c+4}{4(N_c+2)}e^{-g^2(N_c+1)\Gamma\bar{\lambda}}+\frac{N_c-4}{4(N_c-2)}e^{-g^2(N_c-1)\Gamma\bar{\lambda}}\right)\nonumber\\
+d^{adn}d^{bcn}\!\left( \frac{1}{N_c}e^{-g^2N_c\Gamma\bar{\lambda}}-\frac{1}{N_c}e^{-g^2\frac{N_c}{2}\Gamma\bar{\lambda}}+\frac{1}{4}e^{-g^2(N_c+1)\Gamma\bar{\lambda}}-\frac{1}{4}e^{-g^2(N_c-1)\Gamma\bar{\lambda}}\right)\!,
\end{align}
where we still omit dependencies. Note that in order to use these results in the calculation of $D^{ij;kl}_{ab;cd}(x_{\pperp},y_{\pperp},x_{\pperp},y_{\pperp})$ and $C^{ij;kl}_{ab;cd}(x_{\pperp},y_{\pperp},x_{\pperp},y_{\pperp})$ we need to permute the indices $b$ and $c$ (see \eqref{DisconResult}).

\acknowledgments

We thank Tuomas Lappi and Jean-Yves Ollitrault for illuminating discussions on various aspects of this project. The work of JLA and PGR is partially funded by a FP7-PEOPLE-2013-CIG Grant of the European Commission, reference QCDense/631558, and by the MINECO project FPA2016-78220 of the Spanish Government. The work of CM was supported in part by the Agence Nationale de la Recherche under the project ANR-16-CE31-0019-02. PGR also acknowledges financial support from the `La Caixa' Banking Foundation.

%\paragraph{Note added.} This is also a good position for notes added after the paper has been written.

\bibliographystyle{JHEP-2modM}
%\bibliography{Refs}{}

\end{document}